\documentclass{article}

\usepackage{arxiv}
\usepackage[utf8]{inputenc} 
\usepackage[T1]{fontenc}    
\usepackage{hyperref}       
\usepackage{url}            
\usepackage{booktabs}       
\usepackage{amsfonts}       
\usepackage{nicefrac}       
\usepackage{microtype}      
\usepackage{lipsum}		
\usepackage{graphicx}
\usepackage{natbib}
\usepackage{doi}

\usepackage{amsmath}
\DeclareMathOperator*{\argmax}{arg\,max}

\usepackage{multirow}
\usepackage{algorithm}
\usepackage{algpseudocode}
\algnewcommand{\Inputs}[1]{%
  \State \textbf{Inputs:}
  \Statex \hspace*{\algorithmicindent}\parbox[t]{.8\linewidth}{\raggedright #1}
}
\algnewcommand{\Outputs}[1]{%
  \State \textbf{Outputs:}
  \Statex \hspace*{\algorithmicindent}\parbox[t]{.8\linewidth}{\raggedright #1}
}
\algnewcommand{\Initialize}[1]{%
  \State \textbf{Initialize:}
  \Statex \hspace*{\algorithmicindent}\parbox[t]{.8\linewidth}{\raggedright #1}
}
\newcommand*\xor{\oplus}
\newtheorem{definition}{Definition}

\newtheorem{lemma}{Lemma}
\newtheorem{proof}{Proof}

\title{Privacy-Preserving Record Linkage for Cardinality Counting}

\author{ Nan Wu\\
	Macquarie University, CSIRO's Data61\\
	Sydney\\
	Australia \\
	\texttt{nan.wu5@hdr.mq.edu.au} \\
	\And
	Dinusha Vatsalan \\
	Macquarie University\\
	Sydney\\
	Australia \\
	\texttt{dinusha.vatsalan@mq.edu.au} \\
	\And
	Mohamed Ali Kaafar \\
	Macquarie University\\
	Sydney\\
	Australia \\
	\texttt{dali.kaafar@mq.edu.au} \\
	\And
	Sanath Kumar Ramesh \\
	Open Treatments Foundations \\
	Seattle \\
        United States\\
	\texttt{sanath@gpx4.org} \\
}

\begin{document}
\maketitle

\begin{abstract}
    
{

Several applications require counting the number of distinct items in the data, which is known as the cardinality counting problem. 
Example applications include health applications such as rare disease patients counting for adequate awareness and funding, and counting the number of cases of a new disease for outbreak detection, marketing applications such as counting the visibility reached for a new product, 
and cybersecurity applications such as tracking the number of unique views of social media posts. 
The data needed for the counting is 
however 
often personal and sensitive, and need to be processed using privacy-preserving techniques.
The quality of data in different databases, for example typos, errors and variations, poses additional challenges for accurate cardinality estimation.
While privacy-preserving cardinality counting has gained much attention in the recent times and a few privacy-preserving algorithms have been developed for cardinality estimation, 
no work has so far been done on privacy-preserving cardinality counting using record linkage techniques with fuzzy matching and provable privacy guarantees. 
We propose a novel privacy-preserving record linkage algorithm using unsupervised clustering techniques to link and count the cardinality of individuals in multiple datasets without compromising their privacy or identity. 
In addition, existing Elbow methods to find the optimal number of clusters as the cardinality are far from accurate as they do not take into account the purity and completeness of generated clusters. We propose a novel method to find the optimal number of clusters in unsupervised learning. Our experimental results on real and synthetic datasets are highly promising in terms of significantly smaller error rate of less than $0.1$ with a privacy budget $\epsilon=1.0$ compared to the state-of-the-art fuzzy matching and clustering method. 
}

\end{abstract}





\maketitle

\section{Introduction}
\label{sec:Introduction}

Cardinality counting problem has become of tremendous interest in many different applications to enable a variety of analytics of dispersed data. However, the privacy concerns of sharing or revealing individuals' data containing personal information for analytics purposes require privacy-preserving processing of counting. In most cases, the records of the same individual in different databases do not contain a unique identifier and and the quasi identifiers in the records, such as names, addresses, and ages, are often prone to data errors, inconsistencies and variations. Accurately estimating the cardinality of individuals or items represented by records in multiple different databases without compromising the privacy of the individuals is hence a challenging research problem. 

Gaining insight into the number of unique records from multiple data sources is crucial in many  applications. 
A promising real-world application is rare disease patients counting. Rare diseases in general do not receive sufficient funding for treatment~\cite{pmid29866013} and this disparity is created in part because funders measure the impact of their investments based on the size of patient population affected by a given disease. Unfortunately, for a majority of rare diseases, this data is at best a wild guess and at worst non-existent. 
Another example in the health domain is disease outbreak detection where the number of unique cases of a new disease needs to be continuously monitored from multiple different hospitals and clinics to predict the likelihood of an outbreak and to make preventive measures. 

Similarly, national security or cybersecurity applications monitor the number of views of videos or posts in online or social media in order to predict the potential threats of any video/post that becomes viral within a short time period (for example, fake news with phishing links~\cite{Sha17}) and make any timely decision.
Online businesses need to monitor the number of unique views by customers of a new product in order to make decisions on the marketing strategies to manage the marketing costs.
Web search log analysis may require calculating the number of distinct queries in a list of queries from many users (e.g., the number of distinct queries made to a search engine over a week) to improve the performance of the search engine in terms of estimating the selectivity of queries and designing good strategies for executing a query~\cite{Jan06,Wha90,Bar02a}. Social game industry and e-commerce applications use count distinct metrics, such as the daily active users (DAU) and monthly active users (MAU) metrics~\cite{Wan17}, to estimate the workload for those online applications. 
In all these example applications, the data needed to derive such insights is personal and sensitive, and must be processed private.

While there have been several methods proposed for the cardinality counting problem in general~\cite{Gol19,Fla07,Bar02a,Cha10,Heu13,Ert17}, privacy aspects of cardinality counting 
have only recently received attention in the research literature. Some recent works developed privacy-preserving algorithms for cardinality counting using different probabilistic data structures (KMV, FM-Sketch, or HyperLogLog)~\cite{Sta17,Spa18,Von19}. 
A recent study has shown that probabilistic cardinality estimators like HyperLogLog do not preserve privacy as achieving accurate and private cardinality estimation is impossible, and therefore they can be sensitive as raw data~\cite{Des19}. 
In addition, they are not robust or tolerant to errors and variations in data.
Privacy-preserving record linkage is hence required to link or de-duplicate records corresponding to the same individual based on fuzzy matching of personal identifying information (PII) contained in the quasi-identifiers (e.g. names and addresses) to count the cardinality of individuals. 

In this work, we propose a novel privacy-preserving record linkage algorithm for linking and counting unique individuals or items from multiple databases using a combination of Bloom filter encoding, local Differential privacy, and machine learning techniques.
Specifically, the database owners locally encode and perturb the PII in their records using Bloom filters and local Differential privacy. The encoded and perturbed records from all the databases are then input to a clustering algorithm that aims to link and group records corresponding to the same individual/item into the same cluster and different individuals/item into different clusters.

The optimal number of clusters is then computed to calculate the cardinality of records.
Since ground-truth data is not available in real applications and is not trivial to manually label data due to privacy and confidentiality concerns, finding the optimal number of clusters for such unsupervised machine learning tasks is highly challenging. Existing Elbow methods based on metrics like silhoutte coefficient and Calinski-Harabasz score measure the inter and intra cluster distances to find the optimal number of clusters~\cite{Rou87,Din19,Cal74,Wan19}. However, they are not accurate and optimal, especially for linking or deduplicating records, and thereby counting the correct cardinality. The main limitation is that they do not account for the purity and completeness of clusters which are necessary for accurate record linkage. 
Hence, we calculate the optimal number of clusters or cardinality by proposing a novel method to measure the purity and completeness of generated clusters. 
While we propose an algorithm for the distinct-counting problem, our proposed method for finding the optimal number of clusters can be used with any unsupervised clustering techniques that do not have labelled data for fine-tuning and/or evaluation. 

\noindent
The main contributions of this paper are:
\noindent
\begin{enumerate}
    \item We study the problem of privacy-preserving cardinality counting of individuals or items from multiple different databases in the presence of data errors and variations. 
    \item We introduce a novel privacy-preserving record linkage algorithm for cardinality counting with provable privacy guarantees. Our algorithm uses Bloom filter encoding and local Differential privacy for data encoding and unsupervised clustering on the encoded data to estimate the cardinality.
    \item We propose a novel clustering algorithm to find the optimal number of clusters in the absence of labelled data to predict the accurate cardinality. We develop two variations of our clustering method and evaluate their accuracies for cardinality estimation.
    \item We provide formal proof of privacy guarantees of our proposed method and theoretical analysis of utility of our proposed method. 
    \item We conduct experimental evaluation on real and synthetic North Carolina voter registration (NCVR) datasets to validate the accuracy of our method. Since existing cardinality estimators do not allow fuzzy matching for cardinality counting, we compare only with a state-of-the-art baseline method for fuzzy matching and clustering~\cite{Rou87} to provide a fair comparison. The experimental results show that our methods can achieve a very small error rate closer to $0.0$ with a small privacy budget of $\epsilon=1.0$ or $\epsilon=2.0$ even on highly corrupted datasets, and significantly outperform the existing methods.
\end{enumerate}

\noindent
\textbf{Outline:}
We provide preliminaries in the following section and describe our methodology in Section~\ref{sec:methodology}. In Section~\ref{sec:experiments} we present the results of our experimental study and in Section~\ref{sec:rw} we review the literature of privacy-preserving counting techniques. Finally we summarise, discuss limitations, and provide directions to future research in Section~\ref{sec:conclusion}.

\section{Background}
\label{sec:preliminaries}

In this section, we describe the preliminaries 
of this work. 
We first provide preliminaries for the Bloom filter encoding in Section~\ref{subsec:BF}.
We then describe the system architecture and threat model of the research problem we address in this paper in Section~\ref{subsec:threat}, and finally we describe Differential privacy in Section~\ref{subsec:DP}.

\begin{figure*}[!th]
    \centering
    \includegraphics[width= 0.45\linewidth, keepaspectratio]{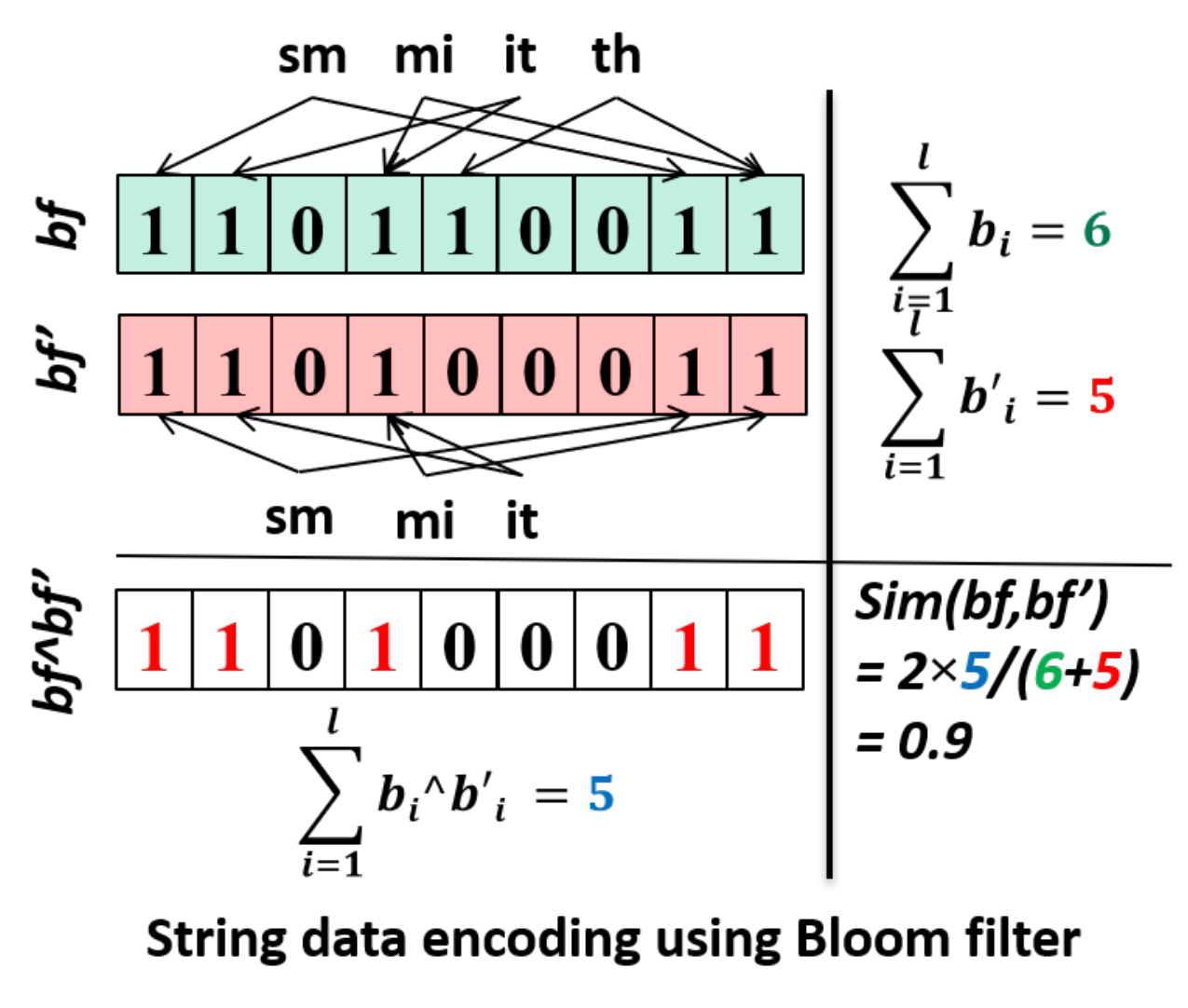}
    \includegraphics[width= 0.45\linewidth, keepaspectratio]{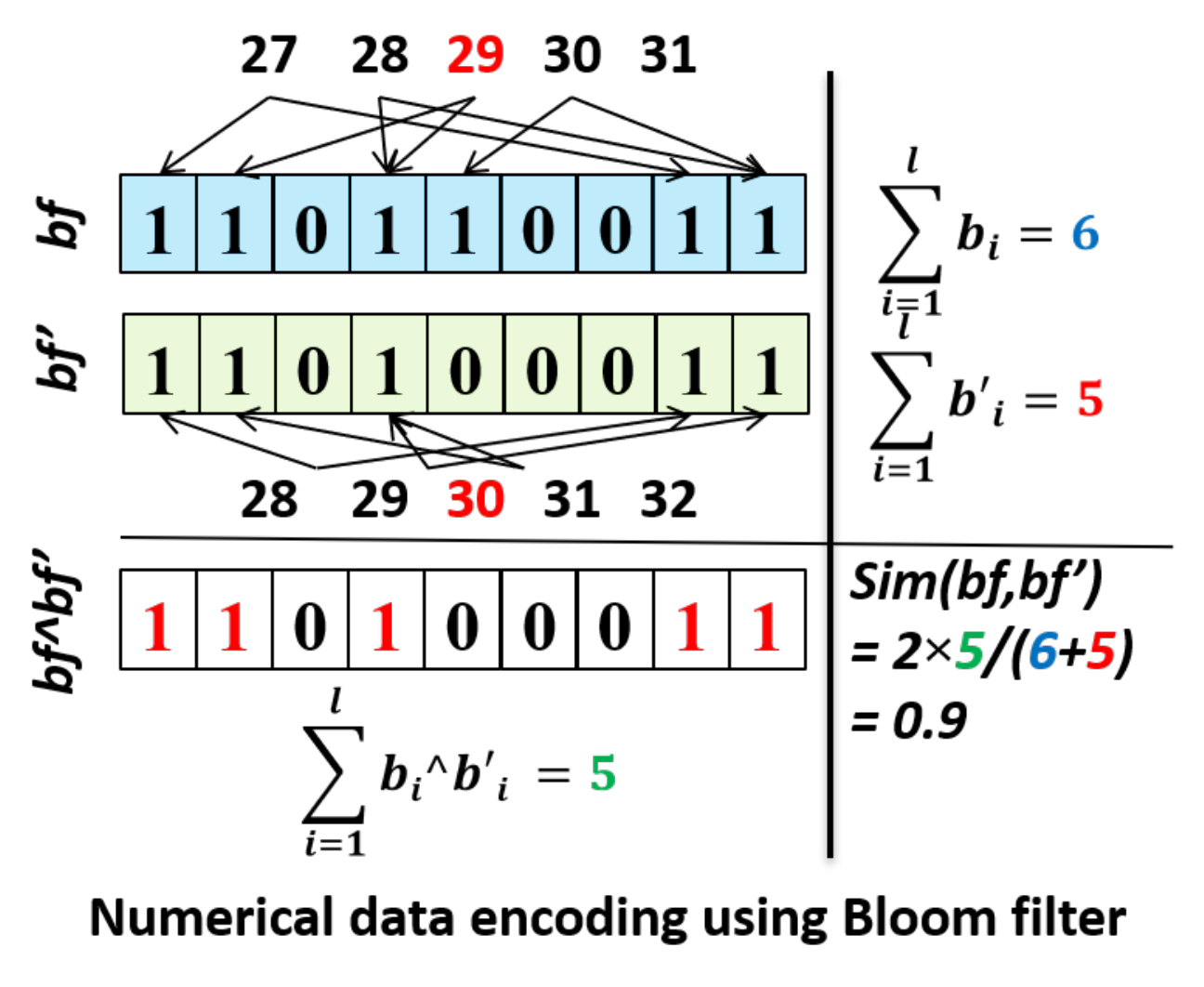}

    \caption{Bloom filter encoding of string values (left) and numerical values (right)~\cite{Sch15,Vat16}, and fuzzy matching using Dice-coefficient similarity function, as described in Section~\ref{subsec:BF}}
    \label{fig:BFs}
\end{figure*}

\subsection{Bloom filter encoding}
\label{subsec:BF}

Bloom filters are probabilistic data structures that are highly efficient for storing, processing, and computation. Essentially, Bloom filters are bit vectors that initially contain $0$ in all the bit positions. $k$ independent hash functions $h_i(\cdot)$ (with $1 \le i \le k$) are used to hash-map an element $x$ by setting the corresponding bit positions in the Bloom filter $b$ to $1$ (i.e. $\forall_i~ b[h_i(x)] = 1)$. A Bloom filter allows a tunable false positive rate $fpr$ so that a query returns either “definitely not” (with no error), or “probably yes” (with probability $fpr$ of being wrong). The lower $fpr$ is, the better utility is, but the more space the filter requires.
The false positive probability for encoding $n$ elements into a Bloom filter of length $\ell$ bits using $k$ hash functions is $fpr = (1 - e^{-kn/\ell})^k$, which is controllable by tuning the parameters $k$ and $\ell$. 

The main feature of Bloom filter encoding that makes it applicable to efficient fuzzy matching of encoded records is that it preserves the similarity/distance between records in the Bloom filter space (with a negligible utility loss)~\cite{Sch15,Vat16}.
For example, with string values the $q$-grams (sub-strings of length
$q$) of string values can be hash-mapped into the Bloom filter $bf$ using $k$ independent hash
functions~\cite{Sch15}, 
while for numerical values, the neighbouring values (within a certain interval to allow fuzzy matching) of values can be hash-mapped into the Bloom filter~\cite{Vat16}. Fig.~\ref{fig:BFs} illustrates an example of fuzzy matching of string and numerical values using Bloom filters~\cite{Sch15,Vat16}.

The matching of Bloom filters can be determined by calculating the similarity value using a token-based similarity function, such as Jaccard, Dice, or Hamming~\cite{Vat13}. For example, Dice-coefficient similarity metric is calculated for the example pairs of Bloom filter encoded strings and integers in Fig.~\ref{fig:BFs} as $2 \times \frac{\sum(bf_1 \cap bf_2)}{\sum(bf_1)+\sum(bf_2)}$, where $bf_1$ and $bf_2$ are the two Bloom filters. Collision of different elements being mapped to the same bit position can occur during the hash-mapping (depending on the parameter setting), resulting in false positives with matching Bloom filter encoded records. However, with appropriate parameter settings, Bloom filters have shown to be successful in providing high matching results while being highly efficient~\cite{Sch15,Ran13,Vat16}.

\begin{figure}[!t]
    \centering
\includegraphics[width=0.6\linewidth, keepaspectratio]{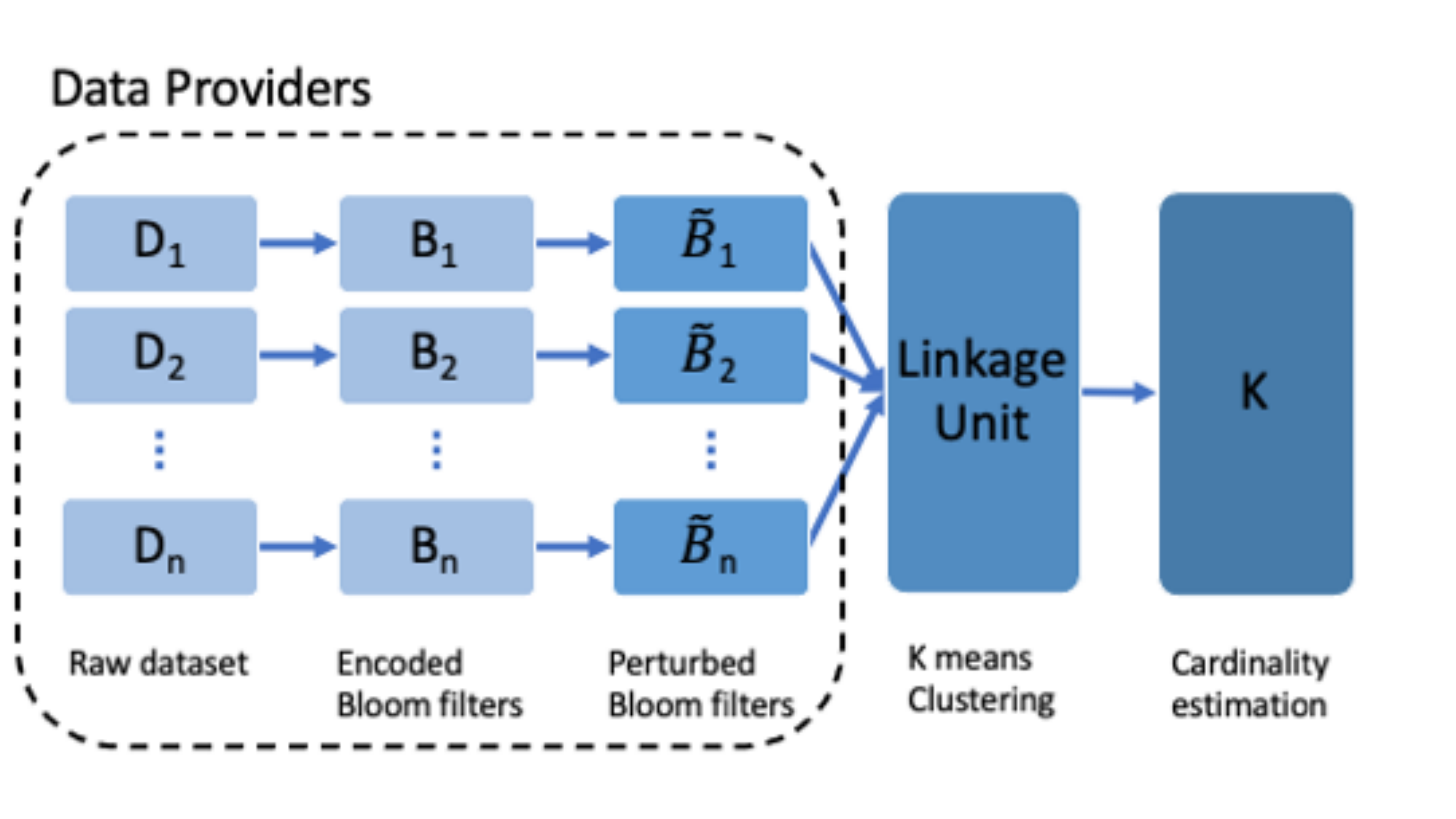}
\caption{An outline of our system model for privacy-preserving cardinality estimation.}
    \label{fig:sys_model}
\end{figure}

\subsection{System architecture and threat model}
\label{subsec:threat}

The system architecture of our proposed method for privacy-preserving cardinality counting is illustrated in Fig.~\ref{fig:sys_model}. At each data owner side, the PII in records are encoded into Bloom filters first and then perturbed using local Differential Privacy. The encoded and perturbed records from multiple different data owners are sent to a linkage unit that applies our proposed clustering algorithm on the Bloom filters such that similar Bloom filters corresponding to the same individual/patient are grouped into one cluster. The optimal number of clusters is estimated as the cardinality of records from multiple databases and reported.

Only the Bloom filters are shared with the linkage unit. Bloom filter encoding does provide some inherent privacy guarantees due to the collision of different elements being hash-mapped to same bits in the Bloom filters, providing uncertainty in decoding. However, as shown in the recent research, Bloom filters can be vulnerable to cryptanalysis attacks that map bits to $q$-grams or elements based on the frequency of bits or bit patterns~\cite{Chr18,Chr18b}. As another layer of privacy to provide provable privacy guarantees, we combine Bloom filter encoding with local Differential privacy. Differential privacy noise is added to the Bloom filters using the randomised response technique, as will be described in detail in the following sub-section, in order to make the bits in the Bloom filters Differentially private so that bits cannot be distinguished based on their presence or frequency information.

\subsection{Differential Privacy}
\label{subsec:DP}

Differential privacy~\cite{Dwo06,Dwo08,Dwo10} 
guarantees for each individual
in a dataset that any information that could be discovered about an individual with their data in
the dataset could also, with high probability, be discovered without their data in the
dataset. That is, the output of any query $f$ performed on dataset $x$ will be
indistinguishable from the output of the same query $f$ performed on dataset $y$, where $y$
differs from $x$ by at most one record (the record of any individual). 
Moreover, it promises that any supplementary data an adversary might have about the individual is irrelevant; the adversary is unable to identify any additional information about an individual from the data regardless of the auxiliary knowledge about the individual with a high probability.


\begin{definition}
[Differential Privacy~\cite{Dwo06}] A randomized 
function $\mathcal{A}$
(i.e. a function with a randomized component) is $\epsilon$-Differentially private if for all outputs
$y \subseteq Range(A)$ and for all data $x$, $x' \in \mathcal{D}^n$ such that $||x - x'||_1 \le 1$:

\begin{equation}
    Pr(\mathcal{A}(x) = y) \le e^{\epsilon} \times Pr(\mathcal{A}(x') = y).
\end{equation}

\end{definition}

Local Differential Privacy (LDP) is a Differential privacy model developed specifically to provide guarantees such that even if an adversary has access to the personal responses of an individual in the dataset, the adversary is still unable to learn additional information about the individual from the personal data with high probability~\cite{Evf03}. LDP has become the de-facto privacy standard around the world in recent years, with the technology companies Google and Apple implementing LDP in their latest operating systems and applications~\cite{Erl14,Dp17,Gre16}.

A widely used mechanism specifically for designing LDP algorithms is the randomized response technique~\cite{War65}. The primary idea is that the data owners respond to binary questions (e.g. $0$ or $1$) in a randomized manner. 
We use randomised response technique to add noise to Bloom filters such that the Bloom filters are Differentially private and robust against cryptanalysis attacks that exploit the presence or frequency of bits in the Bloom filters.
%
%
%


\section{Methodology}
\label{sec:methodology}

In this section, we describe our proposed method for privacy-preserving distinct-counting of individuals/entities from multiple different databases.
Our method consists of two main modules: 1) data encoding and 2) linkage and clustering. The former is conducted at the local data owner side, and the latter is conducted by the central linkage unit.

\begin{algorithm}[!t]
\footnotesize
    \caption{\textit{Privacy-preserving Bloom filter encoding with $\epsilon$-local Differential Privacy}\label{alg:pp bf encoding}}
    \begin{algorithmic}[1]
    \Inputs{Raw dataset from one data provider: $D$,\\ Privacy budget: $\epsilon$}
    \Outputs{Perturbed Bloom filters: $B'$}
    \Initialize{$B' \gets \Phi$ }
    \For{$i=1,\cdots,n_{|D|}$}\Comment{Do for each record in raw dataset}
    \State $bf_i = \text{encode}(\text{record}_i)$, $\text{record}_i \in D$
    \For{$j=1,\cdots, \ell$}\Comment{For each bit in the Bloom filter}
    \State $bf_i^\prime \gets \phi$
    \State $\eta = \frac{1}{1+e^\epsilon}$
    \State $p=random[0,1]$
    \If{$p\le \eta$} \Comment{flip $b_j$ with probability $\eta$}
       \State{ $b_j^\prime = b_j$}
    \Else   
       \State{$b_j^\prime = b_j\xor 1$}
    \EndIf
    \State $bf_i^\prime=bf_i^\prime\cup b_j^\prime$
    \EndFor
    \State $B^\prime = B\cup bf_i^\prime$
    \EndFor
    \end{algorithmic}
\end{algorithm}

\subsection{Data encoding}
Data providers encode their datasets into Bloom filters, one Bloom filter per record. The PII in each of the records are hash-mapped into one record-level Bloom filter per record, as described in Section~\ref{subsec:BF}. Then, the Bloom filters are perturbed using the randomised response method to make the Bloom filters Differentially private.
Randomized response method is utilized to provide $\epsilon$-local Differential privacy (LDP) guarantees by flipping each bit in the Bloom filter of encoded records locally by the data providers with probability $\eta = \frac{1}{1+e^{\epsilon}}$. Then, the perturbed Bloom filters are sent to the linkage unit for linking and clustering. 

\begin{definition}[Adjacent Bloom filters]
Adjacent Bloom filters are two Bloom filters $bf$ and $bf'$ of length $\ell$ bits that differ by only one bit position, i.e. $\forall_{i, 1 \le i \le \ell~and~ i \neq j} b_i = b'_i~ and~ b_j \neq b'_j$.
\end{definition}

\begin{lemma}[$\epsilon$-LDP for Bloom filters]
Flipping the bits in Bloom filters with 
$\frac{1}{1+e^{\epsilon}}$ probability makes the bits in the Bloom filters $\epsilon$-local Differentially private.
\end{lemma}

\begin{proof}
Let us assume two adjacent Bloom filters $bf$ and $bf'$ 
of two records, each containing $\ell$ bits that differ in only one bit position $j$. 
Let $\mathcal{A}: \{0, 1\}^\ell \rightarrow \{0, 1\}^\ell$ be a random noise function
such that $\mathcal{A}(i) = i$ with probability 
$\frac{e^{\epsilon}}{1+e^{\epsilon}}$, and 
$\mathcal{A}(i) = 1-i$ with probability 
$\frac{1}{1+e^{\epsilon}}$, where $i \in \{0,1\}$.

This gives us the expression:

{\footnotesize
\begin{equation}
    \frac{Pr[\mathcal{A}(bf, \epsilon) = \tilde{v}]}{Pr[\mathcal{A}(bf', \epsilon) = \tilde{v}]} = \prod_{i=1}^\ell{l} \frac{Pr[\mathcal{A}(b_i) = \tilde{v_i}]}{Pr[\mathcal{A}(b'_i) = \tilde{v_i}]}
\label{eq:2}
\end{equation}}

\noindent Note that any two adjacent Bloom filters $bf, bf' \in \{0,1\}^\ell$ can only differ in one bit position. Without loss of generality, let us assume that the differing bit position is the first bit position ($j=1$) in the two Bloom filters, i.e. $b_1 \neq b'_1$ and $b_i = b'_i$ with $2 \le i \le \ell$.

This simplifies the ratio in~(\ref{eq:2}) by considering only the first bit position.

{\footnotesize
\begin{equation}
    \frac{Pr[\mathcal{A}(bf, \epsilon) = \tilde{v}]}{Pr[\mathcal{A}(bf', \epsilon) = \tilde{v}]} = \frac{Pr[\mathcal{A}(b_j) = \tilde{v_j}]}{Pr[\mathcal{A}(b'_j) = \tilde{v_j}]},
\label{eq:30}
\end{equation}}

where $j=1$. This ratio is maximised when $j^{th}$ bit position is flipped in only one of the two Bloom filters (maximum ratio).

{\footnotesize
\begin{equation}
    e^{-\epsilon} \le 
    \frac{Pr[\mathcal{A}(bf, \epsilon) = \tilde{v}]}
    {Pr[\mathcal{A}(bf', \epsilon) = \tilde{v}]} \le  \frac{\frac{e^{\epsilon}}{1+e^{\epsilon}}}{\frac{1}{1+e^{\epsilon}}}
    \le e^\epsilon
\label{eq:3}
\end{equation}}

Bounding the above ratio, we get

{\footnotesize
\begin{equation}
    -\epsilon \leq \text{ln} \left(\frac{Pr[\mathcal{A}(bf, \epsilon) = 
    \tilde{v}]}{Pr[\mathcal{A}(bf', \epsilon) = \tilde{v}]} \right) \leq \epsilon
\end{equation}}

\end{proof}

By making the bits in the Bloom filters Differentially private, we make them robust against cryptanalysis attacks based on sensitive bits~\cite{Chr18b}.
The Bloom filter encoding function with local Differential privacy is outlined in Algorithm~\ref{alg:pp bf encoding}.
The local Differentially private Bloom filters of records are then sent to the linkage unit for clustering and calculating the unique individual counts based on the number of clusters. At the linkage unit, clustering algorithm is used, as will be described in detail in the following sub-section, to link and group records which are likely to correspond to the same entity into the same cluster. The number of clusters is the number of unique individuals across multiple datasets from different data providers.

\begin{figure}[!t]
    \centering
\includegraphics[width=0.6\linewidth, keepaspectratio]{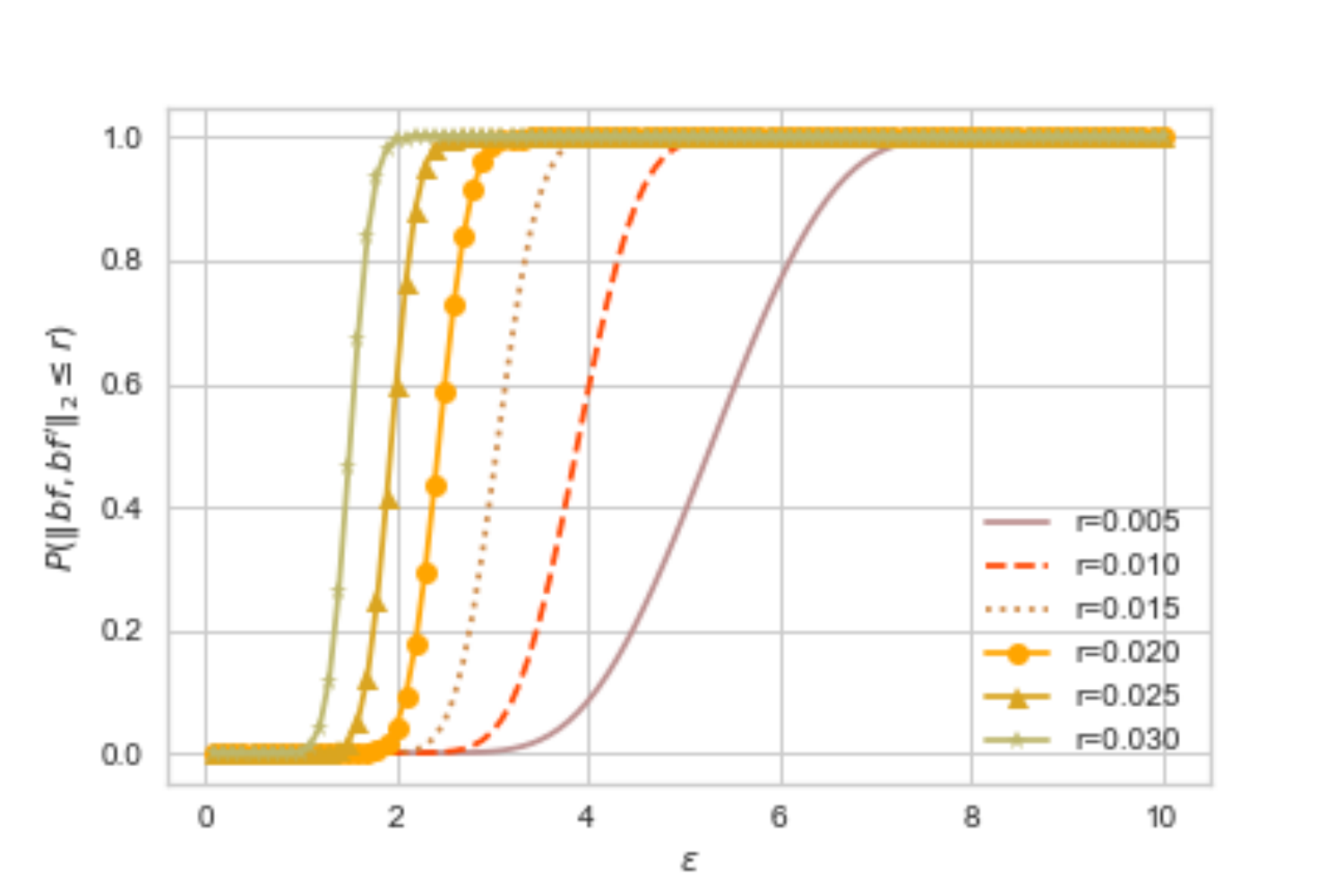}
\caption{Probability of $bf$ and $bf^\prime$ being grouped into the same cluster versus privacy budget $\epsilon$ in Equation~(\ref{equ:probability_epsilon}), with cluster size $r\in[0.005,0.010,0.015,0.020,0.025,0.030]$, and $\ell=200$.}
    \label{fig:theoretical_1}
\end{figure}

\begin{figure}[!t]
    \centering
\includegraphics[width=.6\linewidth, keepaspectratio]{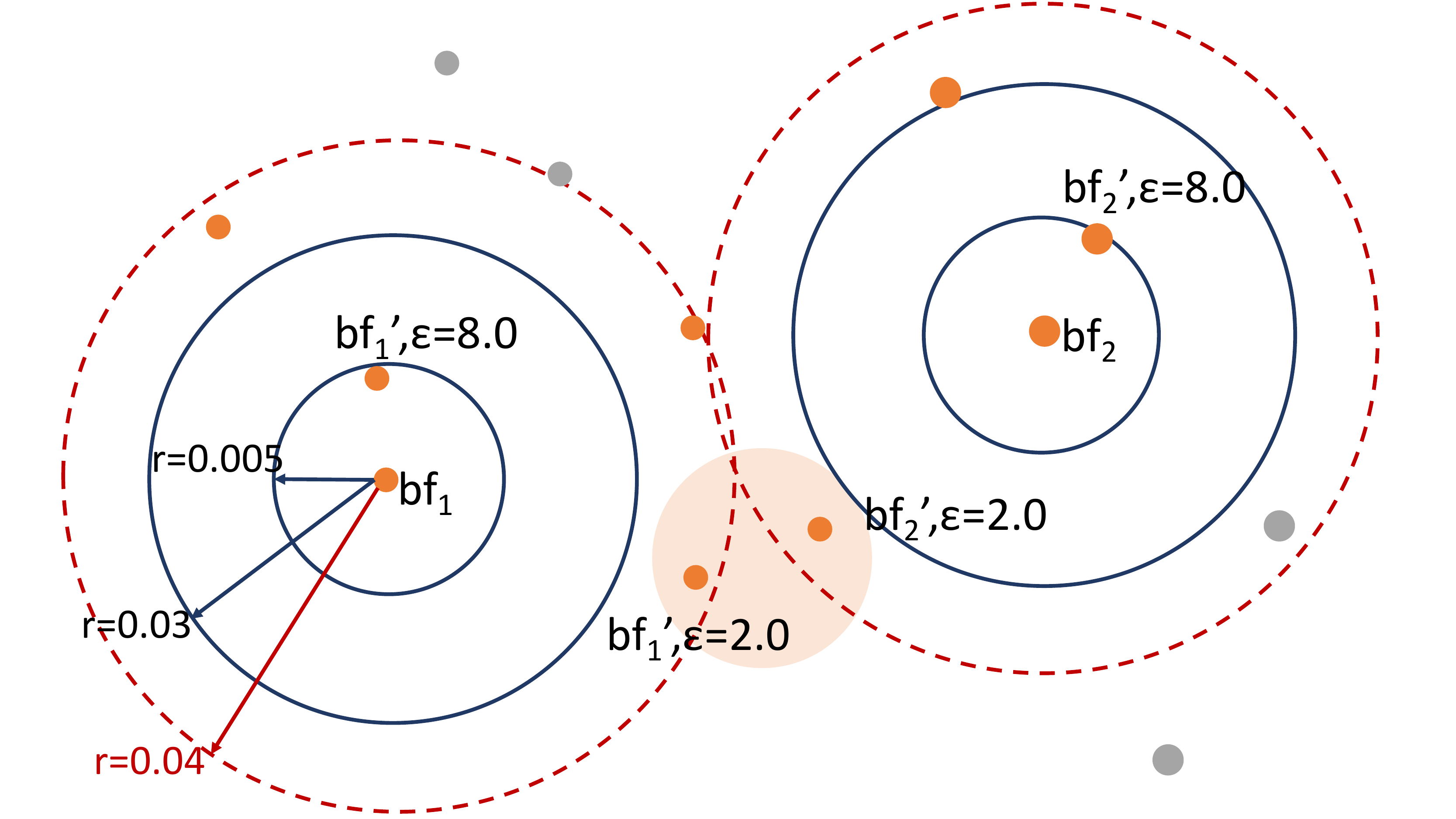}
\caption{Bloom filters belonging to the same entity being grouped into the same cluster versus privacy budget $\epsilon$, with cluster size $r\in[0.005, 0.030,0.040]$, and $\ell=200$.}
    \label{fig:theoretical_2}
\end{figure}


The perturbed Bloom filters are generated by randomly flipping each bit in the Bloom filters with the probability $\eta = \frac{1}{1+e^{\epsilon}}$. The Euclidean distance between an original Bloom filter $bf$ and its perturbed Bloom filter ${bf}^{\prime}$ is:

{
\footnotesize
\begin{align}
    \Vert bf,bf^\prime \Vert _2 = \sqrt{\sum_{i=1}^{\ell} (b_{i}-b^\prime_{i})^2},
\end{align}}

\noindent where $\ell$ is the length of a Bloom filter, $b_{i}$ is the $i^{th}$ bit in original Bloom filter $bf$ and $b^\prime_{i}$ is the $i^{th}$ bit in its perturbed Bloom filter $bf^\prime$. 
The value of $\Vert bf,bf^\prime\Vert_2^2$ follows Normal Distribution with a large number of $\ell$ and a probability $\eta$, $\Vert bf,bf^\prime\Vert_2^2\approx N(\mu, \sigma)$, $\mu = \ell \eta$, $\sigma=\sqrt{\ell \eta (1-\eta)}$. 
Assume if the Euclidean distance $\Vert bf,bf^\prime\Vert_2$ is less than a constant integer value (threshold) $r\in[0,\ell]$, then the original Bloom filter and the perturbed Bloom filter are grouped into same cluster. The probability of $bf$ and $bf^\prime$ being classified as the same entity is:

{\footnotesize
\begin{align}\label{equ:probability_epsilon}
    P(\Vert bf,bf^\prime\Vert_2\le r)&=P\left(\sqrt{\sum_{i=1}^{\ell} (b_{i}-b^\prime_{i})^2}\le r\right)\\\nonumber
    &=\frac{1}{2}+\frac{1}{2} \operatorname{erf} \left( \frac{r^2-\ell\eta}{\sqrt{2 \ell\eta (1-\eta)}}\right)
\end{align}}

As shown in 
Fig.~\ref{fig:theoretical_1}, with the increasing privacy budget $\epsilon$, the probability to flip bits $\eta$ decreases (less noise) and thus the probability of $bf$ and $bf^\prime$ being grouped into the same cluster increases. With a larger value of threshold $r$, a smaller value of privacy budget is required to keep $bf$ and $bf^\prime$ in the same cluster. 
There is a trade-off between privacy budget $\epsilon$ and distance between Bloom filters $bf_1$ and $bf_2$ from two unique person as illustrated in Fig.~\ref{fig:theoretical_2}. If $\epsilon$ is too small, for example $\epsilon<2$, $bf_1^\prime$ and $bf_2^\prime$ are grouped into same cluster. If $r$ is too large, the outlayers of $bf_1$ and $bf_2$ are overlapped. Therefore, it is a challenge to group Bloom filters from the same entity into unique clusters while to 
ensure Bloom filters from different entities are grouped into different clusters.

\begin{figure}[!t]
    \centering
\includegraphics[width=0.7\linewidth, keepaspectratio]{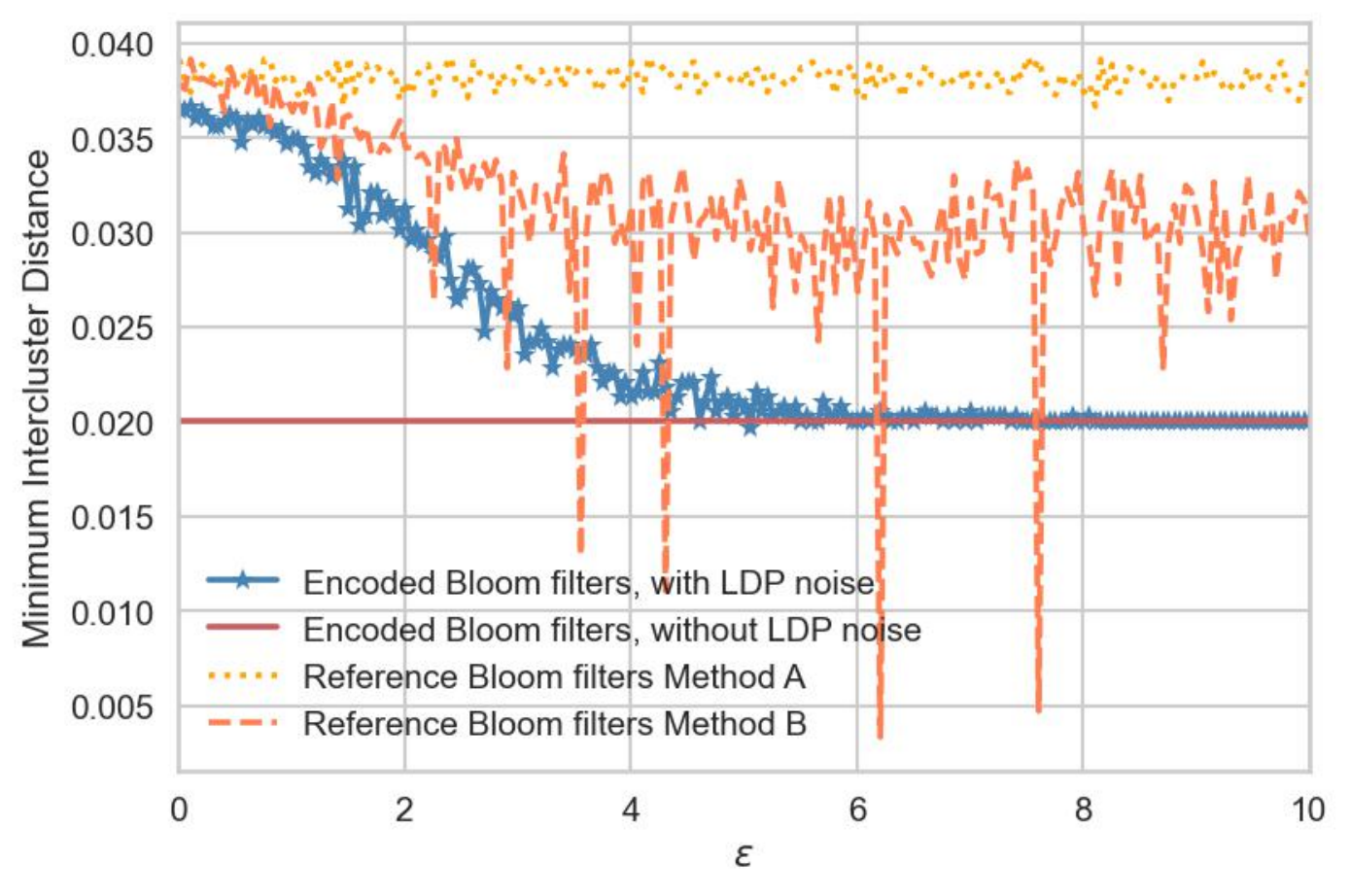}
\caption{Minimum inter-cluster distance of Bloom filters versus privacy budget $\epsilon$, $\ell=200$}
    \label{fig:intercluster}
\end{figure}

\subsection{Unsupervised clustering}
\label{subsec:clustering}

With the noisy encoded Bloom filter files from different data providers as input to the linkage unit, a clustering algorithm is used to group similar Bloom filters into clusters. Due to privacy constraints, it is not trivial to generate labelled data to train a supervised machine learning algorithm for cardinality estimation in the privacy-preserving context. 
Hence an unsupervised clustering algorithm is used to do the clustering without training labels, such as $k$-means clustering or Hierarchical clustering. In addition, Elbow methods with inter and/or intra-cluster distance metrics like Silhouette Score and Calinski-Harabasz score are used in the literature to find the optimal number of clusters $k$ in $k$-means clustering~\cite{Rou87,Din19,Cal74,Wan19}. However, the traditional Elbow methods are far from accurate due to the limitation of data distribution, impurity, incompleteness and uncertainty of generated clusters. 
Therefore, we propose a new algorithm to find the optimal $k$ value in the unsupervised $k$-means clustering.

A set of reference Bloom filters with known training labels is generated to evaluate the clustering performance. Our proposed clustering algorithm first generates random Bloom filters or randomly selects a subset of Bloom filters as reference Bloom filters and then generates corresponding dummy Bloom filters for each reference Bloom filter.
The reference Bloom filters can be generated in two methods:

\begin{itemize}
    \item Method A: randomly creates a number of fake Bloom filters
    \item Method B: selects a subset of all Bloom filters from different data providers.
\end{itemize}

\noindent As shown in Fig.~\ref{fig:intercluster}, the reference Bloom filters are designed to be distinguishable from original encoded Bloom filters. It is noted that when private budget is too small $\epsilon<1.0$, the data distribution of encoded Bloom filters with LDP noise applied is similar to pure noise as reference Bloom filters with Method A.

For each reference Bloom filter $bf_\text{ref}$, a number of dummy Bloom filters $b_{\text{ref},\text{dum}_i}$, $i\in [1,n_\text{dum}]$ are then generated by randomly flipping each bit in reference Bloom filter $bf_\text{ref}$ with the flipping probability $p_\text{flip}$.

The aim of these reference and dummy Bloom filters is to provide some form of labelled data to fine-tune and evaluate unsupervised clustering techniques. In real applications, like rare disease patient records linking, ground-truth data is not available and/or accessible due to privacy constraints. Current methods to evaluate and choose the optimal number of clusters $k$ in unsupervised clustering techniques, like $k$-means clustering, are based on the inter and intra similarities/distances between generated clusters, which do not provide an accurate method to evaluate how pure and complete are the generated clusters.

Our proposed method checks whether the reference Bloom filters are grouped with their corresponding dummy Bloom filters to evaluate the optimal $k$ in the $k$-means clustering algorithm. 
Each Bloom filter is classified into a cluster with a label $c\in[0,k-1]$.
In order to evaluate the clustering quality and find the best optimal $k$, we introduce a new purity function for each reference Bloom filter $i$ at cluster $c$:

{\footnotesize\begin{equation}
    purity_i = \frac{n_{i,\text{dum},c}}{n_{i,\text{dum}}+n_{c}-1-n_{i,\text{dum},c}},
    \label{eq:purity}
\end{equation}}

\noindent where $n_{i,\text{dum},c}$ is the number of dummy records for $i^{th}$ reference Bloom filter that are grouped in the same cluster with label $c$,  $n_{c}$ is the number of Bloom filters that are grouped in to the cluster with label $c$, $n_{i,\text{dum}}$ is the total number of dummy records for $i^{th}$ reference Bloom filter. This purity function measures how accurate the clustering is in terms of grouping all the dummy Bloom filters of each reference Bloom filter in to the same cluster as the reference Bloom filter.

Based on this purity function, we calculate the purity of all reference Bloom filters with different values of $k$, and then find the optimal $k$ as similar to the current Elbow methods with silhoutte score~\cite{Rou87} or Calinski-Harabasz~\cite{Cal74}. Our proposed clustering method is outlined in Algorithm~\ref{alg:proposed method A}.

\begin{algorithm}[!t]
\footnotesize
    \caption{\textit{Linking and clustering records for cardinality counting  }}\label{alg:proposed method A}
    \begin{algorithmic}[1]
    \Inputs{ Encoded noisy datasets from $\mathcal{N}$ multiple data providers: $B^\prime_i$, $i\in [1,\mathcal{N}]$, \\
    Flipping probability for generating dummy records for reference Bloom filters: $p_\text{flip}$}
    \Outputs{ $K^*$}
    \State Obtain $n_\text{ref}$
    \For{$i=1,\cdots,n_\text{ref}$}
    \State Create a reference Bloom filter $bf_\text{ref,i}$ \Comment{obtained from either Method A or Method B}
    \State $B_\text{ref} = B_\text{ref} \cup bf_\text{ref,i}$
    \State Obtain the number of dummy records required for reference Bloom filter $n_\text{ref,dum}$ 
    \For{$j=1,\cdots,n_\text{ref,dum}$}
    \State Create dummy record $bf_\text{ref,dum,j}$
    \State $B_\text{ref,dum}=B_\text{ref,dum}\cup bf_\text{ref,dum,j}$
    \EndFor
    \EndFor
    \State $X=B_\text{ref}+B_\text{ref,dum}+\sum_1^\mathcal{N} B_i $ \Comment{X is the training dataset}
    \For{$k=1,\cdots,n_{|D|}$}
    \State $k,X \to k-means$ and train
    \State Obtain $purity_i, \forall i\in [1,\cdots, n_\text{ref}]$ by  Equation~(\ref{eq:purity}) 
    \State $Purity_k=\sum_1^{n_\text{ref}} purity_i$
    \EndFor
    \State $K^*=\argmax_{k\in[1,\cdots,n_{|B_i|}]} Purity_k$
    \end{algorithmic}
\end{algorithm}

The optimal clustering outcome is subject to:

{
\footnotesize\begin{align}
&\Vert bf_{i},bf_{i}'\Vert _2 \le r, \forall i \in [1,\cdots,n_{|D|}], \\
&\Vert bf_{i},bf_{j}\Vert _2 > r, \forall i,j \in [1,\cdots,n_{|D|}], i\neq j.
\end{align}}

\noindent where $n_{|D|}$ is the size of all input datasets, $bf_{i}$ and $bf_{j}$ belong to any two unique entities in the dataset, and $bf_{i}$ and $bf_{i}^\prime$ belong to the same individual in the dataset. In the ideal case, the Bloom filters corresponding to the same individual are grouped into the same cluster, while the Bloom filters corresponding to two different individuals are grouped into different clusters.

\section{Experimental Evaluation}
\label{sec:experiments}

In this section we present and discuss the results of experimental study of our proposed method. 

\vspace{2.5mm}
\noindent
\textbf{Datasets:}
We used three sets of datasets extracted from the North Carolina Voter Registration (NCVR) database~\footnote{Available from
\url{http://dl.ncsbe.gov/data/}}. This database contains records of voters in the North Carolina State, USA. Ground-truth
is available based on the voter registration identifiers to evaluate the accuracy of our proposed cardinality estimator in our experiments. 
We used given name (string), surname (string), suburb (string), postcode (string), and gender (categorical) attributes as PII for the linkage. 
The ground-truth cardinality is $171$ in all three sets of datasets, i.e. the datasets contain records corresponding to $171$ unique voters.

\begin{figure*}[!ht]
\centering
 \includegraphics[width=0.32\textwidth]{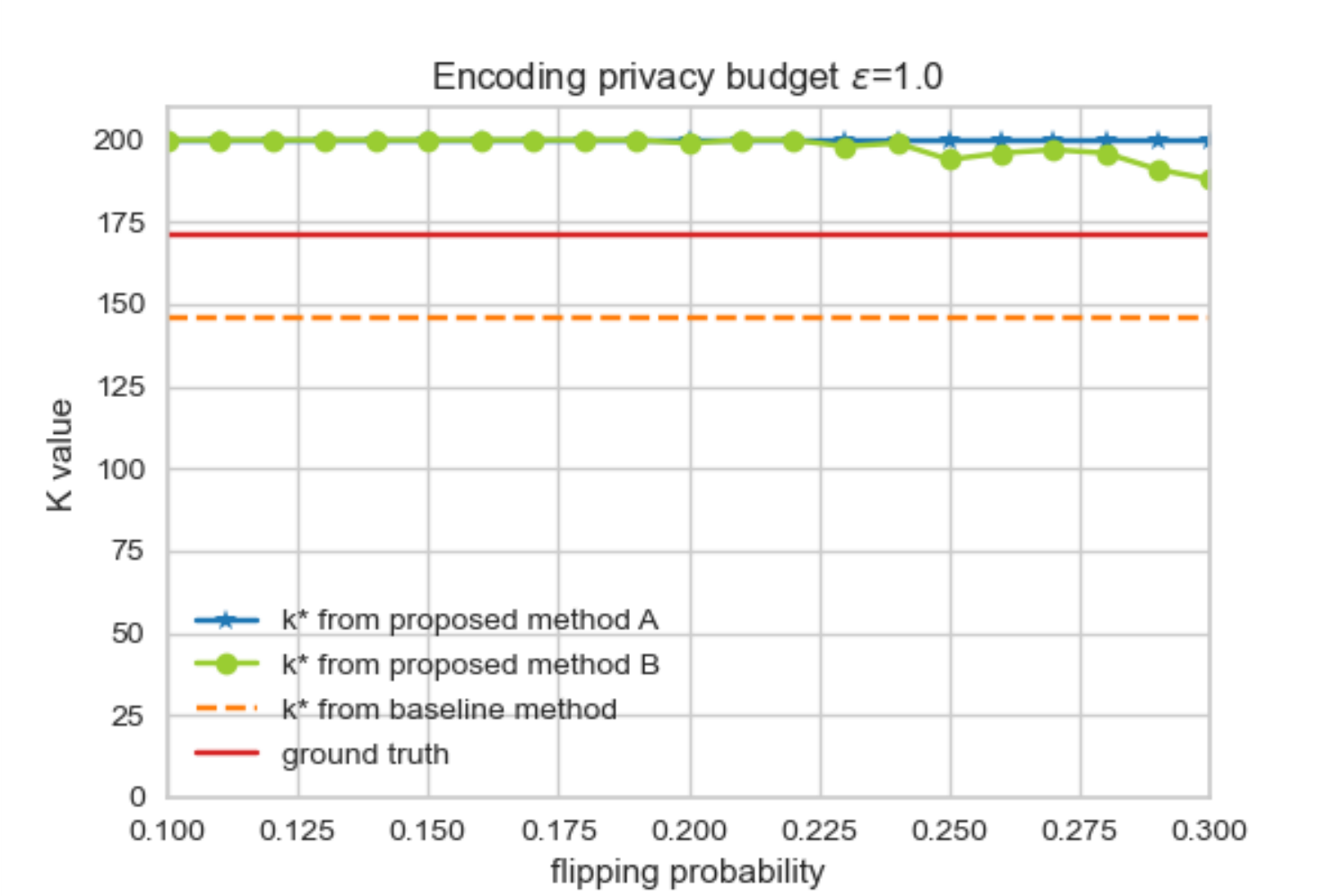}
 \includegraphics[width=0.32\textwidth]{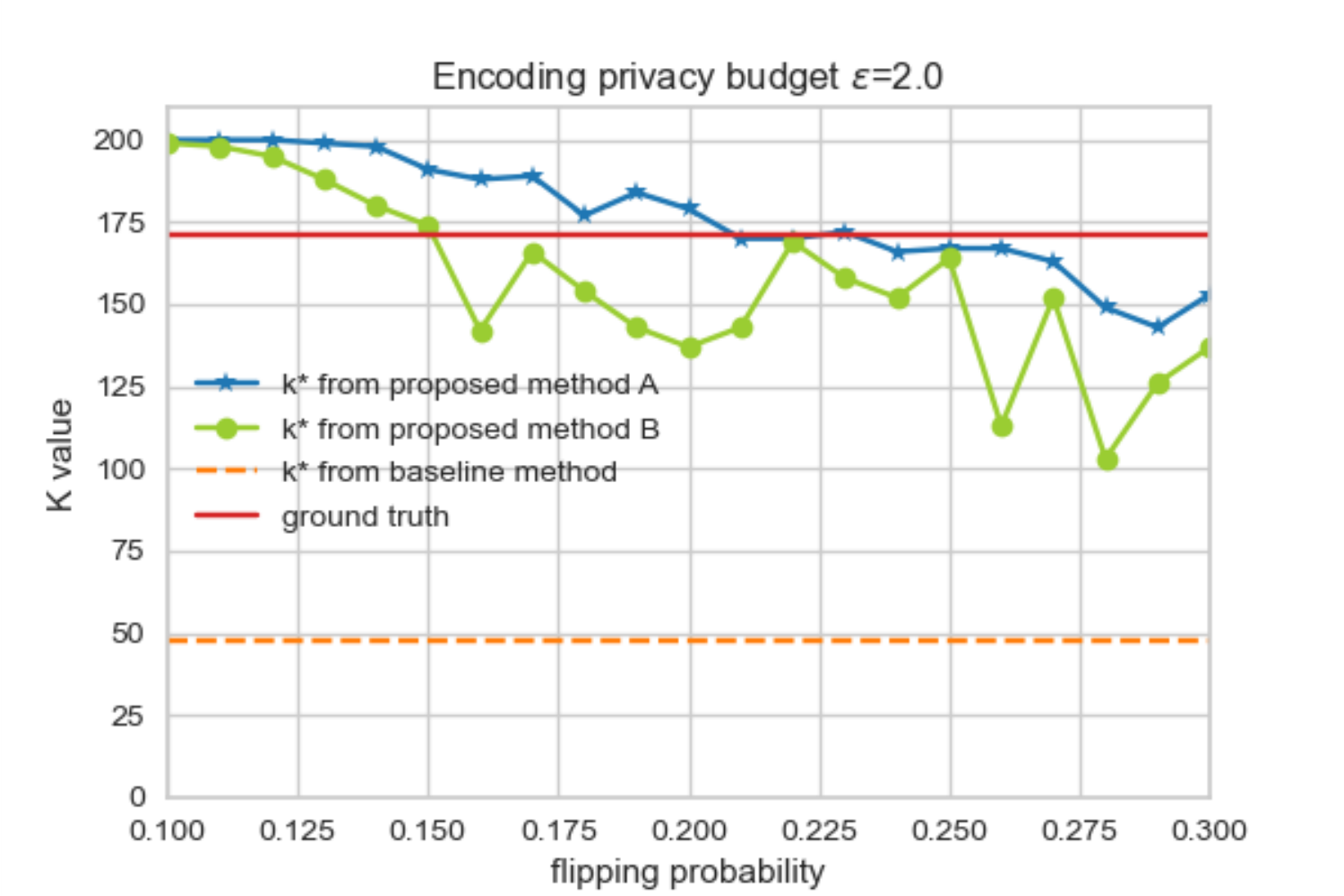}
 \includegraphics[width=0.32\textwidth]{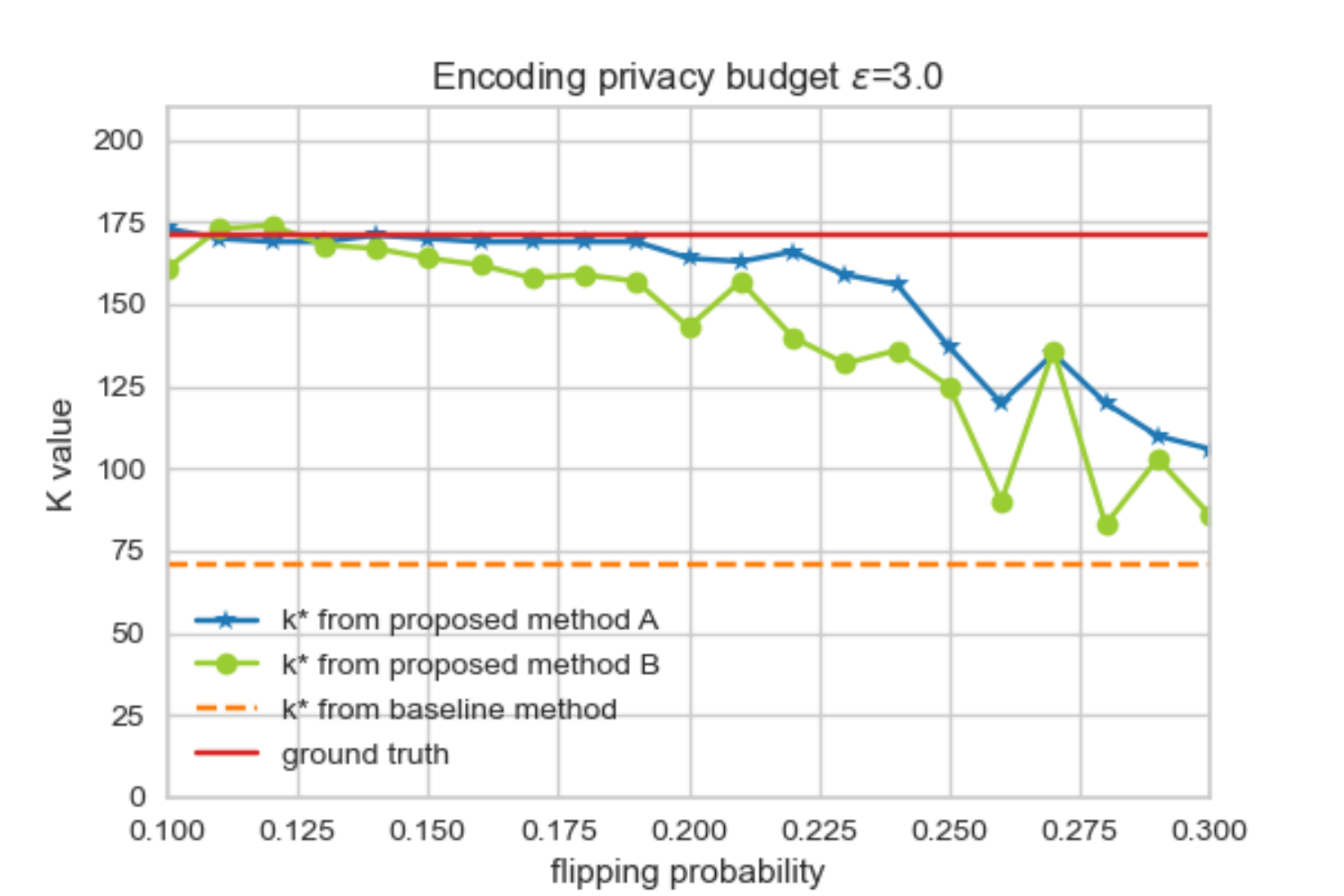}
 \includegraphics[width=0.32\textwidth]{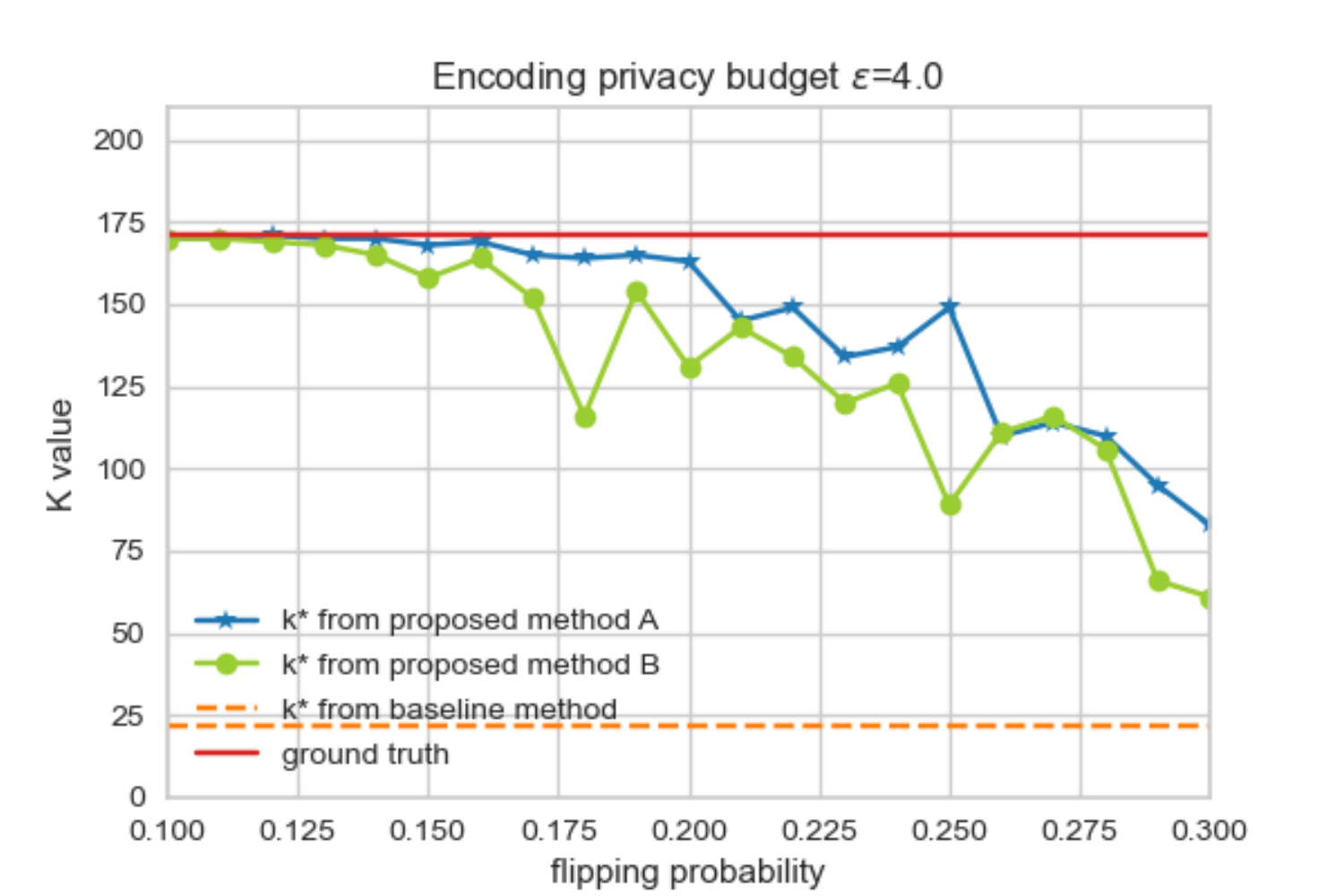}
 \includegraphics[width=0.32\textwidth]{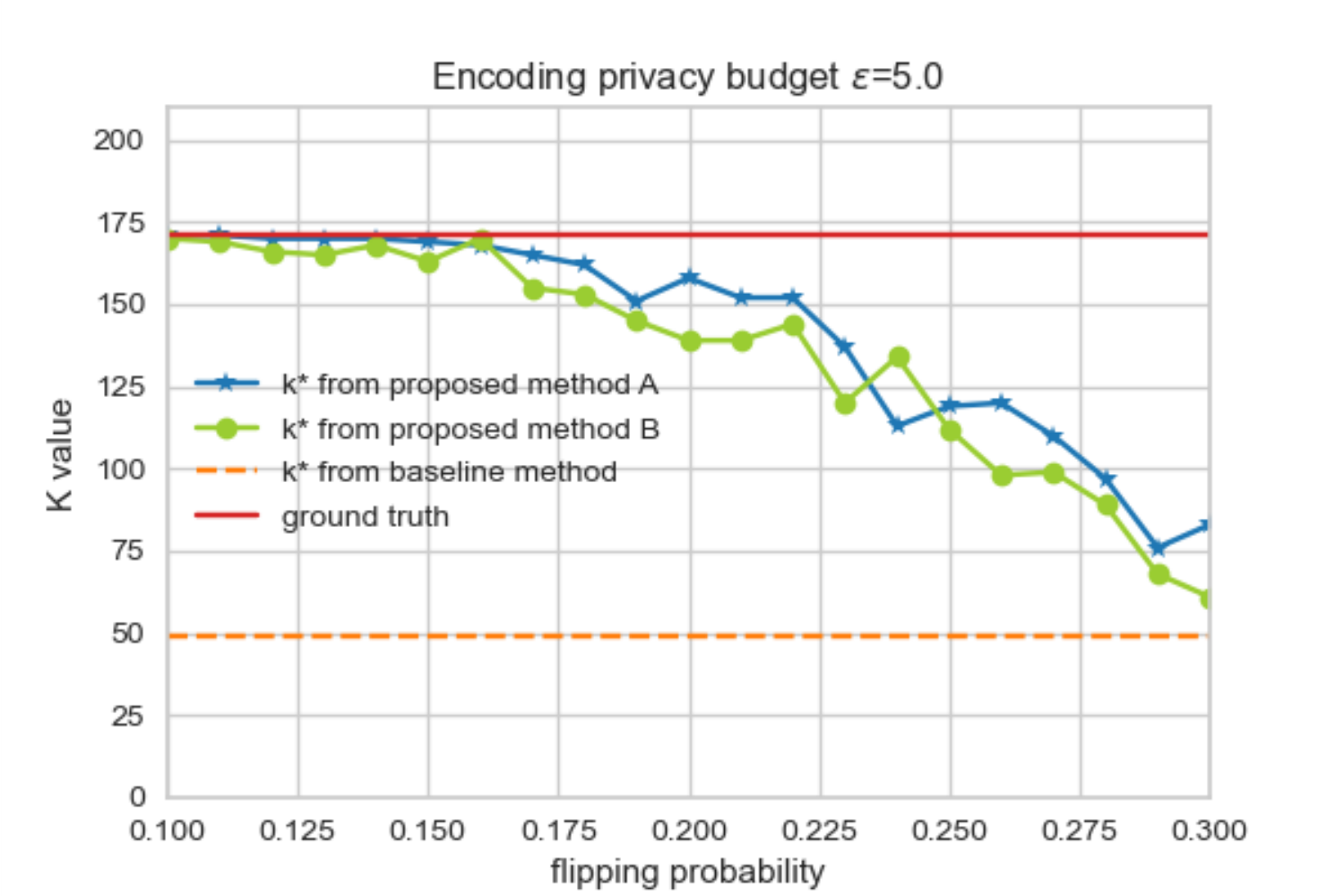}
 \includegraphics[width=0.32\textwidth]{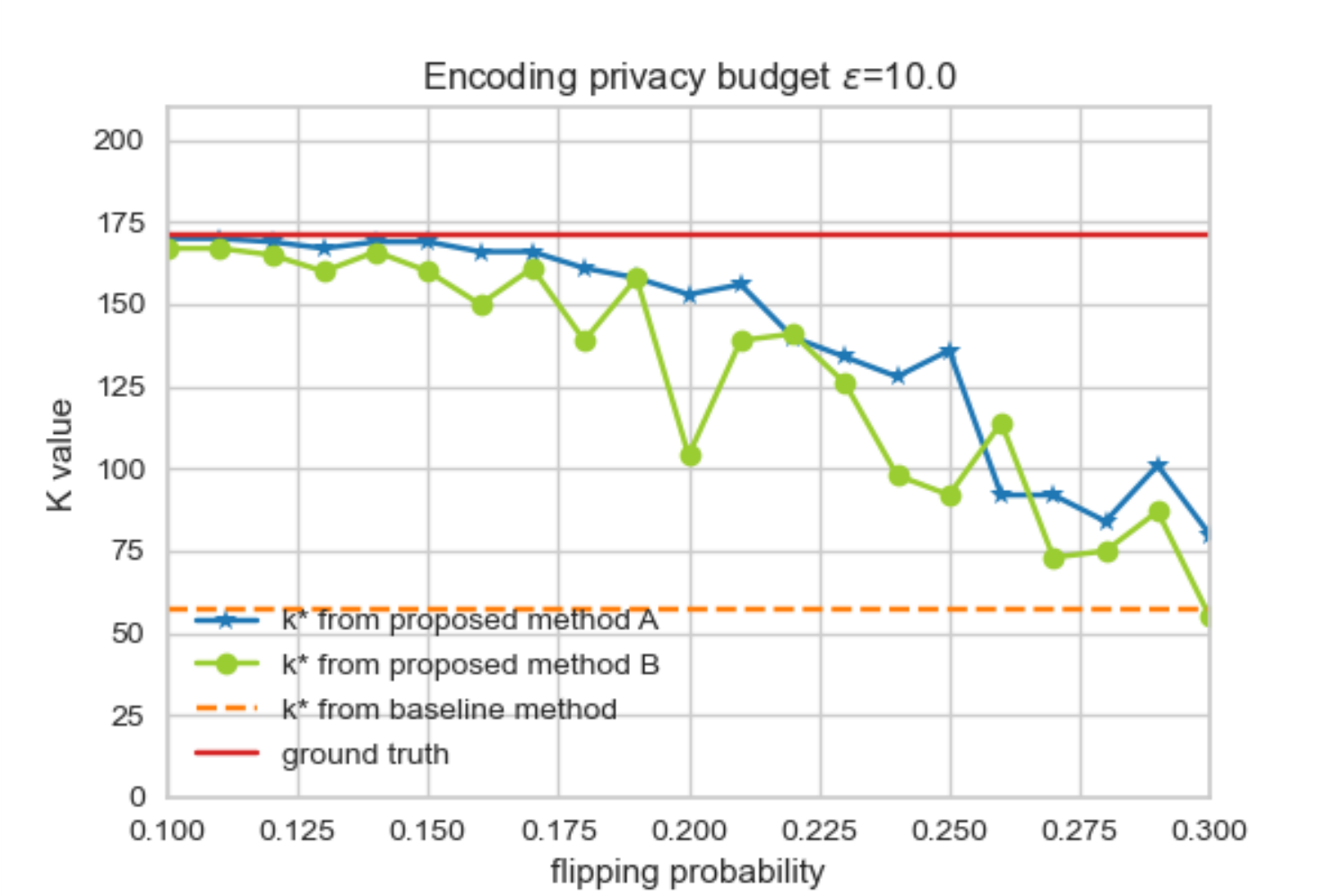}
  \caption{\small{Estimated cardinality (k value) of Method A and Method B with different flipping probabilities compared with the baseline method~\cite{Rou87} on the clean datasets with $\epsilon=[1.0,2.0,3.0,4.0,5.0,10.0]$. The reference Bloom filters pick ratio is 0.1 and dummy Bloom filters ratio is 0.1 in these experiments.
  }
    }
\label{fig:card_kval_clean}
\end{figure*}

\begin{figure*}[ht!]
\centering
 \includegraphics[width=0.32\textwidth]{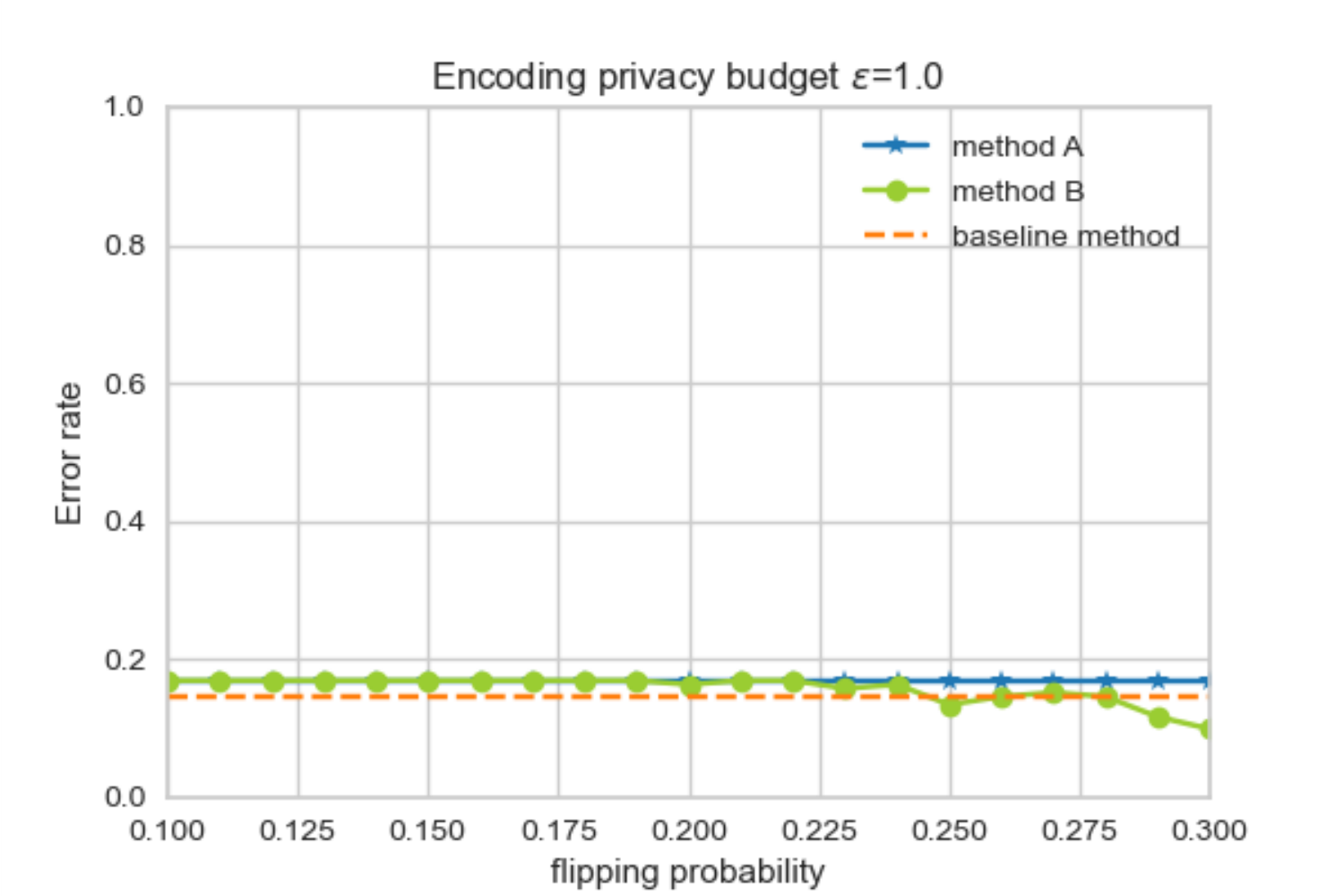}
 \includegraphics[width=0.32\textwidth]{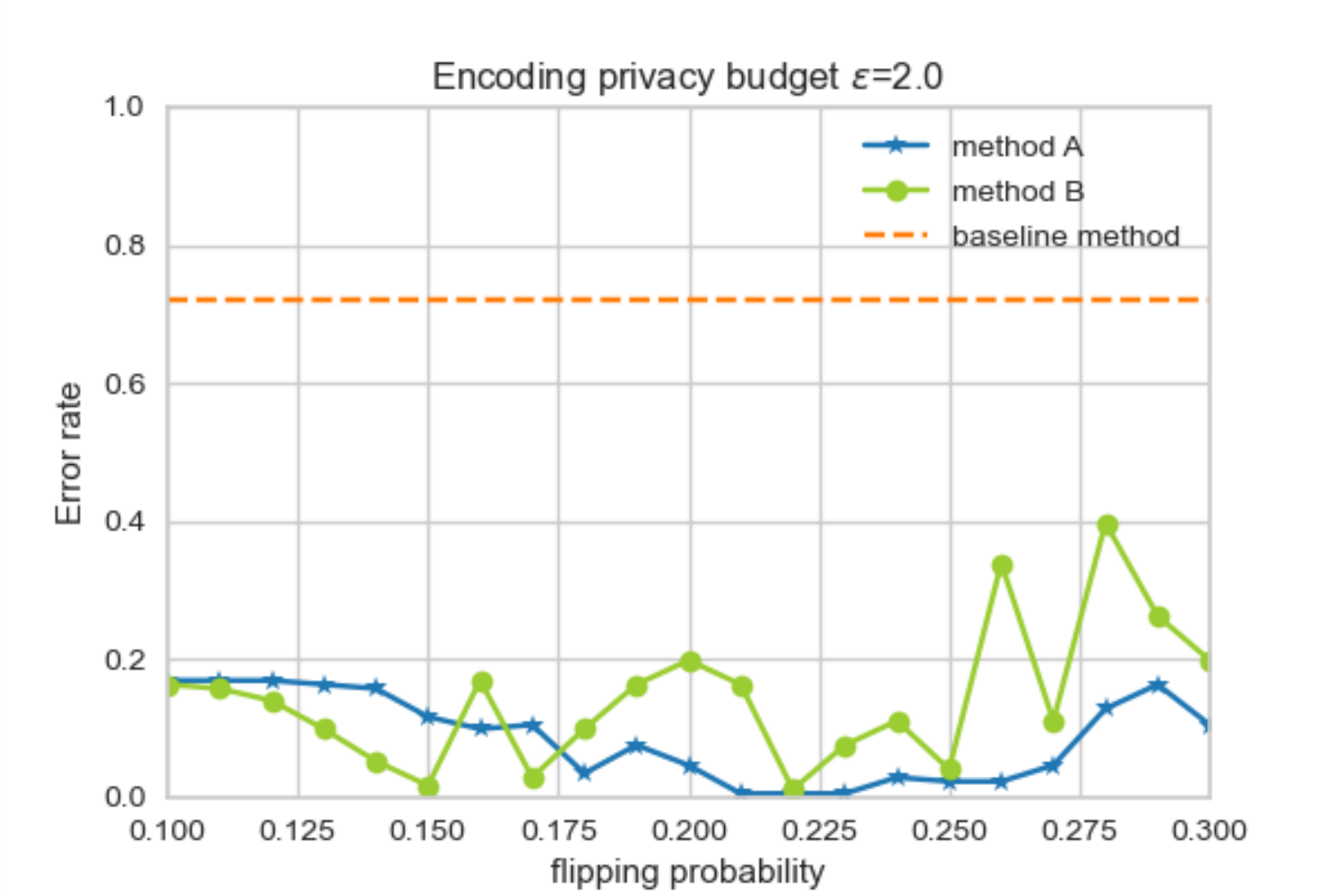}
 \includegraphics[width=0.32\textwidth]{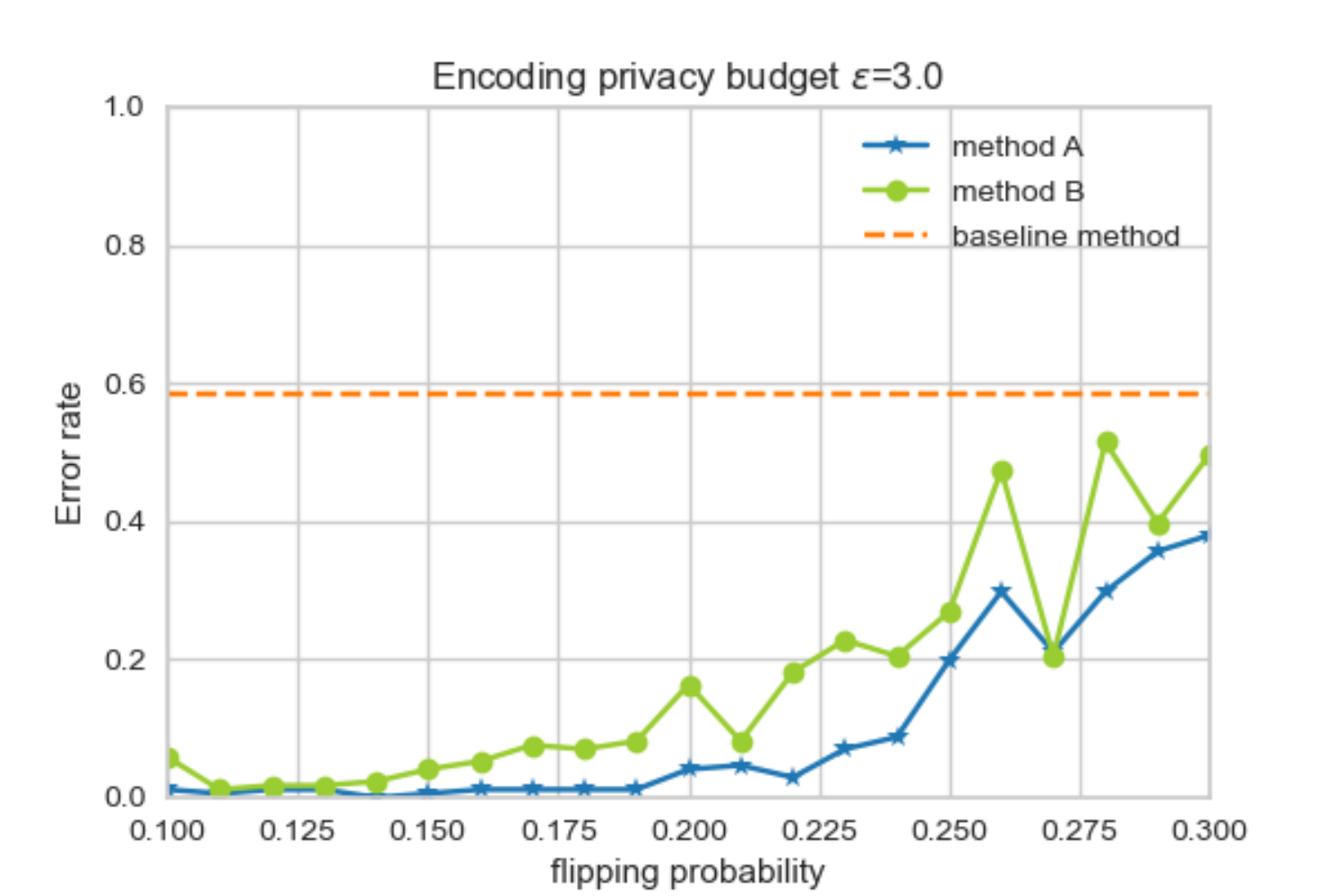}
 \includegraphics[width=0.32\textwidth]{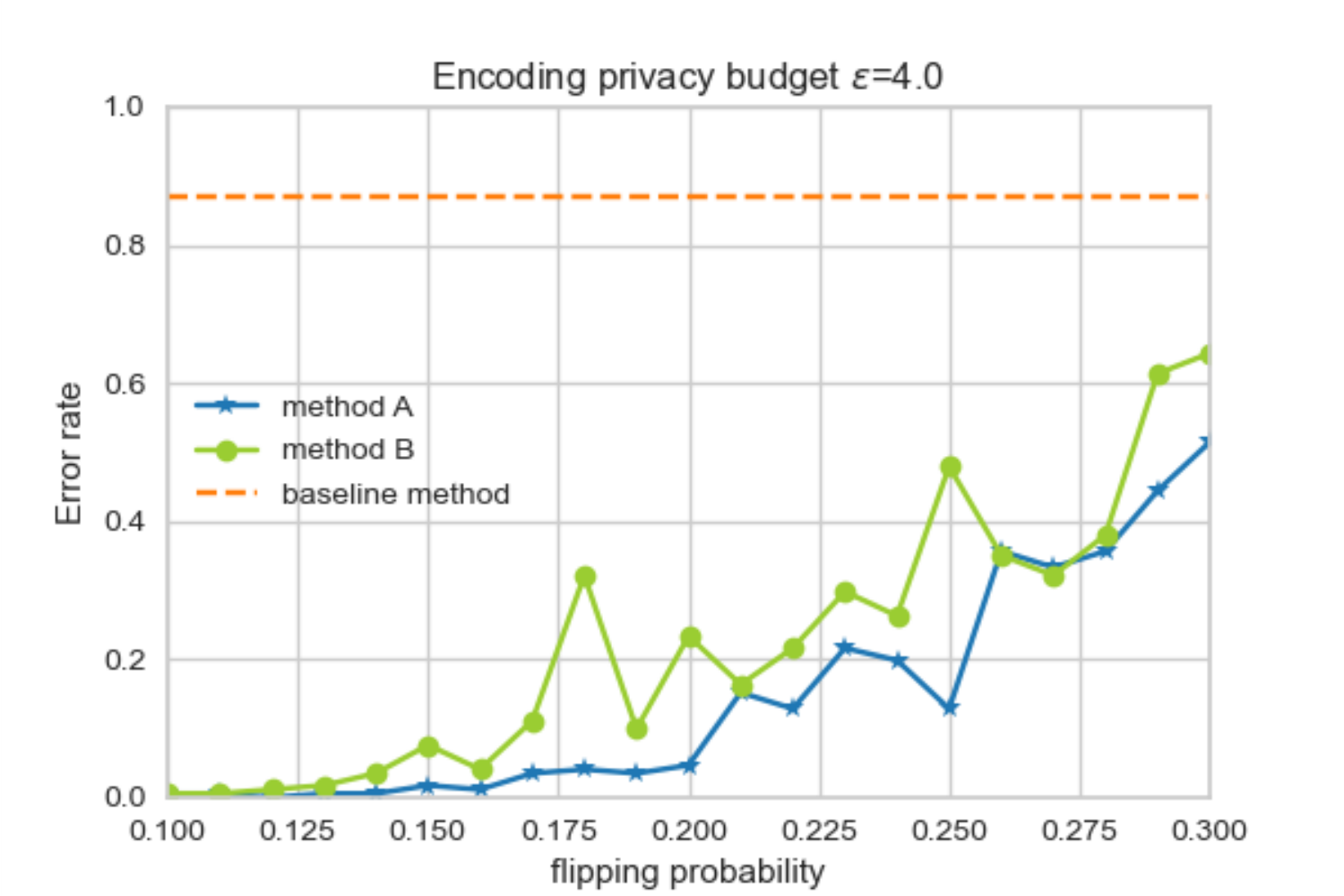}
 \includegraphics[width=0.32\textwidth]{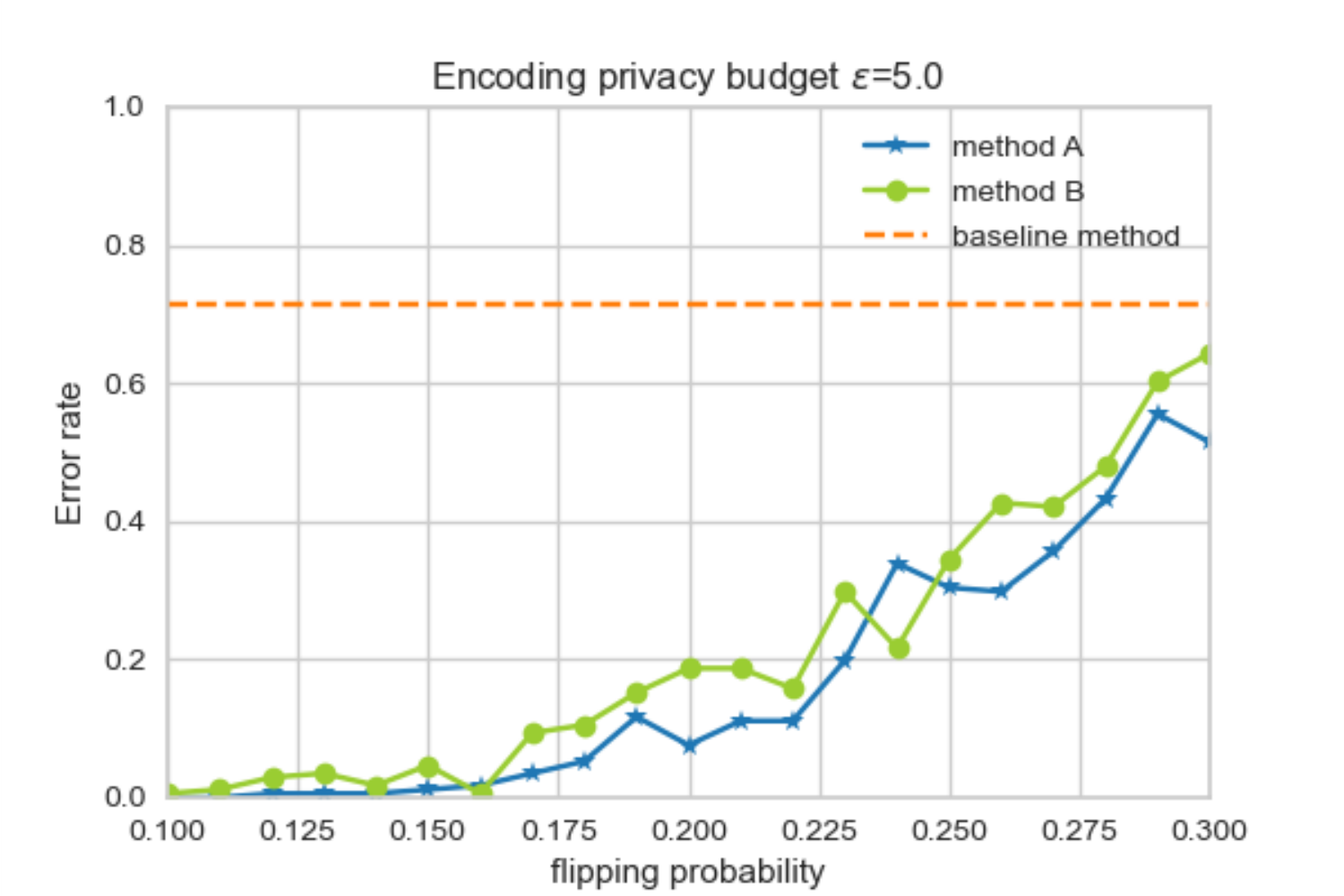} 
 \includegraphics[width=0.32\textwidth]{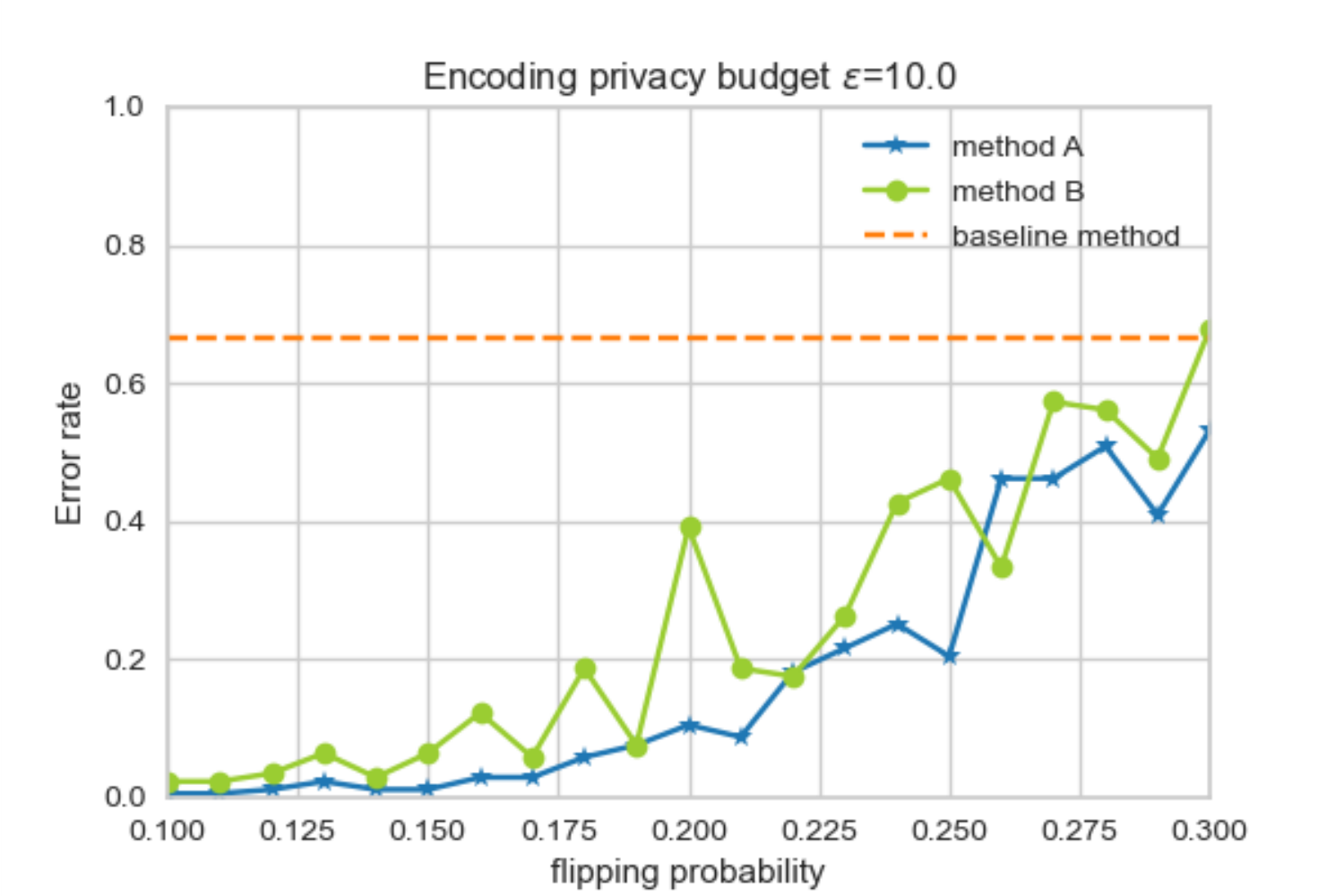}
  \caption{\small{Error rate of cardinality estimation of Method A and Method B with different flipping probabilities compared with the baseline method~\cite{Rou87} on the clean datasets with $\epsilon=[1.0,2.0,3.0,4.0,5.0,10.0]$. The reference Bloom filters pick ratio is 0.1 and dummy Bloom filters ratio is 0.1 in these experiments.
  }
    }
\label{fig:error_rate_clean_methodAB}
\end{figure*}

The first set contains duplicate records of the same person with no modified or corrupted PII values. The second set contains duplicate records with modified or corrupted PII values (20\% of records) to reflect real-world data errors and variations, while the third set contains highly corrupted PII values (40\% of records) to evaluate how real data errors impact the accuracy of cardinality estimation. We used the GeCo tool~\cite{Tra13} to generate the synthetically corrupted/modified duplicate records. We applied various corruption functions from the GeCo tool~\cite{Tra13} on randomly selected attribute values, including character edit operations (insertions, deletions, substitutions, and transpositions), and optical character recognition and phonetic modifications based on look-up tables and corruption rules~\cite{Tra13}. 
We implemented the prototype of our proposed algorithm in Python 3.5.2, and ran all experiments on a server with four 2-core 64-bit Intel Core I7 2.6 GHz CPUs, 8 GBytes of memory and running Ubuntu 16.04. The programs and test datasets are available from the authors.

\begin{figure*}[ht!]
\centering
 \includegraphics[width=0.32\textwidth]{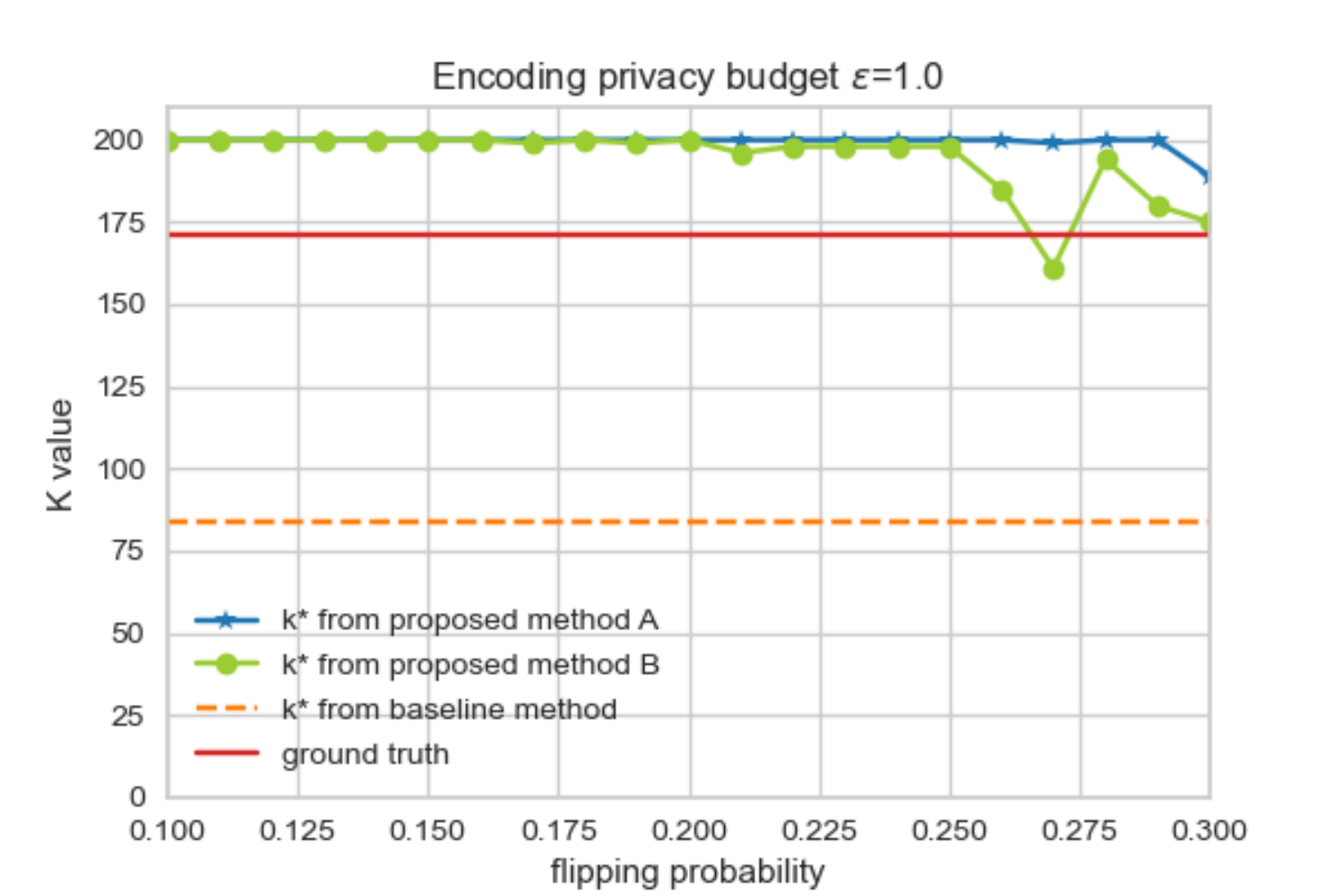}
 \includegraphics[width=0.32\textwidth]{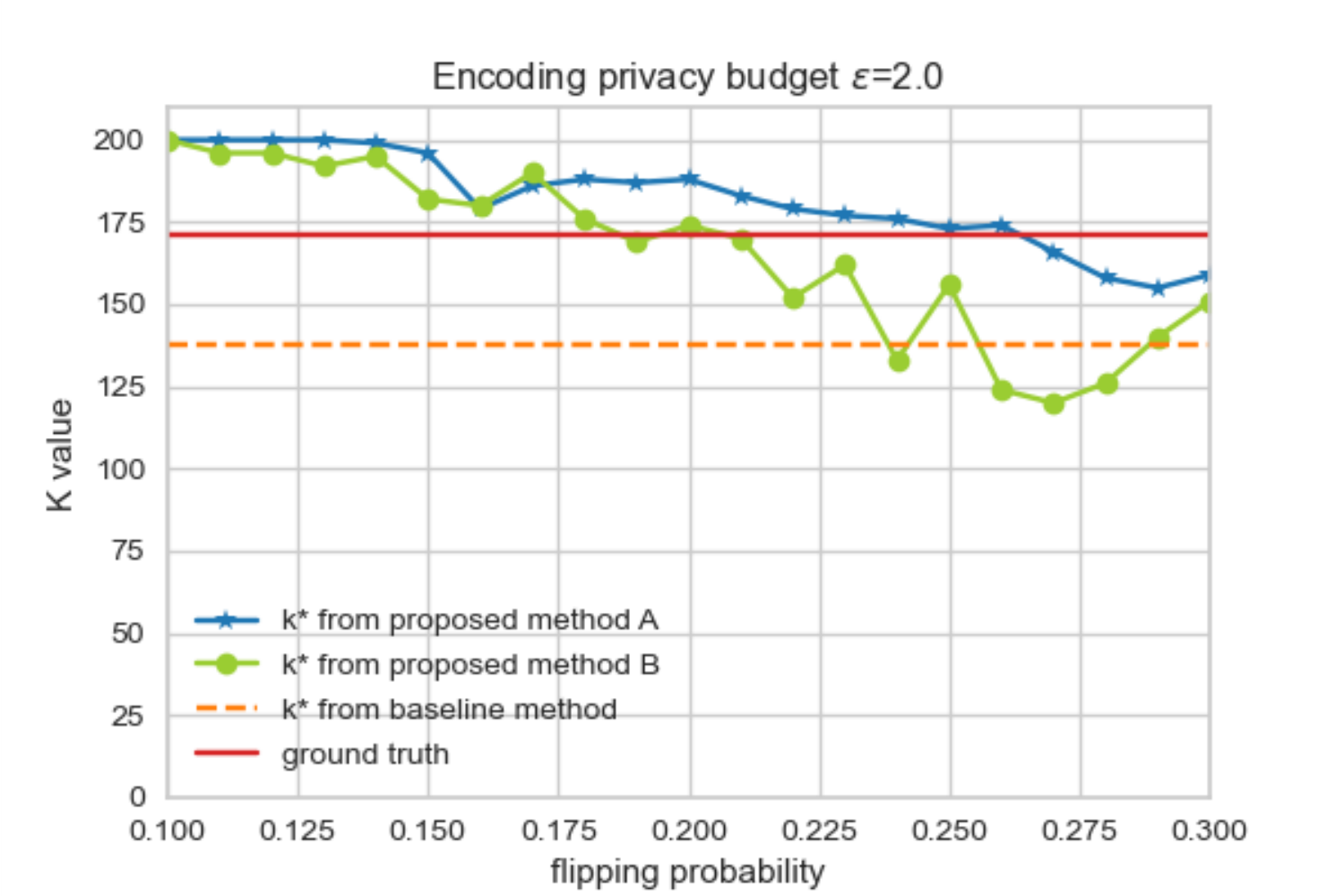}
 \includegraphics[width=0.32\textwidth]{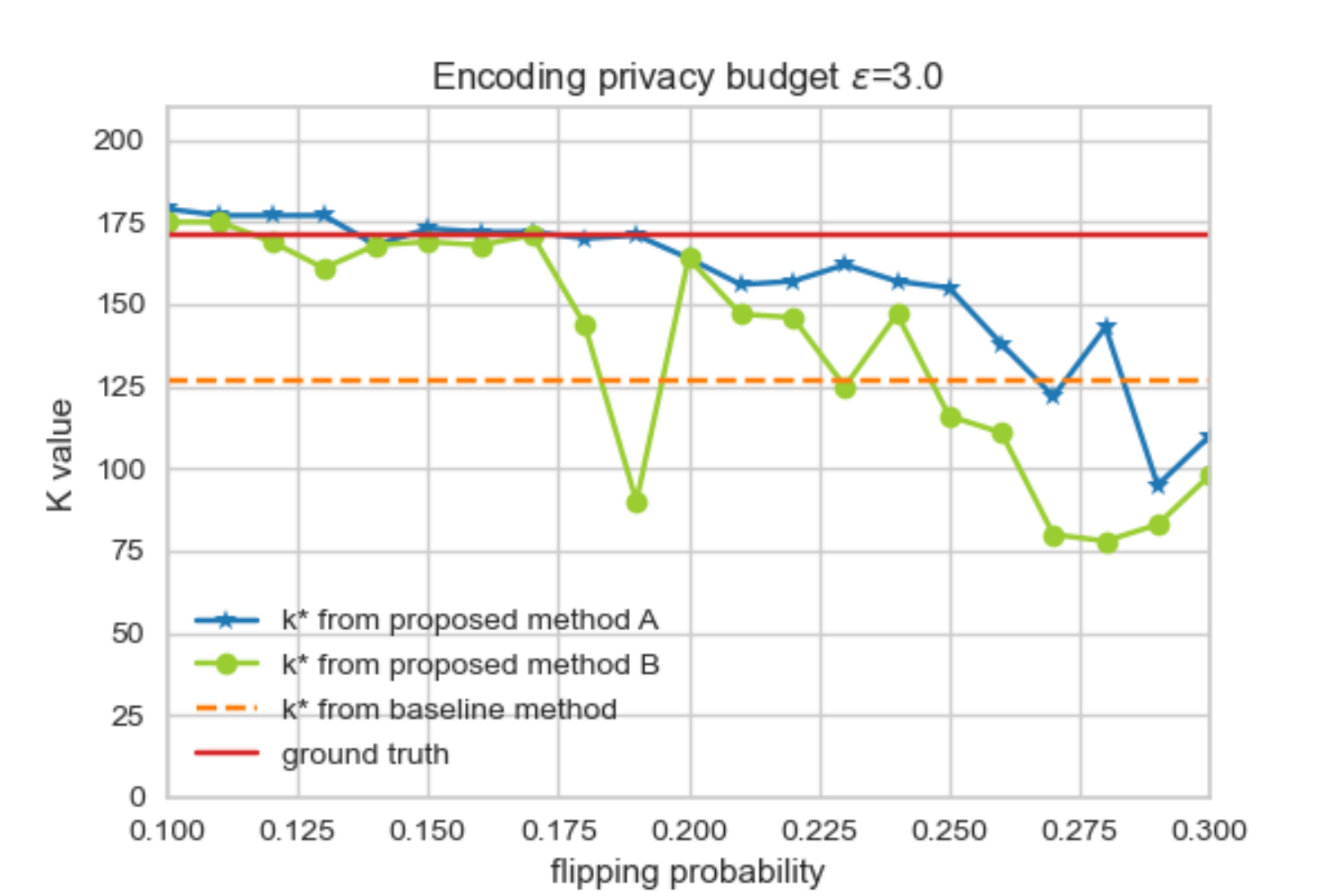}
 \includegraphics[width=0.32\textwidth]{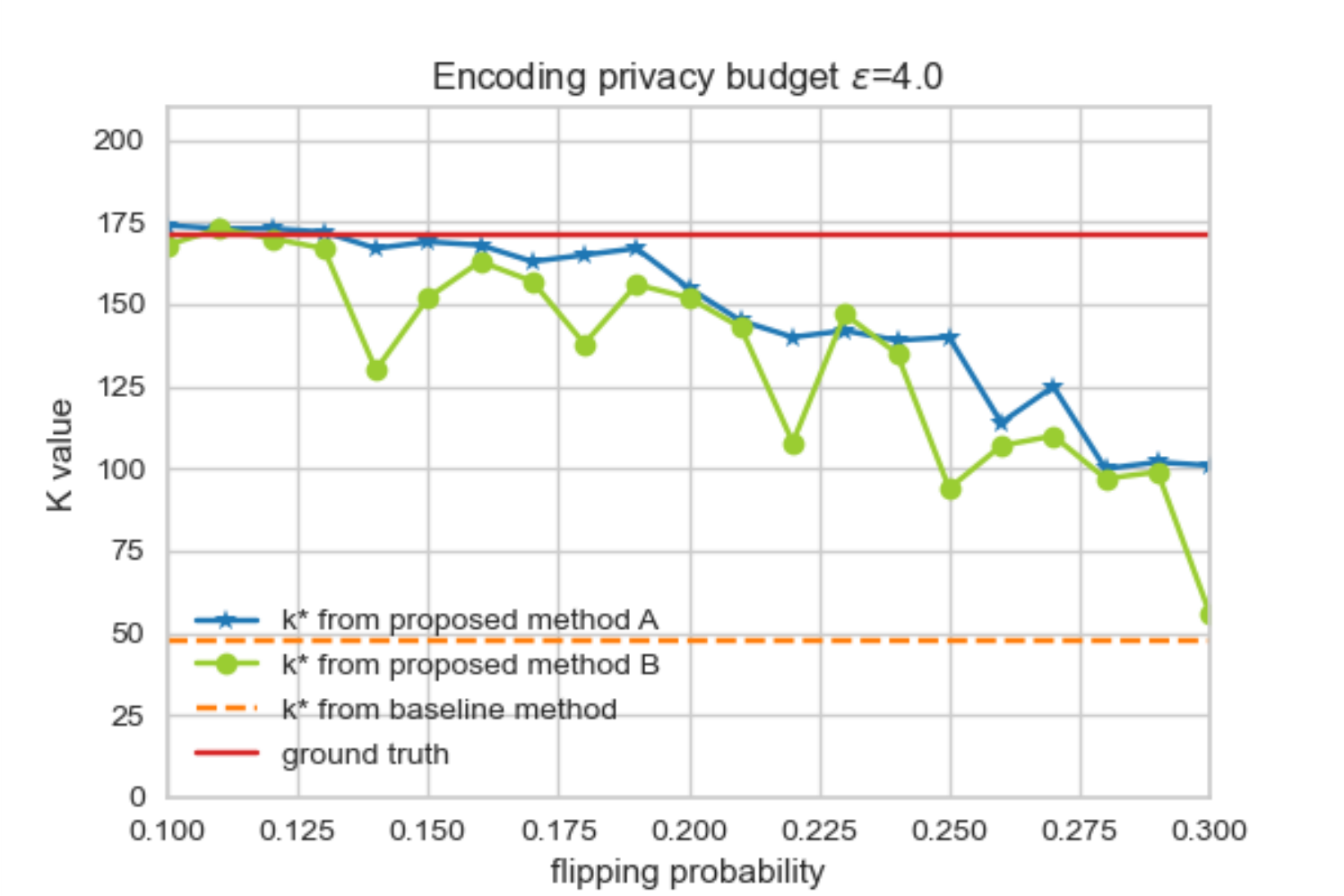}
 \includegraphics[width=0.32\textwidth]{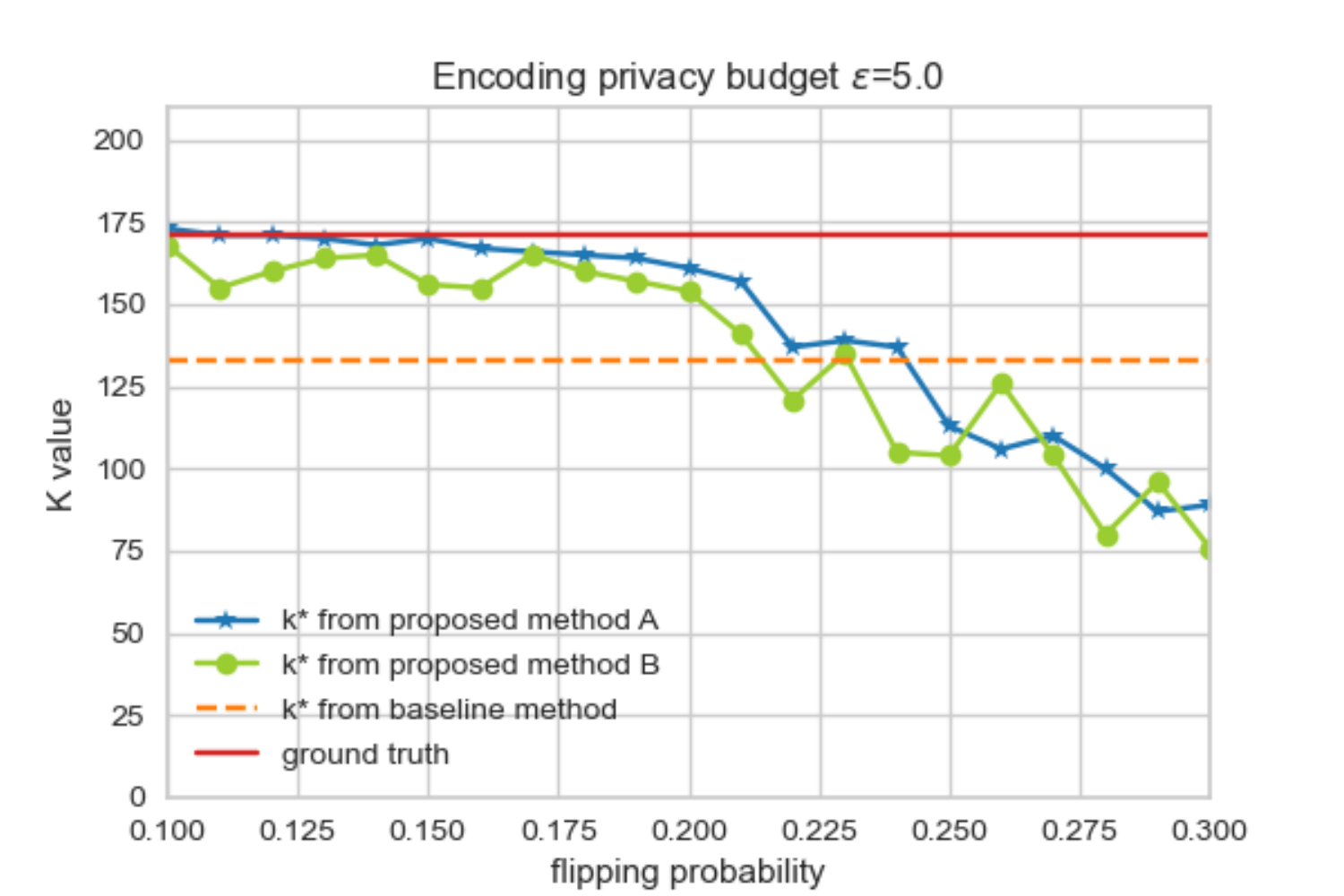}
 \includegraphics[width=0.32\textwidth]{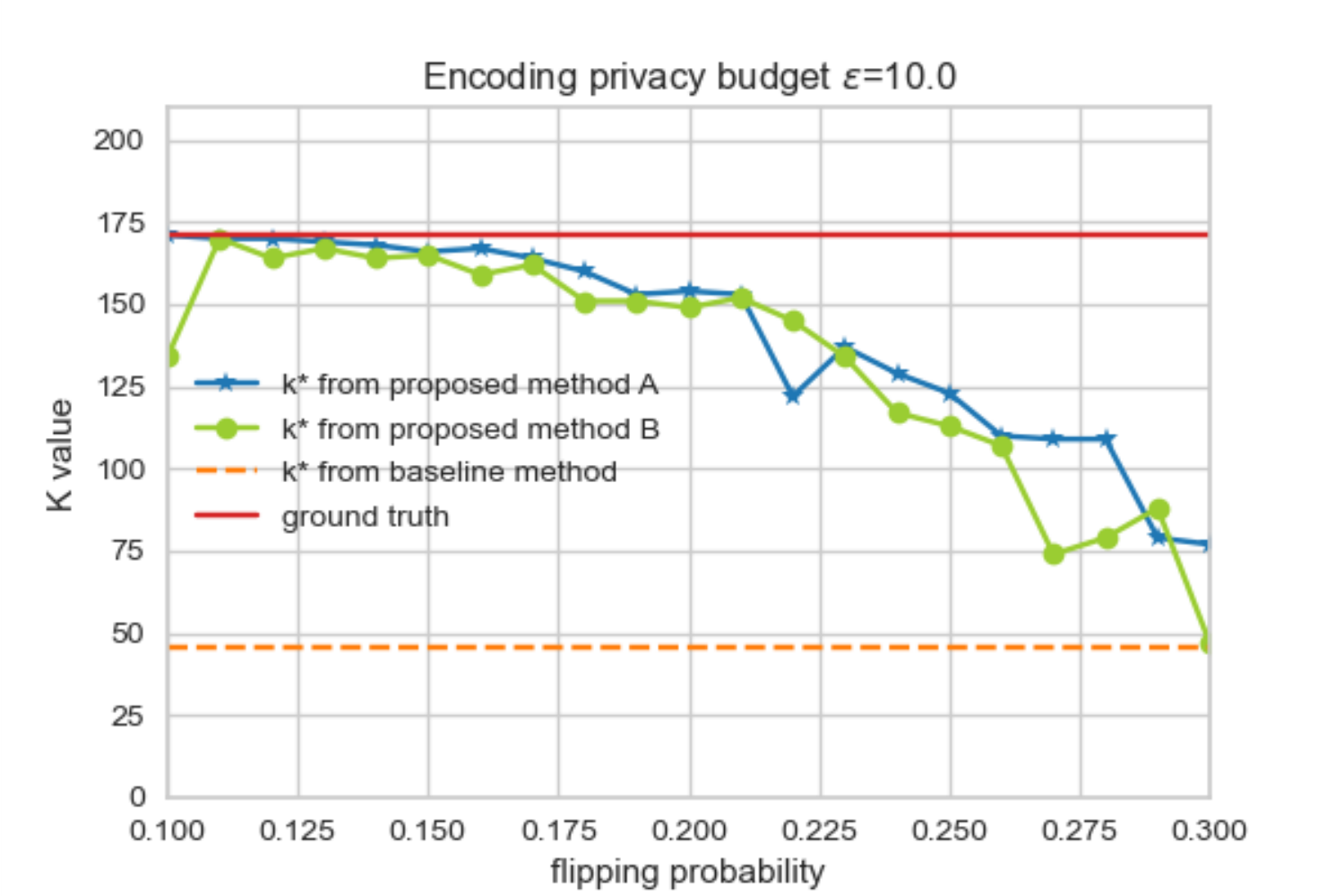}
  \caption{\small{Estimated cardinality (k value) of Method A and Method B with different flipping probabilities compared with the baseline method~\cite{Rou87} on the corrupted datasets with $\epsilon=[1.0,2.0,3.0,4.0,5.0,10.0]$. The reference Bloom filters pick ratio is 0.1 and dummy Bloom filters ratio is 0.1 in these experiments.
  }
    }
\label{fig:card_kval_corr20}
\end{figure*}

\begin{figure*}[ht!]
\centering
 \includegraphics[width=0.32\textwidth]{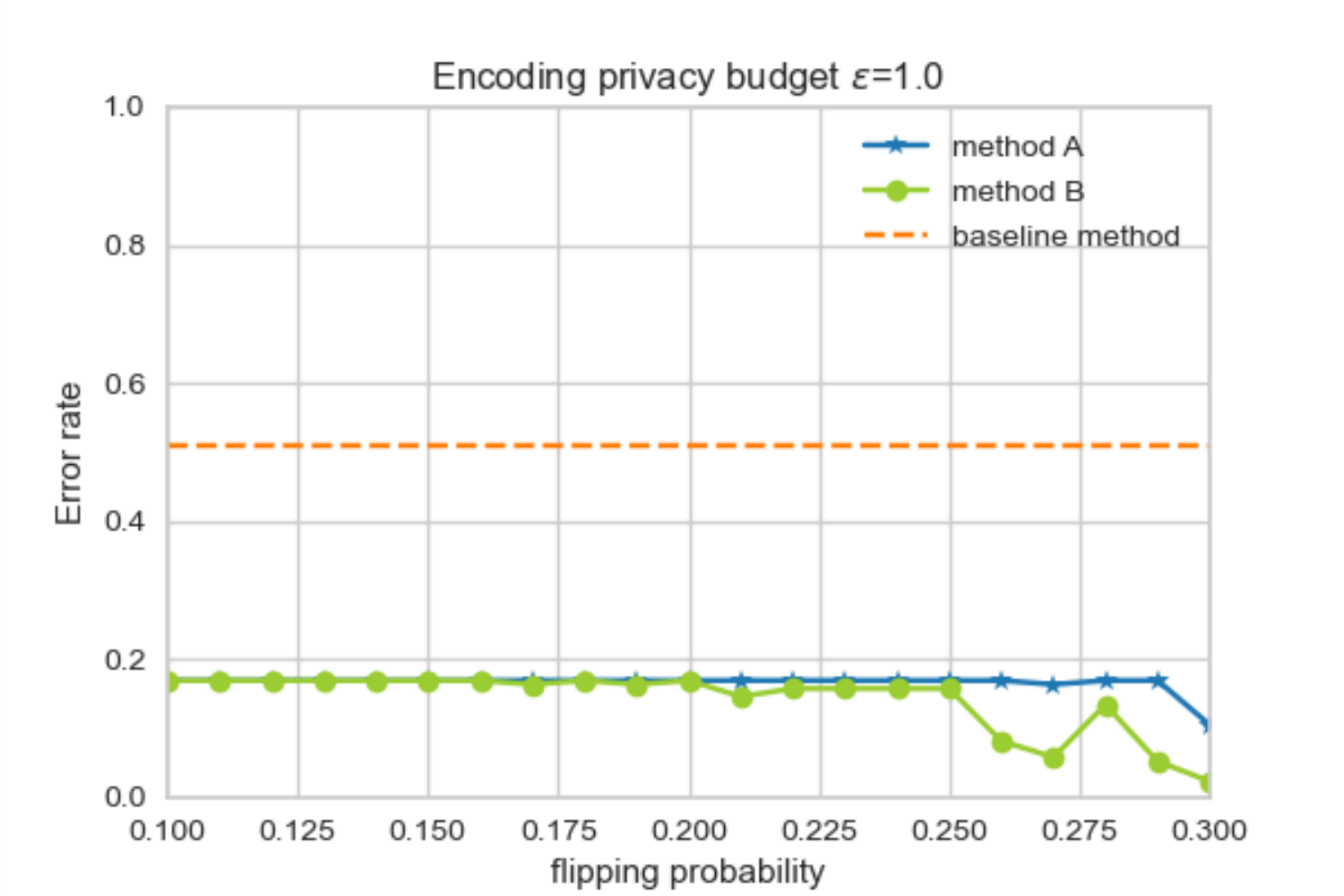}
 \includegraphics[width=0.32\textwidth]{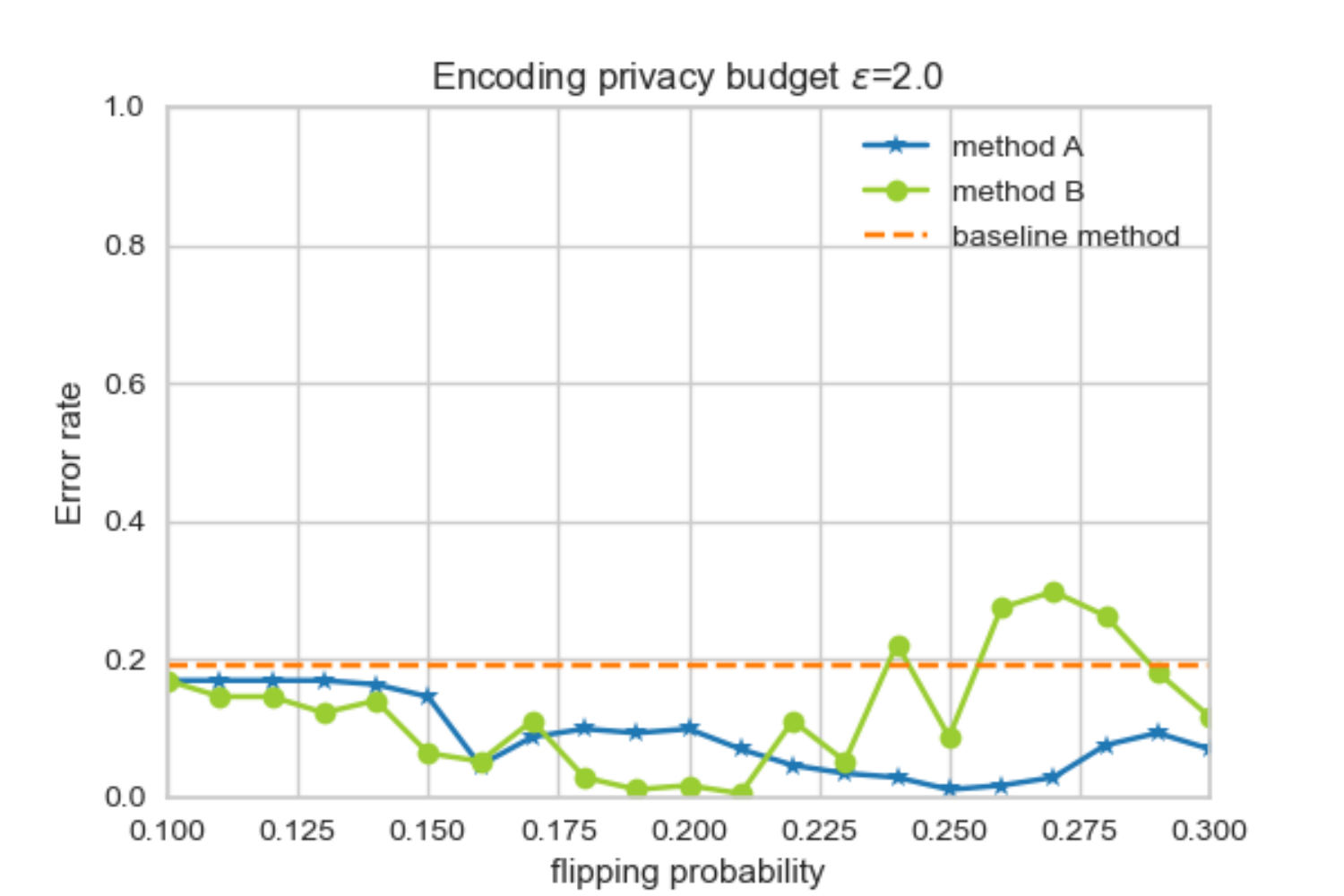}
 \includegraphics[width=0.32\textwidth]{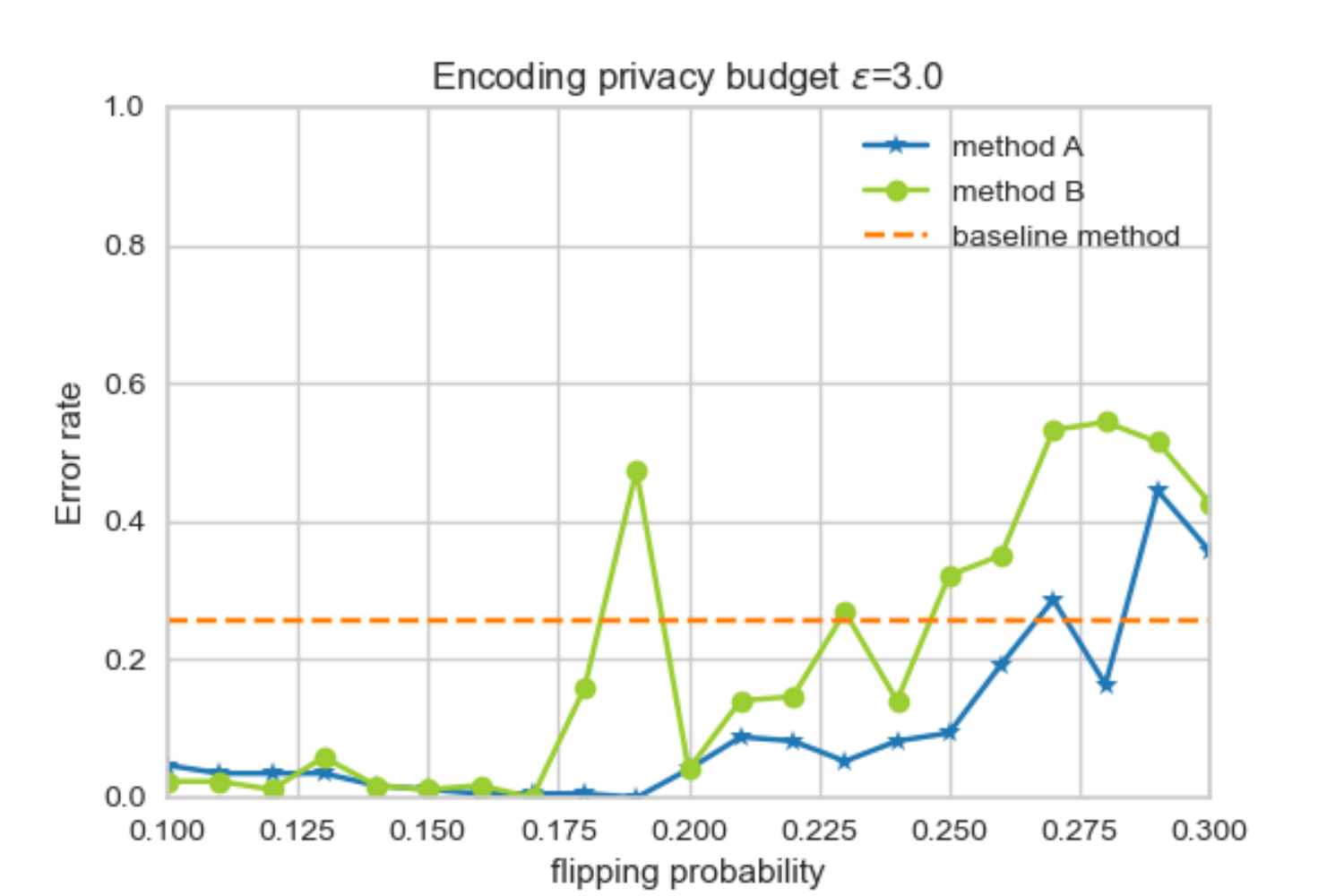}
 \includegraphics[width=0.32\textwidth]{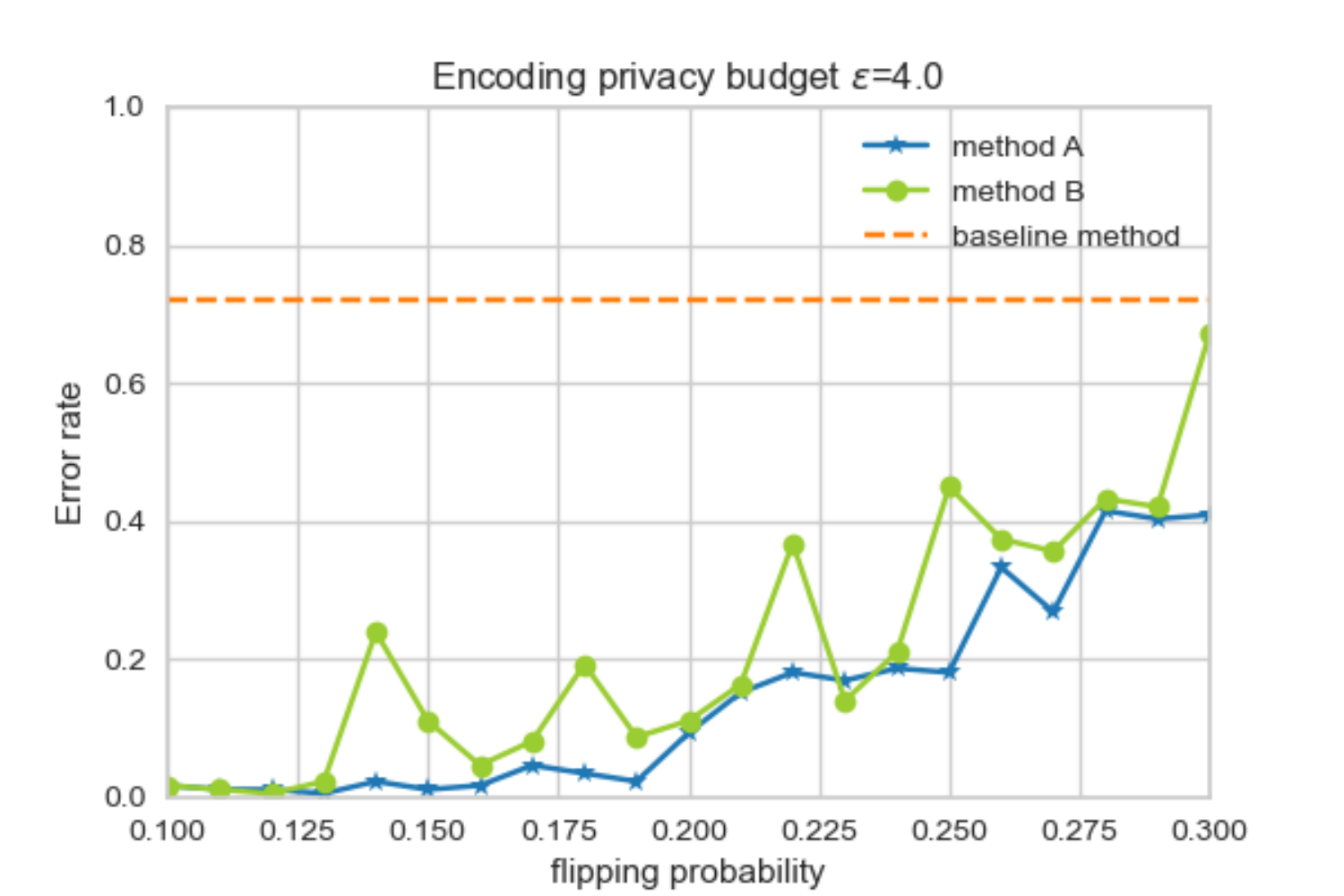}
 \includegraphics[width=0.32\textwidth]{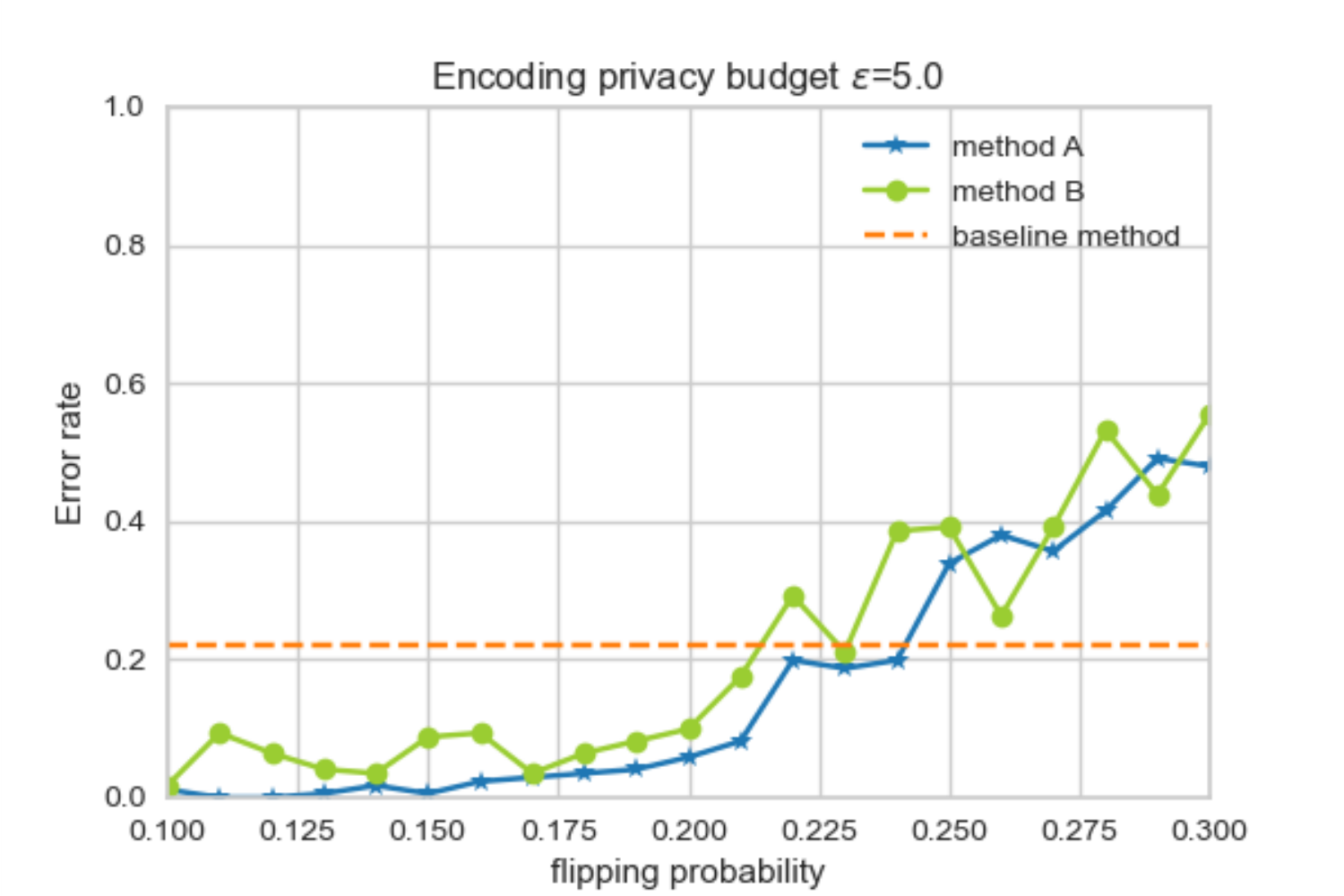} 
 \includegraphics[width=0.32\textwidth]{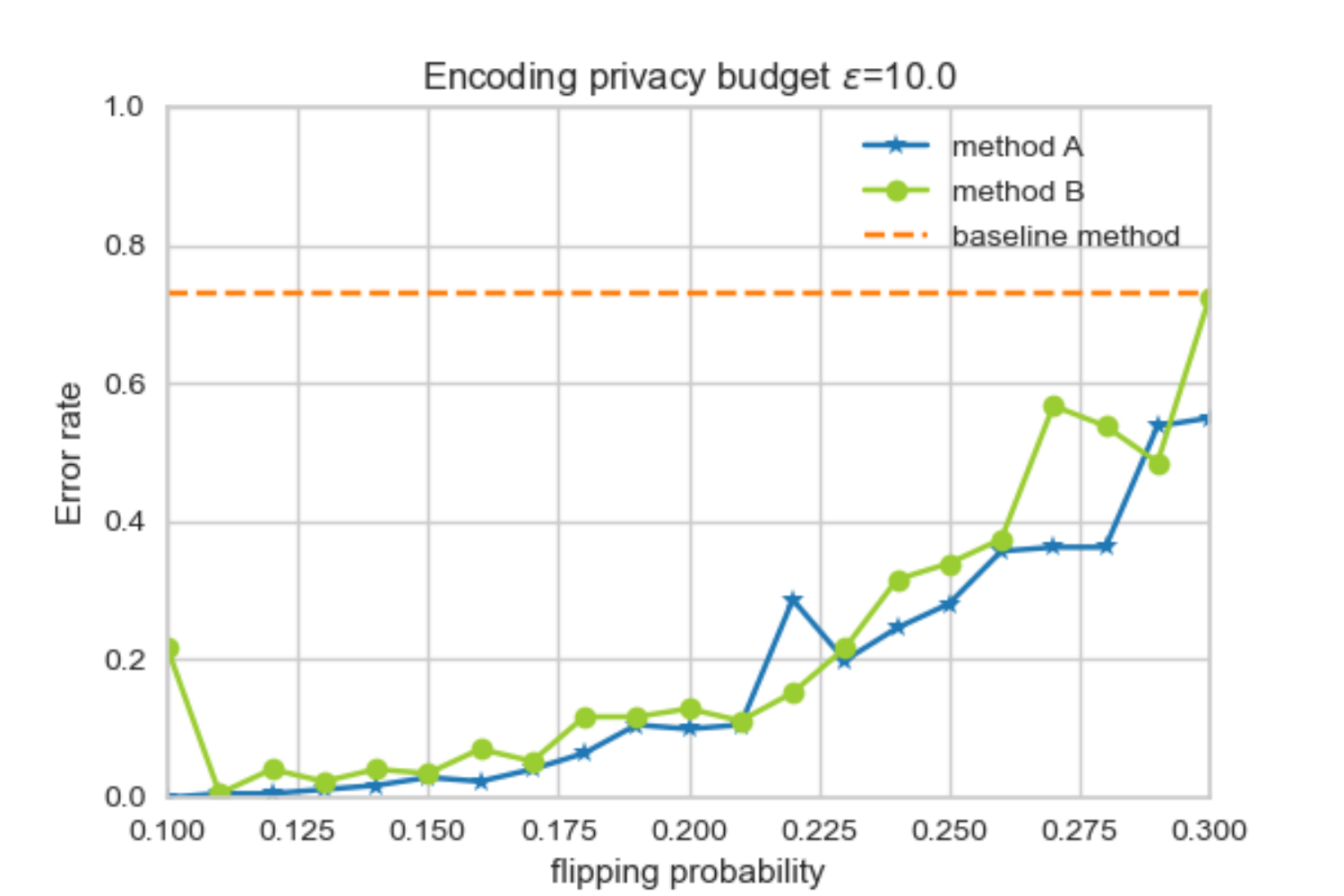}
  \caption{\small{Error rate of cardinality estimation of Method A and Method B with different flipping probabilities compared with the baseline method~\cite{Rou87} on the corrupted datasets with $\epsilon=[1.0,2.0,3.0,4.0,5.0,10.0]$. The reference Bloom filters pick ratio is 0.1 and dummy Bloom filters ratio is 0.1 in these experiments.
  }
    }
\label{fig:error_rate_corr20_methodAB}
\end{figure*}


\vspace{2.5mm}
\noindent
\textbf{Baseline method:}
We compare our methods with the baseline Elbow method that uses the silhoutte coefficient metric~\cite{Rou87} to find the optimal number of clusters.
We do not compare with other existing cardinality estimators as they do not allow fuzzy matching for counting the cardinality and hence do not provide a fair comparison.
The accuracy of count estimation is measured using the estimation error and error rate, i.e. the difference between true count and estimated count and rate of error. Privacy is measured using the privacy budget $\epsilon$. 

\vspace{2.5mm}
\noindent
\textbf{Parameter setting:}
Default parameter setting for the Bloom filter encoding is $q = 2$ for strings, length of Bloom filters is $\ell=200$, and the number of hash functions is $20$. Privacy budgets used are $\epsilon= [1.0, 2.0, 3.0, 4.0, 5.0, 10.0]$. It is important to note that, unlike with central Differential privacy, with local Differential privacy achieving a high utility with a very small privacy budget ($\le 1$) is non-trivial. When $\epsilon = 0.1$, almost 50\% of the bits in the Bloom filters need to be flipped, which makes the Bloom filters completely non-informative. Other local Differential privacy algorithms proposed, for example by Google for RAPPOR statistics and Apple for mobile usage statistics~\cite{Erl14,Dp17}, also use $\epsilon$ in the range of $\epsilon = [1.0, 2.0, 4.0, 6.0, 8.0]$. $\epsilon=10.0$ is used as a baseline with no privacy guarantees.
For the clustering algorithm, the default reference Bloom filters pick ratio used is $0.1$, the default dummy/noisy Bloom filter ratio is set to $0.1$, and the flipping probability for the dummy/noisy Bloom filters is used in the range $[0.10 - 0.30]$, with a step of $0.01$. We vary the flipping probabilities in the dummy Bloom filters and evaluate the $k$ value and error rate as it impacts the quality of clustering depending on the data quality and privacy budget.

\begin{figure*}[ht!]
\centering
 \includegraphics[width=0.32\textwidth]{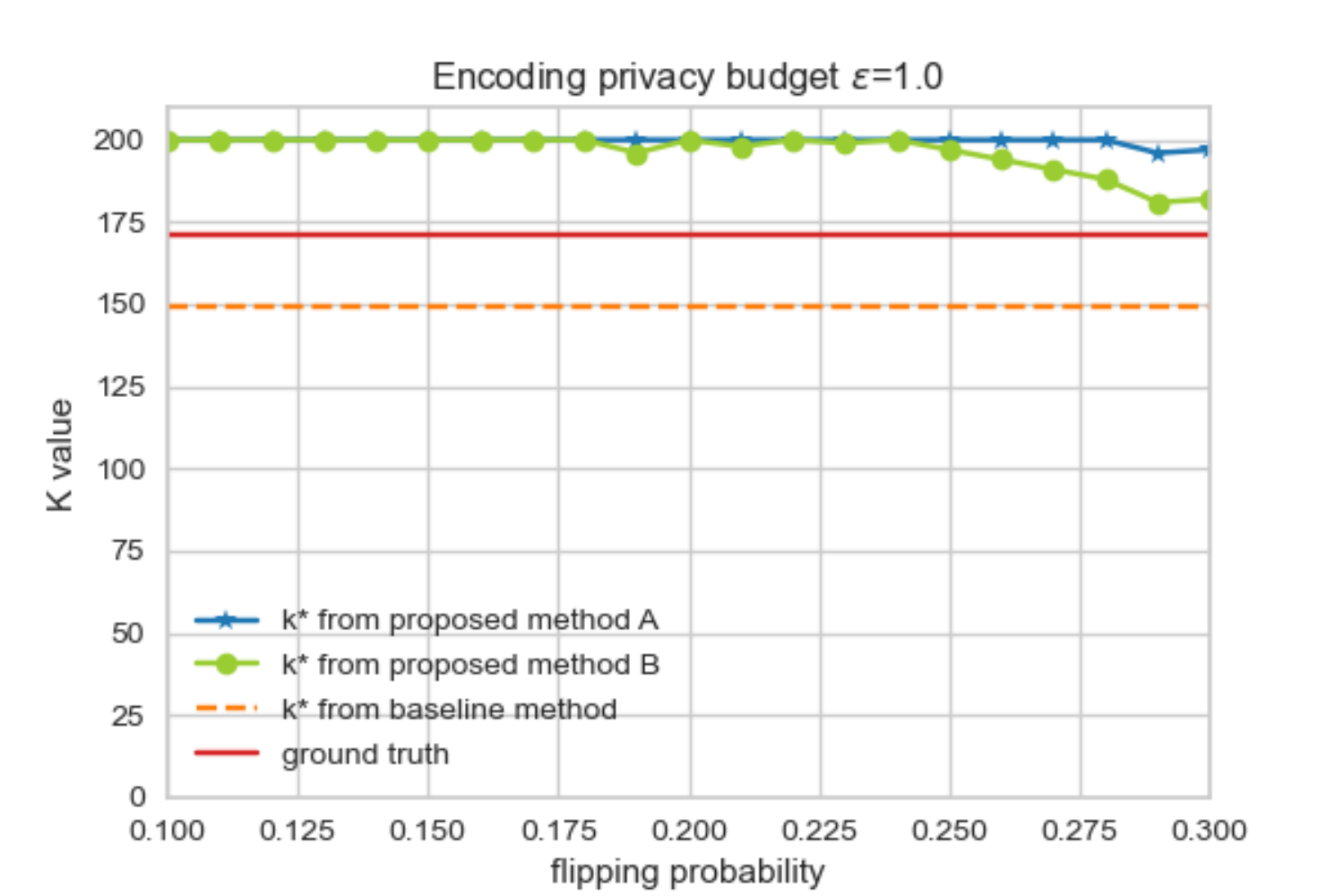}
 \includegraphics[width=0.32\textwidth]{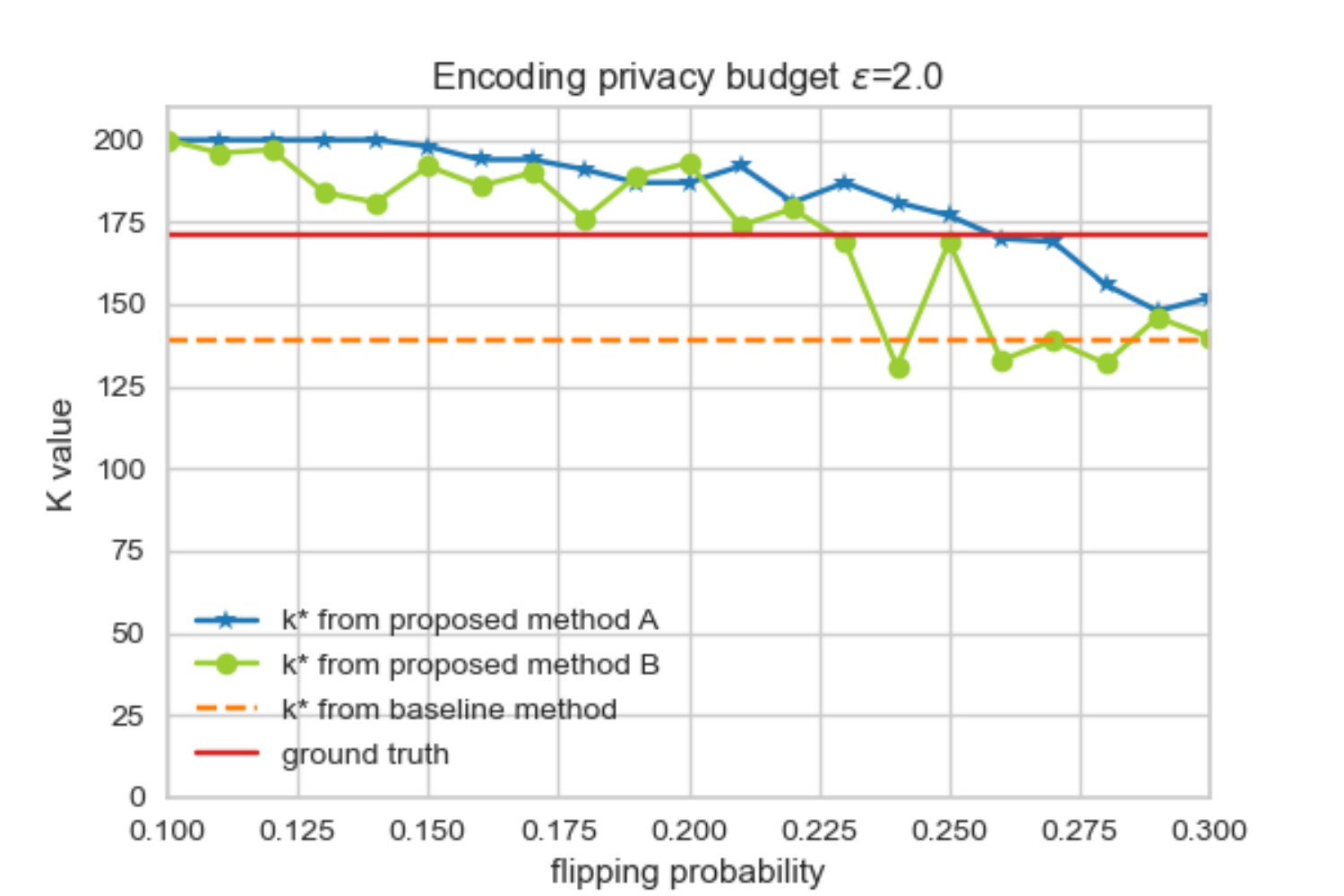}
 \includegraphics[width=0.32\textwidth]{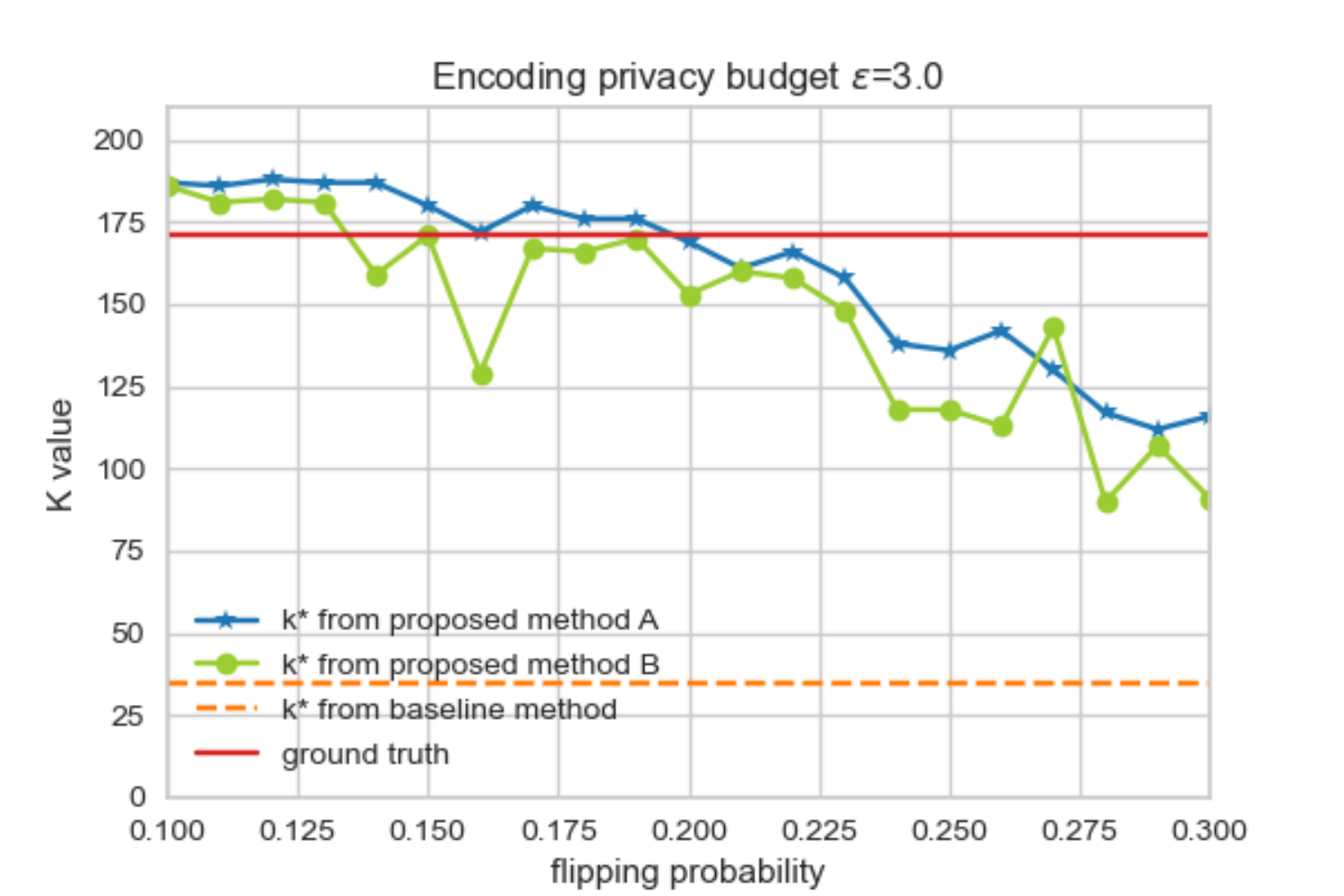}
 \includegraphics[width=0.32\textwidth]{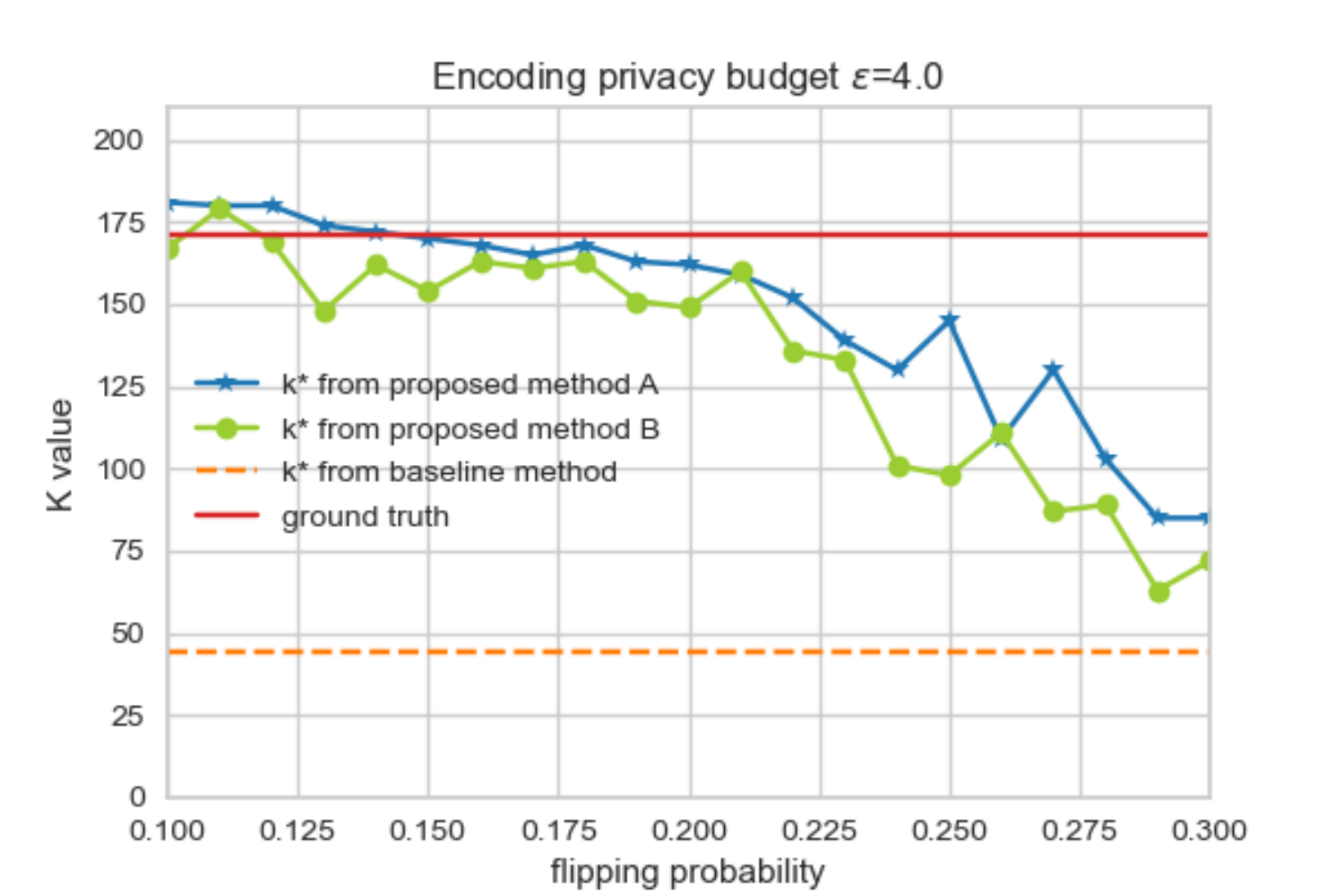}
 \includegraphics[width=0.32\textwidth]{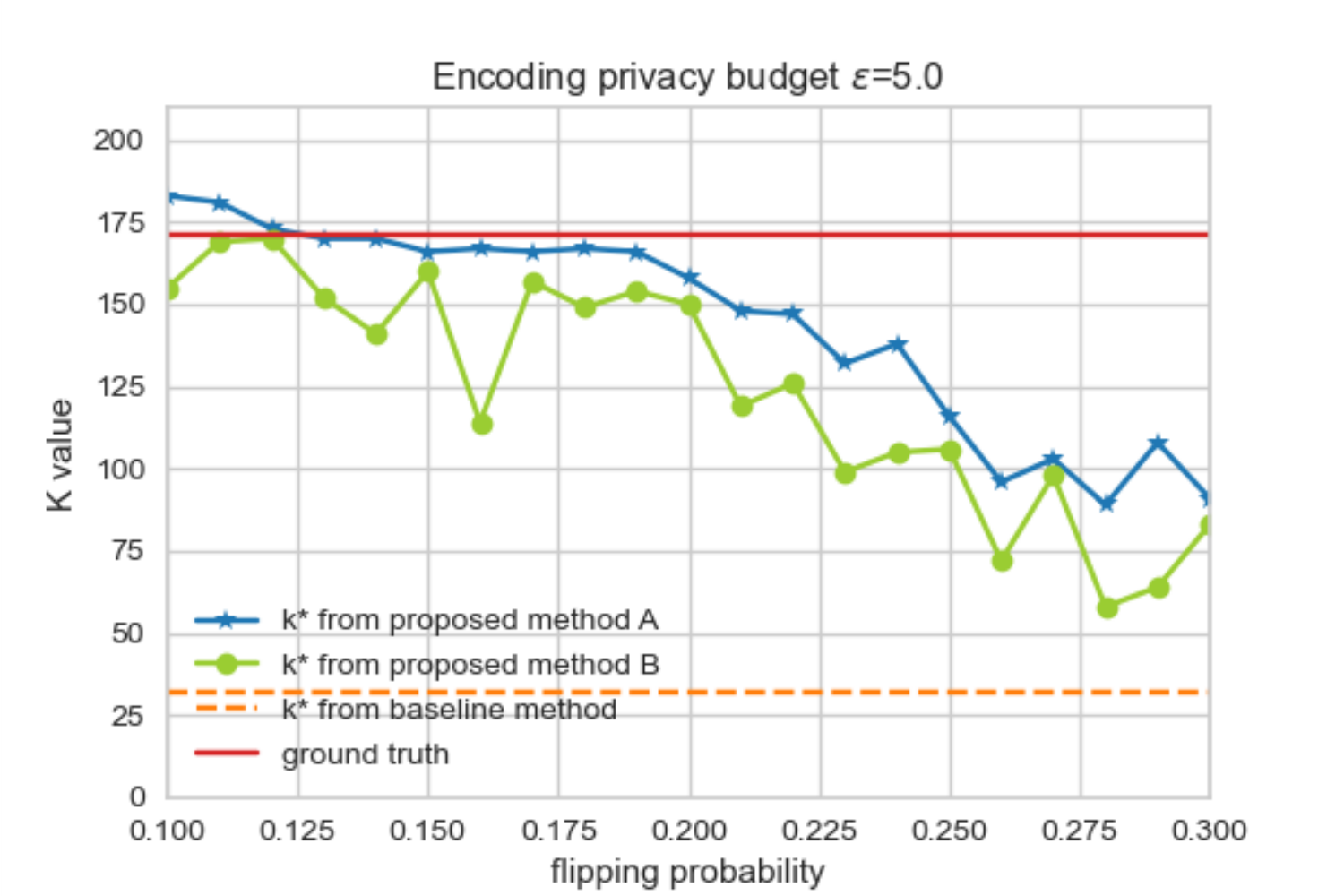}
 \includegraphics[width=0.32\textwidth]{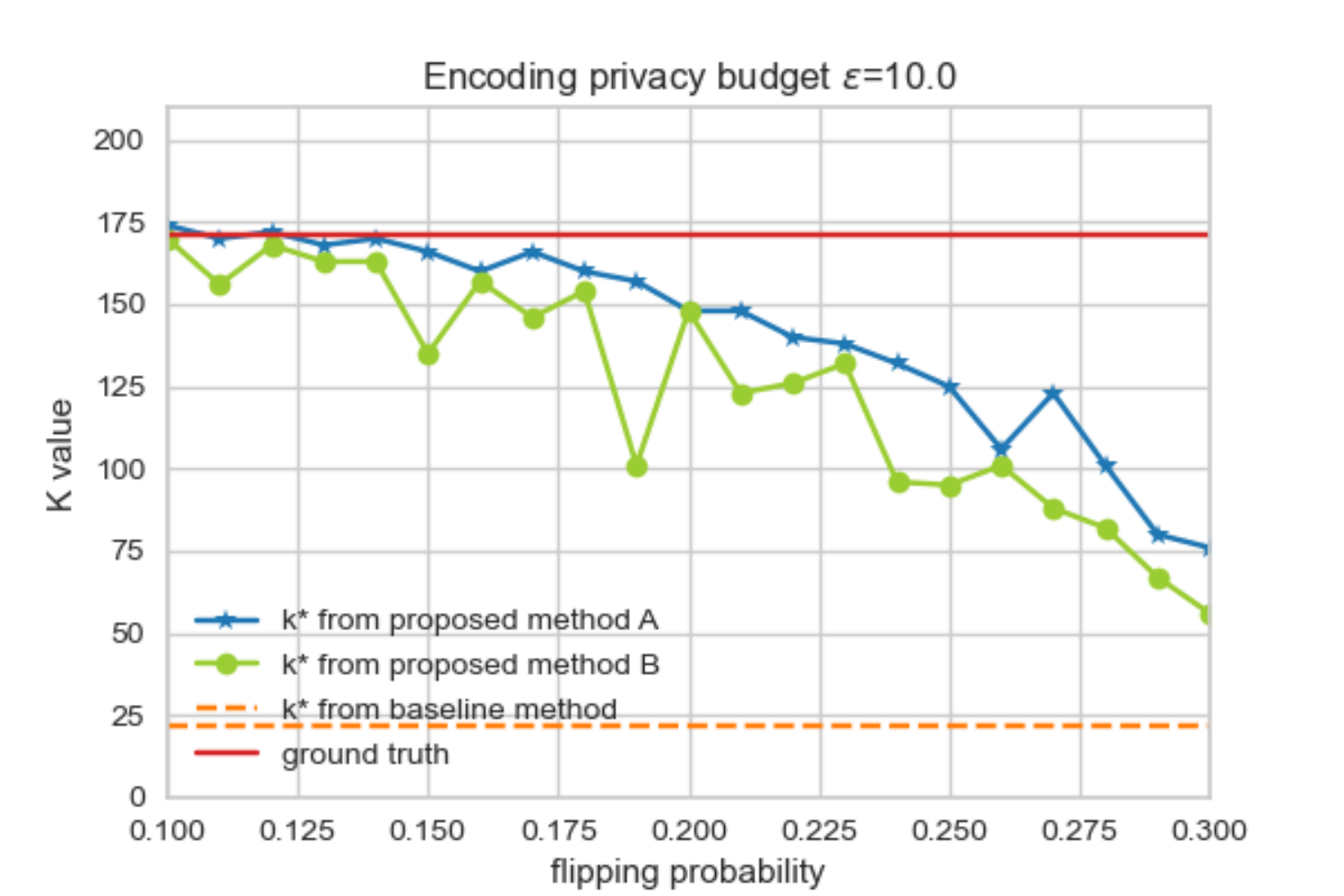}
  \caption{\small{Estimated cardinality (k value) of Method A and Method B with different flipping probabilities compared with the baseline method~\cite{Rou87} on the highly corrupted datasets with $\epsilon=[1.0,2.0,3.0,4.0,5.0,10.0]$. The reference Bloom filters pick ratio is 0.1 and dummy Bloom filters ratio is 0.1 in these experiments.
  }
    }
\label{fig:card_kval_corr40}
\end{figure*}

\begin{figure*}[ht!]
\centering
 \includegraphics[width=0.32\textwidth]{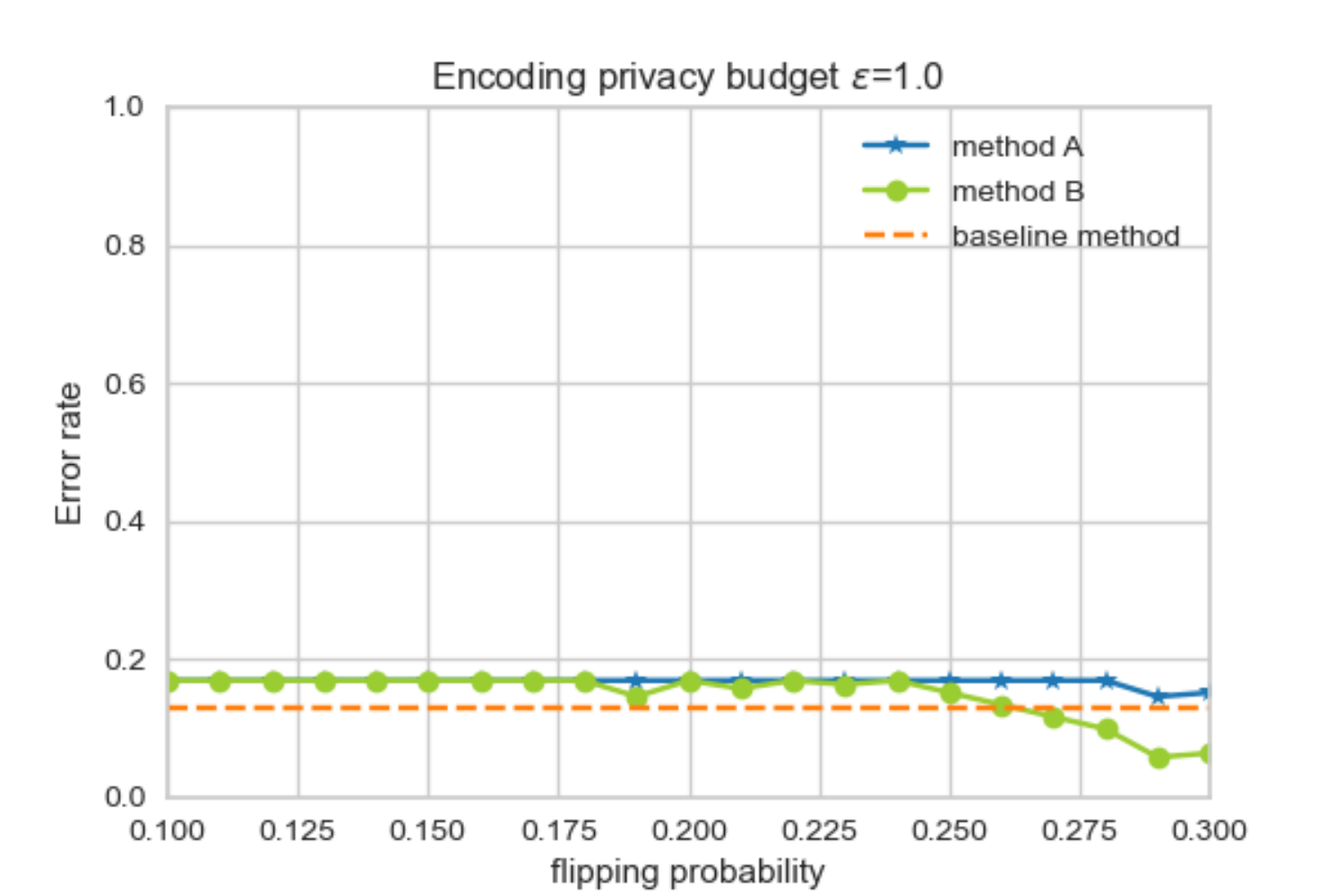}
 \includegraphics[width=0.32\textwidth]{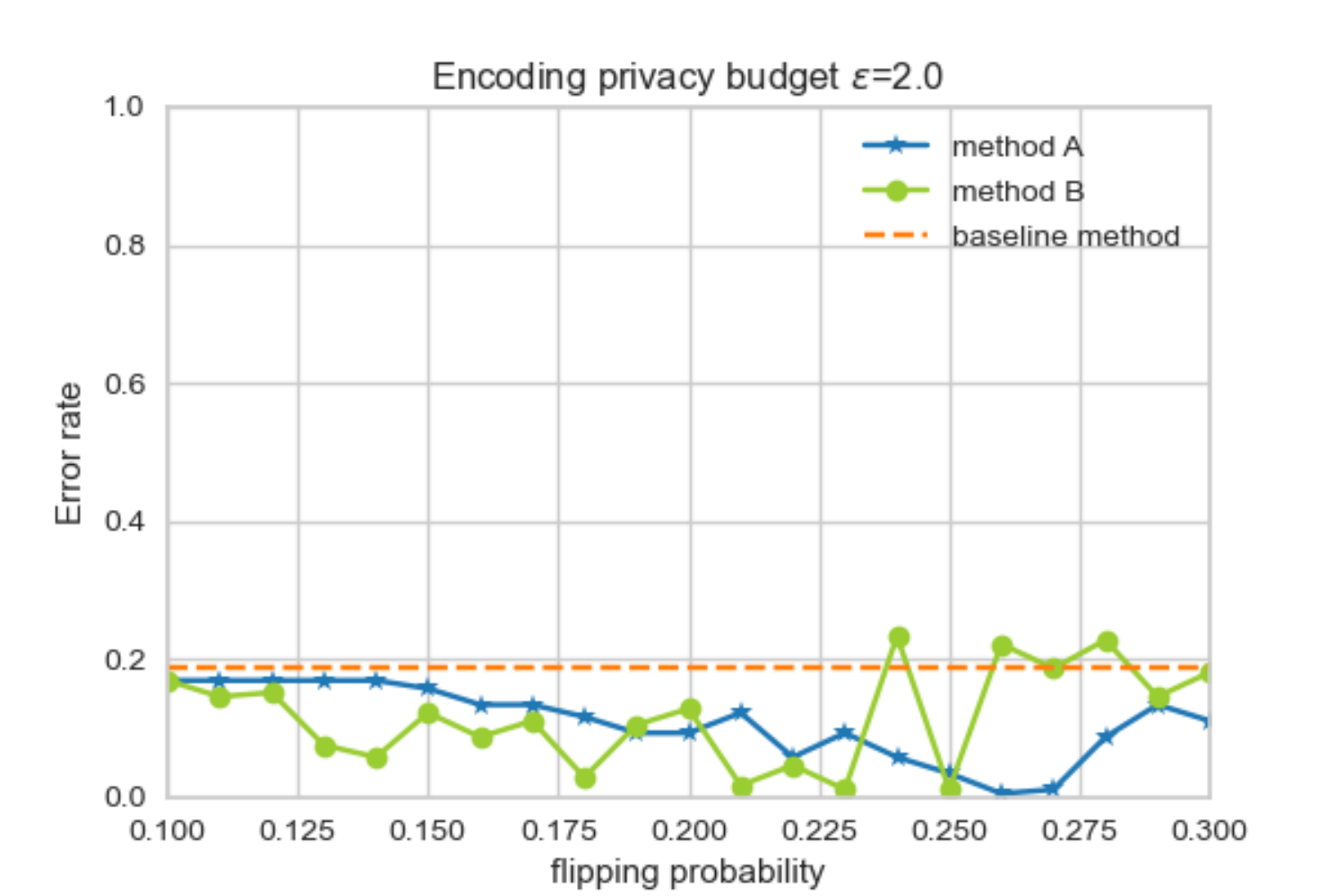}
 \includegraphics[width=0.32\textwidth]{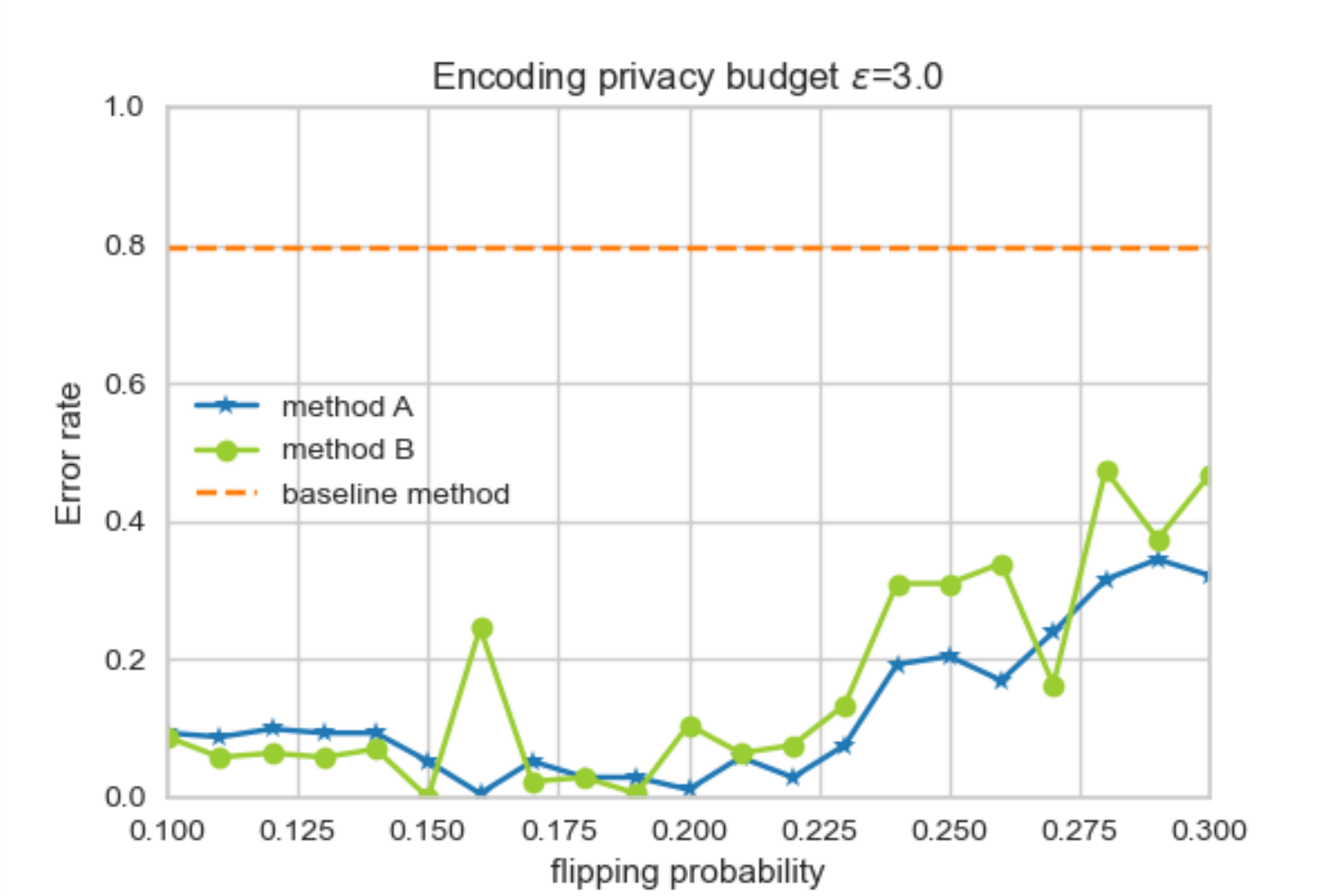}
 \includegraphics[width=0.32\textwidth]{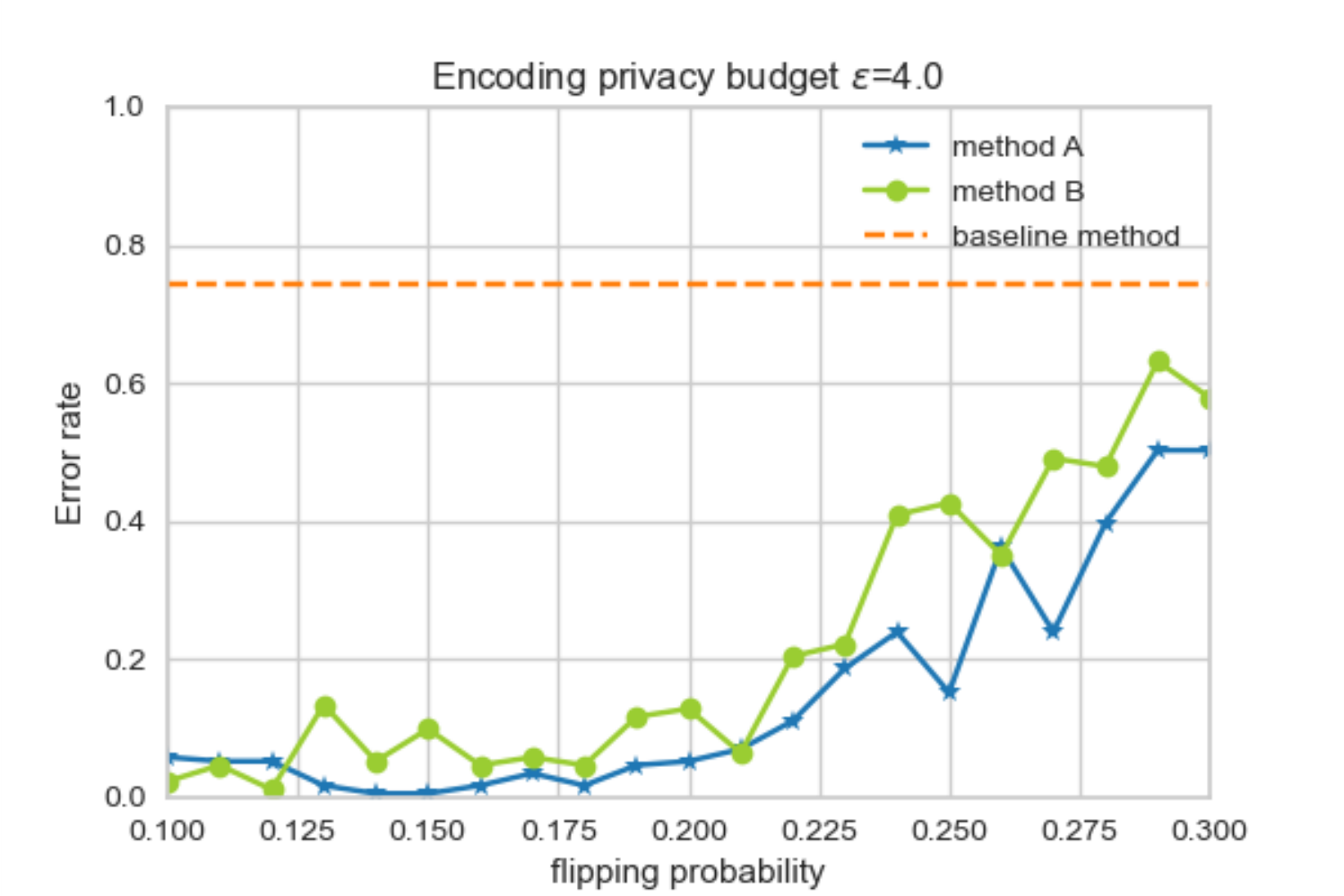}
 \includegraphics[width=0.32\textwidth]{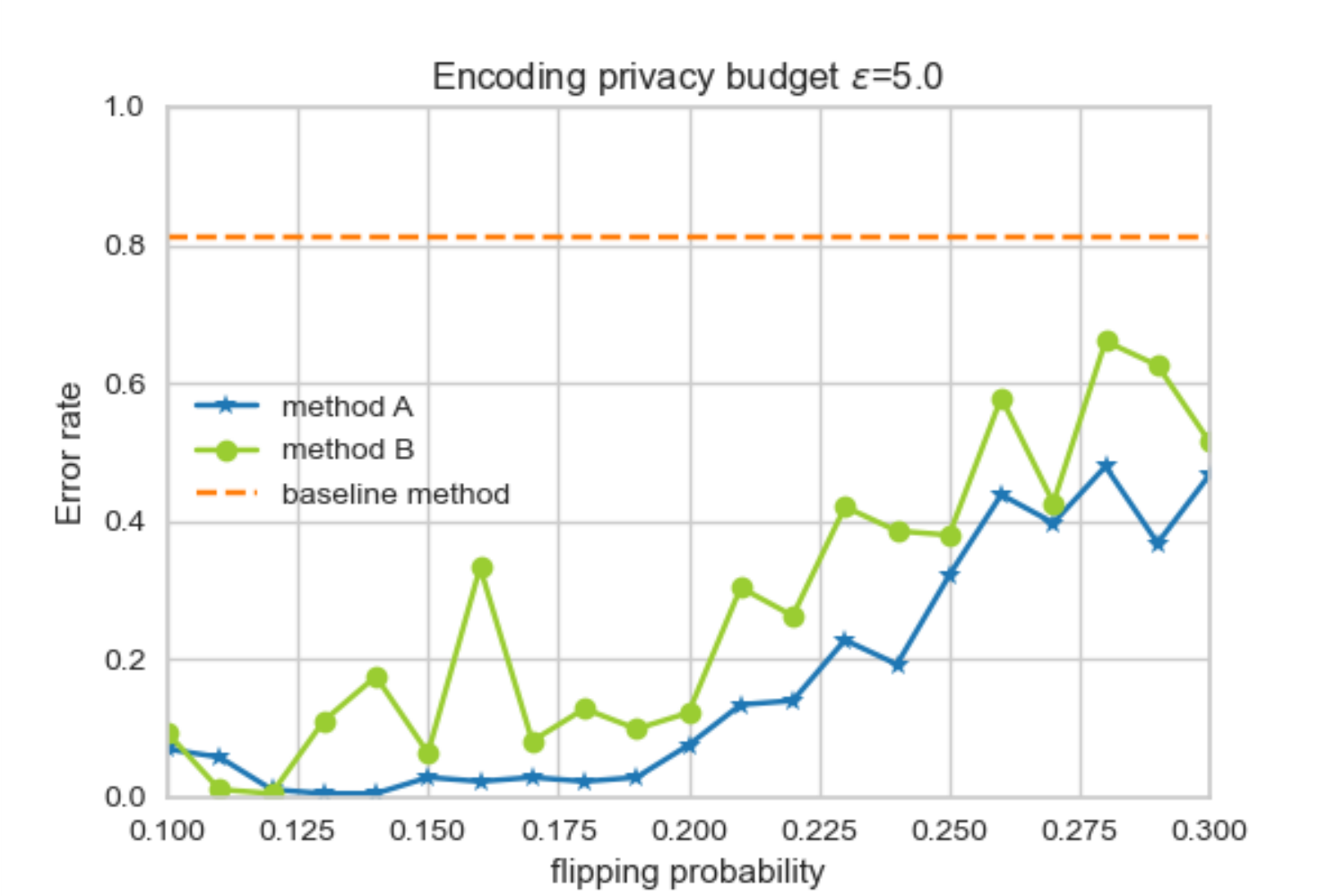} 
 \includegraphics[width=0.32\textwidth]{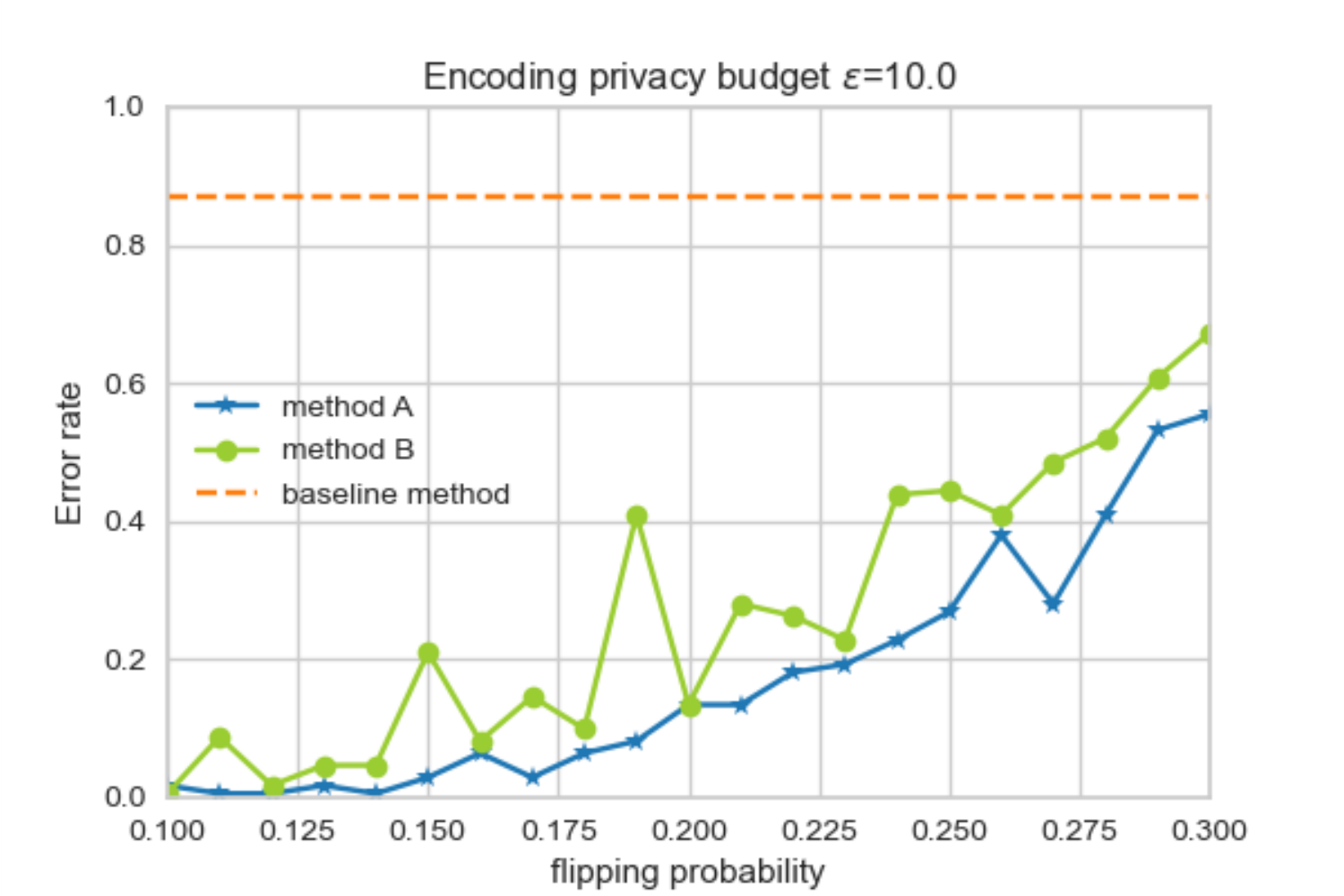}
  \caption{\small{Error rate of cardinality estimation of Method A and Method B with different flipping probabilities compared with the baseline method~\cite{Rou87} on the highly corrupted datasets with $\epsilon=[1.0,2.0,3.0,4.0,5.0,10.0]$. The reference Bloom filters pick ratio is 0.1 and dummy Bloom filters ratio is 0.1 in these experiments.
  }
    }
\label{fig:error_rate_corr40_methodAB}
\end{figure*}

\vspace{2.5mm}
\noindent
\textbf{Discussion:}
We first compare the optimal $k$ value provided by our algorithm with the ground-truth cardinality and evaluate the error in estimation with different flipping probabilities used in the clustering algorithm on the records encoded with different privacy budgets. The results of our methods (Method A and Method B) compared with the baseline method on the clean dataset are shown in Fig.~\ref{fig:card_kval_clean}. As shown, the ground-truth cardinality is {$171$}. When $\epsilon=1.0$, $26\%$ of bits are flipped in the Bloom filters, which means $50$ out of $200$ bits are flipped making most of the Bloom filters unique. Hence, the estimated cardinality by Method A is $200$, leading to an error of $29$. Even when the flipping probability in dummy Bloom filters reduces to $0.0$, the estimated cardinality remains constantly as $200$ with Method A. 
Compared to Method A, Method B has better performance in terms of smaller error of $7$. The reason is that Method B selects a subset of real Bloom filters from all data providers as the reference Bloom filters, whereas Method A randomly generates fake Bloom filters as reference Bloom filters.

With privacy budgets larger than $1.0$ the error in estimation becomes $0.0$, i.e. the optimal $k$ becomes equal to the ground-truth with both our methods. With increasing $\epsilon$, zero or smaller error can be achieved even with larger flipping probabilities. When $\epsilon=10.0$, zero error is achieved with the estimation for flipping probabilities up to $0.15$. However, as can be seen, the baseline method's estimated cardinality (optimal $k$) is far from the ground truth value with all privacy budgets, even with larger $\epsilon$. Our methods significantly outperform the baseline method by providing the optimal $k$ value equal or closer to the ground truth value leading to smaller or zero error rate on all datasets. 
Error rates are shown in Fig.~\ref{fig:error_rate_clean_methodAB}. 
These results show the accuracy of our methods compared to the baseline method.

Given the global and distributed nature of the data provider network, we expect inconsistencies in the PII attributes among others. 
We hence evaluate our proposed methods on corrupted datasets to validate the effectiveness of our methods with inconsistent and low quality data. The estimated cardinalities of Method A and Method B for different flipping probabilities and different privacy budgets compared with the baseline method on corrupted datasets are shown in Fig.~\ref{fig:card_kval_corr20}, and the error rates in estimations are shown in Fig.~\ref{fig:error_rate_corr20_methodAB}.
Both Method A and Method B perform mostly similar on this dataset and achieved similar error rates in cardinality estimation. 
Since the datasets are already corrupted, random generation of reference Bloom filters and sampling a subset of corrupted records' Bloom filters do not make significant difference with these datasets. 
As can be seen in the results, optimal $k$ with a very small error rate closer to $0.0$ can be found with both methods, and with increasing $\epsilon$, optimal error rates of $0.0$ (for smaller flipping probabilities) are achieved. We can observe that achieving a high utility in terms of lower error rate with a small budget becomes challenging with data with errors and variations compared to clean datasets. However, our methods significantly outperform the baseline method and still be able to achieve a small optimal error rate $< 0.02$. For example, when $\epsilon=3.0$, our methods have an optimal error rate of $0.02$ when the flipping probabilities are less than $0.175$, while the Elbow method with silhoutte coefficient metric has an error rate of $0.26$.

Generally, as expected for larger $\epsilon$ (e.g. $\epsilon \ge 3.0$), with increasing flipping probabilities the error rate increases as well. On the other hand, with smaller $\epsilon$ ($\le 2.0$), more Differential privacy noise is added to Bloom filters, and hence, larger flipping probabilities might mean that the added noise in the Bloom filters is reduced, leading to smaller error rate with larger flipping probabilities. This is especially the case with corrupted datasets. Hence, depending on the data quality level and privacy budget, an appropriate flipping probability can be chosen to get the best performance. Optimising the flipping probability based on the privacy and data quality constraints is left as a future work.

We finally evaluate the performances on highly corrupted datasets. The results for Method A and Method B are shown in Fig.~\ref{fig:card_kval_corr40} and Fig.~\ref{fig:error_rate_corr40_methodAB}. Similar to corrupted datasets, both methods provide similar error rates with Method B being slightly better in terms of achieving optimal lower error rates for smaller $\epsilon$ than Method A. 
As can be seen, achieving a small error rate is even more challenging with these highly corrupted datasets compared to corrupted datasets. Our methods still achieve a small error rate, for example less than $0.05$ with $\epsilon=3.0$ in the presence of data errors and variations, whereas the baseline Elbow method achieves $0.8$ error rate.
Providing a higher error rate even when no Differential privacy noise ($\epsilon=10.0$) is added indicates the ineffectiveness of the baseline method. Our methods can achieve a very small error rate even with a small privacy budget and on highly corrupted datasets by fine-tuning the flipping probability parameter depending on the privacy and data quality constraints.

\section{Related Work}
\label{sec:rw}

Different approaches have been proposed to estimate the cardinality of multiple sets, as surveyed in~\cite{Har17}. The naive approach of using a bitmap of size of the universe, where all the bit positions are initialized to $0$ and each item is assigned with a number and therefore corresponding bit position in the bitmap is set to $1$ whenever an item is observed, is not feasible. Sorting is used as another traditional method where the items are sorted to eliminate duplicates in the sets~\cite{Wha90}. However, sorting is an expensive operation for large sets. 
Hashing allows de-duplication of sets in one pass over the sets without sorting them, however, it requires more memory space.

While these methods allow calculating the exact cardinality of sets, they are not only expensive in terms of both memory size and runtime, but also are not effective with real data that contain data errors, typos, and variations. Fuzzy matching for record linkage methods have been investigated~\cite{Chr12}. A Bloom filter is a probabilistic data structure used for efficiently checking set membership~\cite{Bloom70}. This can be used for fuzzy matching problem~\cite{Kum06,Vat11} effectively with appropriate parameter settings for the Bloom filter~\cite{Sch15,Ran13,Vat16}. 
Another general approach is sampling~\cite{Haa95,Gib01,Fla90} which assumes that the sample generally reflects the properties of the whole. Ensuring true randomness is a difficult task, so the success of random sampling may be limited by the selection process and/or the properties of the data itself. Haas et al.~\cite{Haa95} showed that almost all the data need to be sampled in order to bound the estimation error within a small constant, which reflects the problem with sampling-based approaches.

Employing other types of probabilistic data structures, such as Sketches and HyperLogLog, 
is used as an efficient and effective method in several cardinality estimation algorithms. A family of such algorithms are developed by Flajolet and Martin~\cite{Gib16}. HyperLogLog is one of these algorithms that have widely been used in many applications and research~\cite{Che16,Su16,Bal19}. 
%
%
Several recent works have studied privacy-preserving cardinality estimators using probabilistic data structures combined with Differential privacy. Randomized response-based Differentially private algorithms for Bloom filters~\cite{Sta17}, FM-sketch~\cite{Von19}, and K Minimum Values (KMV)-based sketch~\cite{Spa18} have been developed. A recent work has shown that cardinality estimators, such as HyperLogLog and Sketches, do not preserve privacy without impacting the utility to a significant level~\cite{Des19}. 
Further, none of these works allow fuzzy matching to count the cardinalities, making the cardinality estimators not robust or tolerant to data errors and variations in the duplicate records. 

A long line of research has been conducted in privacy-preserving fuzzy matching and linkage over the past three decades, as surveyed in~\cite{Vat13,Vat17,Gko21}. While machine learning-based techniques show promising results in terms of high linkage quality, these are often supervised, i.e they are dependant on significantly large training data and the existence of ground-truth labels. Only few unsupervised techniques have been developed for linkage~\cite{Coh02,Has09c,Sae18,Vat20}. However, most of these techniques either do not consider privacy constraints or are not capable of fine-tuning/optimising the clustering performance due to no labelled data.

\section{Discussion, Limitations and Future Work}
\label{sec:conclusion}

In this paper we have addressed the problem of privacy-preserving cardinality estimation of individuals/entities represented by records from multiple databases. Our proposed method 
uses Bloom filter encoding with local Differential privacy to encode the data and 
unsupervised clustering 
to fuzzy link records and calculate the optimal number of clusters as the cardinality of unique individuals. 
We propose a novel method to calculate the optimal number of clusters in the absence of ground-truth labels of matching and non-matching records, which is often the case with privacy-preserving applications. Our experimental results show that, compared to the baseline Elbow method, our method can achieve a high accuracy of cardinality estimation 
even on corrupted records with a small privacy budget.

In the future, we aim to apply our proposed algorithm for the rare disease patient counting application. Rare disease patient counting application involves small-scale datasets as the number of patients with rare disease is generally small - most rare diseases often have 10, 100 or just 1000 patients spread across the world. However, experimenting on large datasets for other applications of cardinality estimation and improving the scalability to large databases is one important future work. Moreover, optimising the flipping probability constrained on the level of data quality and privacy budget is yet to be investigated and implemented in our algorithm.
Finally, facilitating real-time counting 
and efficient
dynamic updates without requiring to re-do clustering is an important yet challenging research direction.





\bibliographystyle{ACM-Reference-Format}
\bibliography{paper}


\begin{thebibliography}{52}


\ifx \showCODEN    \undefined \def \showCODEN     #1{\unskip}     \fi
\ifx \showDOI      \undefined \def \showDOI       #1{#1}\fi
\ifx \showISBNx    \undefined \def \showISBNx     #1{\unskip}     \fi
\ifx \showISBNxiii \undefined \def \showISBNxiii  #1{\unskip}     \fi
\ifx \showISSN     \undefined \def \showISSN      #1{\unskip}     \fi
\ifx \showLCCN     \undefined \def \showLCCN      #1{\unskip}     \fi
\ifx \shownote     \undefined \def \shownote      #1{#1}          \fi
\ifx \showarticletitle \undefined \def \showarticletitle #1{#1}   \fi
\ifx \showURL      \undefined \def \showURL       {\relax}        \fi
\providecommand\bibfield[2]{#2}
\providecommand\bibinfo[2]{#2}
\providecommand\natexlab[1]{#1}
\providecommand\showeprint[2][]{arXiv:#2}

\bibitem[Apple(2017)]%
        {Dp17}
\bibfield{author}{\bibinfo{person}{Differential Privacy~Team Apple}.}
  \bibinfo{year}{2017}\natexlab{}.
\newblock \showarticletitle{Learning with privacy at scale}.
\newblock \bibinfo{journal}{\emph{Apple Machine Learning Journal - Online at:
  https://machinelearning.apple.com/2017/12/06/learning-with-privacy-at-scale.html}}
  (\bibinfo{year}{2017}).
\newblock


\bibitem[Balasubramaniam and Nandhini(2019)]%
        {Bal19}
\bibfield{author}{\bibinfo{person}{Ramesh Balasubramaniam} {and}
  \bibinfo{person}{K Nandhini}.} \bibinfo{year}{2019}\natexlab{}.
\newblock \showarticletitle{Algorithms Associated with Streaming Data
  Problems}.
\newblock \bibinfo{journal}{\emph{International Journal of Applied Engineering
  Research}} \bibinfo{volume}{14}, \bibinfo{number}{9} (\bibinfo{year}{2019}),
  \bibinfo{pages}{2238--2243}.
\newblock


\bibitem[Bar-Yossef et~al\mbox{.}(2002)]%
        {Bar02a}
\bibfield{author}{\bibinfo{person}{Ziv Bar-Yossef}, \bibinfo{person}{TS
  Jayram}, \bibinfo{person}{Ravi Kumar}, \bibinfo{person}{D Sivakumar}, {and}
  \bibinfo{person}{Luca Trevisan}.} \bibinfo{year}{2002}\natexlab{}.
\newblock \showarticletitle{Counting distinct elements in a data stream}. In
  \bibinfo{booktitle}{\emph{International Workshop on Randomization and
  Approximation Techniques in Computer Science}}. Springer,
  \bibinfo{pages}{1--10}.
\newblock


\bibitem[Bloom(1970)]%
        {Bloom70}
\bibfield{author}{\bibinfo{person}{B.H. Bloom}.}
  \bibinfo{year}{1970}\natexlab{}.
\newblock \showarticletitle{Space/time trade-offs in hash coding with allowable
  errors}.
\newblock \bibinfo{journal}{\emph{Commun. ACM}} \bibinfo{volume}{13},
  \bibinfo{number}{7} (\bibinfo{year}{1970}), \bibinfo{pages}{422--426}.
\newblock


\bibitem[Cali{\'n}ski and Harabasz(1974)]%
        {Cal74}
\bibfield{author}{\bibinfo{person}{Tadeusz Cali{\'n}ski} {and}
  \bibinfo{person}{Jerzy Harabasz}.} \bibinfo{year}{1974}\natexlab{}.
\newblock \showarticletitle{A dendrite method for cluster analysis}.
\newblock \bibinfo{journal}{\emph{Communications in Statistics-theory and
  Methods}} \bibinfo{volume}{3}, \bibinfo{number}{1} (\bibinfo{year}{1974}),
  \bibinfo{pages}{1--27}.
\newblock


\bibitem[Chabchoub and H{\'e}brail(2010)]%
        {Cha10}
\bibfield{author}{\bibinfo{person}{Yousra Chabchoub} {and}
  \bibinfo{person}{Georges H{\'e}brail}.} \bibinfo{year}{2010}\natexlab{}.
\newblock \showarticletitle{Sliding hyperloglog: Estimating cardinality in a
  data stream over a sliding window}. In
  \bibinfo{booktitle}{\emph{International Conference on Data Mining
  Workshops}}. IEEE, \bibinfo{pages}{1297--1303}.
\newblock


\bibitem[Chen et~al\mbox{.}(2016)]%
        {Che16}
\bibfield{author}{\bibinfo{person}{Guoqiang~Jerry Chen},
  \bibinfo{person}{Janet~L Wiener}, \bibinfo{person}{Shridhar Iyer},
  \bibinfo{person}{Anshul Jaiswal}, \bibinfo{person}{Ran Lei},
  \bibinfo{person}{Nikhil Simha}, \bibinfo{person}{Wei Wang},
  \bibinfo{person}{Kevin Wilfong}, \bibinfo{person}{Tim Williamson}, {and}
  \bibinfo{person}{Serhat Yilmaz}.} \bibinfo{year}{2016}\natexlab{}.
\newblock \showarticletitle{Realtime data processing at Facebook}. In
  \bibinfo{booktitle}{\emph{International Conference on Management of Data}}.
  \bibinfo{pages}{1087--1098}.
\newblock


\bibitem[Christen(2012)]%
        {Chr12}
\bibfield{author}{\bibinfo{person}{Peter Christen}.}
  \bibinfo{year}{2012}\natexlab{}.
\newblock \bibinfo{booktitle}{\emph{Data matching - concepts and techniques for
  record linkage, entity resolution, and duplicate detection}}.
\newblock \bibinfo{publisher}{Springer}.
\newblock


\bibitem[Christen et~al\mbox{.}(2018a)]%
        {Chr18b}
\bibfield{author}{\bibinfo{person}{Peter Christen}, \bibinfo{person}{Thilina
  Ranbaduge}, \bibinfo{person}{Dinusha Vatsalan}, {and} \bibinfo{person}{Rainer
  Schnell}.} \bibinfo{year}{2018}\natexlab{a}.
\newblock \showarticletitle{Precise and fast cryptanalysis for {B}loom filter
  based privacy-preserving record linkage}.
\newblock \bibinfo{journal}{\emph{IEEE Transactions on Knowledge and Data
  Engineering}} (\bibinfo{year}{2018}), \bibinfo{pages}{1}.
\newblock


\bibitem[Christen et~al\mbox{.}(2018b)]%
        {Chr18}
\bibfield{author}{\bibinfo{person}{Peter Christen}, \bibinfo{person}{Anushka
  Vidanage}, \bibinfo{person}{Thilina Ranbaduge}, {and} \bibinfo{person}{Rainer
  Schnell}.} \bibinfo{year}{2018}\natexlab{b}.
\newblock \showarticletitle{Pattern-mining based cryptanalysis of {B}loom
  filters for privacy-preserving record linkage}. In
  \bibinfo{booktitle}{\emph{PAKDD, Springer LNAI}}.
  \bibinfo{address}{Melbourne}, \bibinfo{pages}{530--542}.
\newblock


\bibitem[Cohen and Richman(2002)]%
        {Coh02}
\bibfield{author}{\bibinfo{person}{William~W. Cohen} {and}
  \bibinfo{person}{Jacob Richman}.} \bibinfo{year}{2002}\natexlab{}.
\newblock \showarticletitle{Learning to Match and Cluster Large
  High-dimensional Data Sets for Data Integration}. In
  \bibinfo{booktitle}{\emph{ACM SIGKDD}}. \bibinfo{pages}{475--480}.
\newblock


\bibitem[Desfontaines et~al\mbox{.}(2019)]%
        {Des19}
\bibfield{author}{\bibinfo{person}{Damien Desfontaines},
  \bibinfo{person}{Andreas Lochbihler}, {and} \bibinfo{person}{David Basin}.}
  \bibinfo{year}{2019}\natexlab{}.
\newblock \showarticletitle{Cardinality estimators do not preserve privacy}.
\newblock \bibinfo{journal}{\emph{Proceedings on Privacy Enhancing
  Technologies}} \bibinfo{volume}{2019}, \bibinfo{number}{2}
  (\bibinfo{year}{2019}), \bibinfo{pages}{26--46}.
\newblock


\bibitem[Dinh et~al\mbox{.}(2019)]%
        {Din19}
\bibfield{author}{\bibinfo{person}{Duy-Tai Dinh}, \bibinfo{person}{Tsutomu
  Fujinami}, {and} \bibinfo{person}{Van-Nam Huynh}.}
  \bibinfo{year}{2019}\natexlab{}.
\newblock \showarticletitle{Estimating the optimal number of clusters in
  categorical data clustering by silhouette coefficient}. In
  \bibinfo{booktitle}{\emph{International Symposium on Knowledge and Systems
  Sciences}}. Springer, \bibinfo{pages}{1--17}.
\newblock


\bibitem[Dwork(2006)]%
        {Dwo06}
\bibfield{author}{\bibinfo{person}{C. Dwork}.} \bibinfo{year}{2006}\natexlab{}.
\newblock \showarticletitle{Differential privacy}.
\newblock \bibinfo{journal}{\emph{International Colloquium on Automata,
  Languages and Programming}} (\bibinfo{year}{2006}), \bibinfo{pages}{1--12}.
\newblock


\bibitem[Dwork(2008)]%
        {Dwo08}
\bibfield{author}{\bibinfo{person}{Cynthia Dwork}.}
  \bibinfo{year}{2008}\natexlab{}.
\newblock \showarticletitle{Differential privacy: A survey of results}.
\newblock In \bibinfo{booktitle}{\emph{Theory and Applications of Models of
  Computation}}. \bibinfo{publisher}{Springer}, \bibinfo{pages}{1--19}.
\newblock


\bibitem[Dwork et~al\mbox{.}(2010)]%
        {Dwo10}
\bibfield{author}{\bibinfo{person}{Cynthia Dwork}, \bibinfo{person}{Moni Naor},
  \bibinfo{person}{Toniann Pitassi}, \bibinfo{person}{Guy~N Rothblum}, {and}
  \bibinfo{person}{Sergey Yekhanin}.} \bibinfo{year}{2010}\natexlab{}.
\newblock \showarticletitle{Pan-Private Streaming Algorithms.}. In
  \bibinfo{booktitle}{\emph{ICS}}. \bibinfo{pages}{66--80}.
\newblock


\bibitem[Erlingsson et~al\mbox{.}(2014)]%
        {Erl14}
\bibfield{author}{\bibinfo{person}{{\'U}lfar Erlingsson},
  \bibinfo{person}{Vasyl Pihur}, {and} \bibinfo{person}{Aleksandra Korolova}.}
  \bibinfo{year}{2014}\natexlab{}.
\newblock \showarticletitle{Rappor: Randomized aggregatable privacy-preserving
  ordinal response}. In \bibinfo{booktitle}{\emph{SIGSAC conference on computer
  and communications security}}. ACM, \bibinfo{pages}{1054--1067}.
\newblock


\bibitem[Ertl(2017)]%
        {Ert17}
\bibfield{author}{\bibinfo{person}{Otmar Ertl}.}
  \bibinfo{year}{2017}\natexlab{}.
\newblock \showarticletitle{New cardinality estimation algorithms for
  HyperLogLog sketches}.
\newblock \bibinfo{journal}{\emph{arXiv preprint arXiv:1702.01284}}
  (\bibinfo{year}{2017}).
\newblock


\bibitem[Evfimievski et~al\mbox{.}(2003)]%
        {Evf03}
\bibfield{author}{\bibinfo{person}{Alexandre Evfimievski},
  \bibinfo{person}{Johannes Gehrke}, {and} \bibinfo{person}{Ramakrishnan
  Srikant}.} \bibinfo{year}{2003}\natexlab{}.
\newblock \showarticletitle{Limiting privacy breaches in privacy preserving
  data mining}. In \bibinfo{booktitle}{\emph{ACM SIGMOD-SIGACT-SIGART symposium
  on Principles of database systems}}. \bibinfo{pages}{211--222}.
\newblock


\bibitem[Flajolet(1990)]%
        {Fla90}
\bibfield{author}{\bibinfo{person}{Philippe Flajolet}.}
  \bibinfo{year}{1990}\natexlab{}.
\newblock \showarticletitle{On adaptive sampling}.
\newblock \bibinfo{journal}{\emph{Computing}} \bibinfo{volume}{43},
  \bibinfo{number}{4} (\bibinfo{year}{1990}), \bibinfo{pages}{391--400}.
\newblock


\bibitem[Flajolet et~al\mbox{.}(2007)]%
        {Fla07}
\bibfield{author}{\bibinfo{person}{Philippe Flajolet},
  \bibinfo{person}{{\'E}ric Fusy}, \bibinfo{person}{Olivier Gandouet}, {and}
  \bibinfo{person}{Fr{\'e}d{\'e}ric Meunier}.} \bibinfo{year}{2007}\natexlab{}.
\newblock \showarticletitle{Hyperloglog: the analysis of a near-optimal
  cardinality estimation algorithm}. In \bibinfo{booktitle}{\emph{Conference on
  Analysis of Algorithms (AofA)}}. \bibinfo{address}{Nancy, France}.
\newblock


\bibitem[Gibbons(2001)]%
        {Gib01}
\bibfield{author}{\bibinfo{person}{Phillip~B Gibbons}.}
  \bibinfo{year}{2001}\natexlab{}.
\newblock \showarticletitle{Distinct sampling for highly-accurate answers to
  distinct values queries and event reports}. In
  \bibinfo{booktitle}{\emph{VLDB}}, Vol.~\bibinfo{volume}{1}.
  \bibinfo{pages}{541--550}.
\newblock


\bibitem[Gibbons(2016)]%
        {Gib16}
\bibfield{author}{\bibinfo{person}{Phillip~B Gibbons}.}
  \bibinfo{year}{2016}\natexlab{}.
\newblock \showarticletitle{Distinct-values estimation over data streams}.
\newblock In \bibinfo{booktitle}{\emph{Data Stream Management}}.
  \bibinfo{publisher}{Springer}, \bibinfo{pages}{121--147}.
\newblock


\bibitem[Gkoulalas-Divanis et~al\mbox{.}(2021)]%
        {Gko21}
\bibfield{author}{\bibinfo{person}{A. Gkoulalas-Divanis}, \bibinfo{person}{D.
  Vatsalan}, \bibinfo{person}{D. Karapiperis}, {and} \bibinfo{person}{M.
  Kantarcioglu}.} \bibinfo{year}{2021}\natexlab{}.
\newblock \showarticletitle{Modern Privacy-Preserving Record Linkage
  Techniques: An Overview}.
\newblock \bibinfo{journal}{\emph{IEEE TIFS}} (\bibinfo{year}{2021}).
\newblock


\bibitem[Golov et~al\mbox{.}(2019)]%
        {Gol19}
\bibfield{author}{\bibinfo{person}{Nikolay Golov}, \bibinfo{person}{Alexander
  Filatov}, {and} \bibinfo{person}{Sergey Bruskin}.}
  \bibinfo{year}{2019}\natexlab{}.
\newblock \showarticletitle{Efficient Exact Algorithm for Count Distinct
  Problem}. In \bibinfo{booktitle}{\emph{International Workshop on Computer
  Algebra in Scientific Computing}}. Springer, \bibinfo{pages}{67--77}.
\newblock


\bibitem[Greenberg(2016)]%
        {Gre16}
\bibfield{author}{\bibinfo{person}{Andy Greenberg}.}
  \bibinfo{year}{2016}\natexlab{}.
\newblock \showarticletitle{Apple’s ‘differential privacy’is about
  collecting your data—but not your data}.
\newblock \bibinfo{journal}{\emph{Wired, June}}  \bibinfo{volume}{13}
  (\bibinfo{year}{2016}).
\newblock


\bibitem[Haas et~al\mbox{.}(1995)]%
        {Haa95}
\bibfield{author}{\bibinfo{person}{Peter~J Haas}, \bibinfo{person}{Jeffrey~F
  Naughton}, \bibinfo{person}{S Seshadri}, {and} \bibinfo{person}{Lynne
  Stokes}.} \bibinfo{year}{1995}\natexlab{}.
\newblock \showarticletitle{Sampling-based estimation of the number of distinct
  values of an attribute}. In \bibinfo{booktitle}{\emph{VLDB}},
  Vol.~\bibinfo{volume}{95}. \bibinfo{pages}{311--322}.
\newblock


\bibitem[Harmouch and Naumann(2017)]%
        {Har17}
\bibfield{author}{\bibinfo{person}{Hazar Harmouch} {and} \bibinfo{person}{Felix
  Naumann}.} \bibinfo{year}{2017}\natexlab{}.
\newblock \showarticletitle{Cardinality estimation: An experimental survey}.
\newblock \bibinfo{journal}{\emph{Proceedings of the VLDB Endowment}}
  \bibinfo{volume}{11}, \bibinfo{number}{4} (\bibinfo{year}{2017}),
  \bibinfo{pages}{499--512}.
\newblock


\bibitem[Hassanzadeh et~al\mbox{.}(2009)]%
        {Has09c}
\bibfield{author}{\bibinfo{person}{Oktie Hassanzadeh}, \bibinfo{person}{Fei
  Chiang}, \bibinfo{person}{Hyun~Chul Lee}, {and} \bibinfo{person}{Ren{\'e}e~J
  Miller}.} \bibinfo{year}{2009}\natexlab{}.
\newblock \showarticletitle{Framework for evaluating clustering algorithms in
  duplicate detection}.
\newblock \bibinfo{journal}{\emph{Proceedings of the Very Large Database
  Endowment}} \bibinfo{volume}{2}, \bibinfo{number}{1} (\bibinfo{year}{2009}),
  \bibinfo{pages}{1282--1293}.
\newblock


\bibitem[Heule et~al\mbox{.}(2013)]%
        {Heu13}
\bibfield{author}{\bibinfo{person}{Stefan Heule}, \bibinfo{person}{Marc
  Nunkesser}, {and} \bibinfo{person}{Alexander Hall}.}
  \bibinfo{year}{2013}\natexlab{}.
\newblock \showarticletitle{HyperLogLog in practice: algorithmic engineering of
  a state of the art cardinality estimation algorithm}. In
  \bibinfo{booktitle}{\emph{International Conference on Extending Database
  Technology}}. \bibinfo{pages}{683--692}.
\newblock


\bibitem[Jansen(2006)]%
        {Jan06}
\bibfield{author}{\bibinfo{person}{Bernard~J Jansen}.}
  \bibinfo{year}{2006}\natexlab{}.
\newblock \showarticletitle{Search log analysis: What it is, what's been done,
  how to do it}.
\newblock \bibinfo{journal}{\emph{Library \& information science research}}
  \bibinfo{volume}{28}, \bibinfo{number}{3} (\bibinfo{year}{2006}),
  \bibinfo{pages}{407--432}.
\newblock


\bibitem[Kumar et~al\mbox{.}(2006)]%
        {Kum06}
\bibfield{author}{\bibinfo{person}{Abhishek Kumar}, \bibinfo{person}{Jun Xu},
  {and} \bibinfo{person}{Jia Wang}.} \bibinfo{year}{2006}\natexlab{}.
\newblock \showarticletitle{Space-code bloom filter for efficient per-flow
  traffic measurement}.
\newblock \bibinfo{journal}{\emph{IEEE Journal on Selected Areas in
  Communications}} \bibinfo{volume}{24}, \bibinfo{number}{12}
  (\bibinfo{year}{2006}), \bibinfo{pages}{2327--2339}.
\newblock


\bibitem[Randall et~al\mbox{.}(2014)]%
        {Ran13}
\bibfield{author}{\bibinfo{person}{Sean~M Randall}, \bibinfo{person}{Anna~M
  Ferrante}, \bibinfo{person}{James~H Boyd}, {and} \bibinfo{person}{James~B
  Semmens}.} \bibinfo{year}{2014}\natexlab{}.
\newblock \showarticletitle{Privacy-preserving record linkage on large real
  world datasets}.
\newblock \bibinfo{journal}{\emph{Journal of Biomedical Informatics}}
  \bibinfo{volume}{50}, \bibinfo{number}{1} (\bibinfo{year}{2014}),
  \bibinfo{pages}{1}.
\newblock


\bibitem[Rousseeuw(1987)]%
        {Rou87}
\bibfield{author}{\bibinfo{person}{Peter~J Rousseeuw}.}
  \bibinfo{year}{1987}\natexlab{}.
\newblock \showarticletitle{Silhouettes: a graphical aid to the interpretation
  and validation of cluster analysis}.
\newblock \bibinfo{journal}{\emph{Journal of computational and applied
  mathematics}}  \bibinfo{volume}{20} (\bibinfo{year}{1987}),
  \bibinfo{pages}{53--65}.
\newblock


\bibitem[Saeedi et~al\mbox{.}(2018)]%
        {Sae18}
\bibfield{author}{\bibinfo{person}{Alieh Saeedi}, \bibinfo{person}{Markus
  Nentwig}, \bibinfo{person}{Eric Peukert}, {and} \bibinfo{person}{Erhard
  Rahm}.} \bibinfo{year}{2018}\natexlab{}.
\newblock \showarticletitle{Scalable matching and clustering of entities with
  FAMER}.
\newblock \bibinfo{journal}{\emph{Complex Systems Informatics and Modeling
  Quarterly}} \bibinfo{number}{16} (\bibinfo{year}{2018}),
  \bibinfo{pages}{61--83}.
\newblock


\bibitem[Sakate et~al\mbox{.}(2018)]%
        {pmid29866013}
\bibfield{author}{\bibinfo{person}{R. Sakate}, \bibinfo{person}{A. Fukagawa},
  \bibinfo{person}{Y. Takagaki}, \bibinfo{person}{H. Okura}, {and}
  \bibinfo{person}{A. Matsuyama}.} \bibinfo{year}{2018}\natexlab{}.
\newblock \showarticletitle{{{T}rends of {C}linical {T}rials for {D}rug
  {D}evelopment in {R}are {D}iseases}}.
\newblock \bibinfo{journal}{\emph{Curr Clin Pharmacol}} \bibinfo{volume}{13},
  \bibinfo{number}{3} (\bibinfo{year}{2018}), \bibinfo{pages}{199--208}.
\newblock


\bibitem[Schnell(2016)]%
        {Sch15}
\bibfield{author}{\bibinfo{person}{Rainer Schnell}.}
  \bibinfo{year}{2016}\natexlab{}.
\newblock \showarticletitle{Privacy preserving record linkage}.
\newblock In \bibinfo{booktitle}{\emph{Methodological developments in data
  linkage}}, \bibfield{editor}{\bibinfo{person}{Katie Harron},
  \bibinfo{person}{Harvey Goldstein}, {and} \bibinfo{person}{Chris Dibben}}
  (Eds.). \bibinfo{publisher}{Wiley}, \bibinfo{address}{Chichester},
  \bibinfo{pages}{201--225}.
\newblock


\bibitem[Shao et~al\mbox{.}(2017)]%
        {Sha17}
\bibfield{author}{\bibinfo{person}{Chengcheng Shao},
  \bibinfo{person}{Giovanni~Luca Ciampaglia}, \bibinfo{person}{Onur Varol},
  \bibinfo{person}{Alessandro Flammini}, {and} \bibinfo{person}{Filippo
  Menczer}.} \bibinfo{year}{2017}\natexlab{}.
\newblock \showarticletitle{The spread of fake news by social bots}.
\newblock \bibinfo{journal}{\emph{arXiv preprint arXiv:1707.07592}}
  \bibinfo{volume}{96} (\bibinfo{year}{2017}), \bibinfo{pages}{104}.
\newblock


\bibitem[Sparka et~al\mbox{.}(2018)]%
        {Spa18}
\bibfield{author}{\bibinfo{person}{Hagen Sparka}, \bibinfo{person}{Florian
  Tschorsch}, {and} \bibinfo{person}{Bj{\"o}rn Scheuermann}.}
  \bibinfo{year}{2018}\natexlab{}.
\newblock \showarticletitle{P2KMV: a privacy-preserving counting sketch for
  efficient and accurate set intersection cardinality estimations}.
\newblock  (\bibinfo{year}{2018}).
\newblock


\bibitem[Stanojevic et~al\mbox{.}(2017)]%
        {Sta17}
\bibfield{author}{\bibinfo{person}{Rade Stanojevic}, \bibinfo{person}{Mohamed
  Nabeel}, {and} \bibinfo{person}{Ting Yu}.} \bibinfo{year}{2017}\natexlab{}.
\newblock \showarticletitle{Distributed cardinality estimation of set
  operations with differential privacy}. In \bibinfo{booktitle}{\emph{2017 IEEE
  Symposium on Privacy-Aware Computing (PAC)}}. IEEE, \bibinfo{pages}{37--48}.
\newblock


\bibitem[Su et~al\mbox{.}(2016)]%
        {Su16}
\bibfield{author}{\bibinfo{person}{Hong Su}, \bibinfo{person}{Mohamed Zait},
  \bibinfo{person}{Vladimir Barri{\`e}re}, \bibinfo{person}{Joseph Torres},
  {and} \bibinfo{person}{Andre Menck}.} \bibinfo{year}{2016}\natexlab{}.
\newblock \showarticletitle{Approximate aggregates in oracle 12c}. In
  \bibinfo{booktitle}{\emph{ACM International on Conference on Information and
  Knowledge Management}}. \bibinfo{pages}{1603--1612}.
\newblock


\bibitem[Tran et~al\mbox{.}(2013)]%
        {Tra13}
\bibfield{author}{\bibinfo{person}{Khoi-Nguyen Tran}, \bibinfo{person}{Dinusha
  Vatsalan}, {and} \bibinfo{person}{Peter Christen}.}
  \bibinfo{year}{2013}\natexlab{}.
\newblock \showarticletitle{{GeCo}: an online personal data generator and
  corruptor}. In \bibinfo{booktitle}{\emph{ACM Conference in Knowledge
  Management}}. \bibinfo{address}{San Francisco}, \bibinfo{pages}{2473--2476}.
\newblock


\bibitem[Vatsalan and Christen(2016)]%
        {Vat16}
\bibfield{author}{\bibinfo{person}{Dinusha Vatsalan} {and}
  \bibinfo{person}{Peter Christen}.} \bibinfo{year}{2016}\natexlab{}.
\newblock \showarticletitle{Privacy-preserving matching of similar patients}.
\newblock \bibinfo{journal}{\emph{Journal of Biomedical Informatics}}
  \bibinfo{volume}{59} (\bibinfo{year}{2016}), \bibinfo{pages}{285--298}.
\newblock


\bibitem[Vatsalan and Christen(2017)]%
        {Vat17}
\bibfield{author}{\bibinfo{person}{Dinusha Vatsalan} {and}
  \bibinfo{person}{Peter Christen}.} \bibinfo{year}{2017}\natexlab{}.
\newblock \showarticletitle{Scalable privacy-preserving linking of multiple
  databases using counting Bloom filters}.
\newblock \bibinfo{journal}{\emph{arXiv preprint arXiv:1701.01232}}
  (\bibinfo{year}{2017}).
\newblock


\bibitem[Vatsalan et~al\mbox{.}(2020)]%
        {Vat20}
\bibfield{author}{\bibinfo{person}{Dinusha Vatsalan}, \bibinfo{person}{Peter
  Christen}, {and} \bibinfo{person}{Erhard Rahm}.}
  \bibinfo{year}{2020}\natexlab{}.
\newblock \showarticletitle{Incremental clustering techniques for multi-party
  Privacy-Preserving Record Linkage}.
\newblock \bibinfo{journal}{\emph{Data \& Knowledge Engineering}}
  (\bibinfo{year}{2020}).
\newblock


\bibitem[Vatsalan et~al\mbox{.}(2011)]%
        {Vat11}
\bibfield{author}{\bibinfo{person}{D. Vatsalan}, \bibinfo{person}{P. Christen},
  {and} \bibinfo{person}{Vassilios~S. Verykios}.}
  \bibinfo{year}{2011}\natexlab{}.
\newblock \showarticletitle{An Efficient Two-Party Protocol for Approximate
  Matching in Private Record Linkage}. In
  \bibinfo{booktitle}{\emph{Australasian Data Mining Conference}}.
  \bibinfo{address}{Ballarat, Australia}.
\newblock


\bibitem[Vatsalan et~al\mbox{.}(2013)]%
        {Vat13}
\bibfield{author}{\bibinfo{person}{Dinusha Vatsalan}, \bibinfo{person}{Peter
  Christen}, {and} \bibinfo{person}{Vassilios~S. Verykios}.}
  \bibinfo{year}{2013}\natexlab{}.
\newblock \showarticletitle{A Taxonomy of Privacy-Preserving Record Linkage
  Techniques}.
\newblock \bibinfo{journal}{\emph{Information Systems}} \bibinfo{volume}{38},
  \bibinfo{number}{6} (\bibinfo{year}{2013}), \bibinfo{pages}{946--969}.
\newblock


\bibitem[von Voigt and Tschorsch(2019)]%
        {Von19}
\bibfield{author}{\bibinfo{person}{Saskia~Nu{\~n}ez von Voigt} {and}
  \bibinfo{person}{Florian Tschorsch}.} \bibinfo{year}{2019}\natexlab{}.
\newblock \showarticletitle{RRTxFM: Probabilistic Counting for Differentially
  Private Statistics}. In \bibinfo{booktitle}{\emph{Conference on e-Business,
  e-Services and e-Society}}. Springer, \bibinfo{pages}{86--98}.
\newblock


\bibitem[Wang et~al\mbox{.}(2017)]%
        {Wan17}
\bibfield{author}{\bibinfo{person}{Gang Wang}, \bibinfo{person}{Xinyi Zhang},
  \bibinfo{person}{Shiliang Tang}, \bibinfo{person}{Christo Wilson},
  \bibinfo{person}{Haitao Zheng}, {and} \bibinfo{person}{Ben~Y Zhao}.}
  \bibinfo{year}{2017}\natexlab{}.
\newblock \showarticletitle{Clickstream user behavior models}.
\newblock \bibinfo{journal}{\emph{ACM Transactions on the Web (TWEB)}}
  \bibinfo{volume}{11}, \bibinfo{number}{4} (\bibinfo{year}{2017}),
  \bibinfo{pages}{1--37}.
\newblock


\bibitem[Wang and Xu(2019)]%
        {Wan19}
\bibfield{author}{\bibinfo{person}{Xu Wang} {and} \bibinfo{person}{Yusheng
  Xu}.} \bibinfo{year}{2019}\natexlab{}.
\newblock \showarticletitle{An improved index for clustering validation based
  on Silhouette index and Calinski-Harabasz index}. In
  \bibinfo{booktitle}{\emph{IOP Conference Series: Materials Science and
  Engineering}}, Vol.~\bibinfo{volume}{569}. IOP Publishing,
  \bibinfo{pages}{052024}.
\newblock


\bibitem[Warner(1965)]%
        {War65}
\bibfield{author}{\bibinfo{person}{Stanley~L Warner}.}
  \bibinfo{year}{1965}\natexlab{}.
\newblock \showarticletitle{Randomized response: A survey technique for
  eliminating evasive answer bias}.
\newblock \bibinfo{journal}{\emph{J. Amer. Statist. Assoc.}}
  \bibinfo{volume}{60}, \bibinfo{number}{309} (\bibinfo{year}{1965}),
  \bibinfo{pages}{63--69}.
\newblock


\bibitem[Whang et~al\mbox{.}(1990)]%
        {Wha90}
\bibfield{author}{\bibinfo{person}{Kyu-Young Whang}, \bibinfo{person}{Brad~T
  Vander-Zanden}, {and} \bibinfo{person}{Howard~M Taylor}.}
  \bibinfo{year}{1990}\natexlab{}.
\newblock \showarticletitle{A linear-time probabilistic counting algorithm for
  database applications}.
\newblock \bibinfo{journal}{\emph{ACM Transactions on Database Systems (TODS)}}
  \bibinfo{volume}{15}, \bibinfo{number}{2} (\bibinfo{year}{1990}),
  \bibinfo{pages}{208--229}.
\newblock


\end{thebibliography}

\end{document}